\documentclass[a4paper,11pt]{article}

\usepackage{jheppub}
\usepackage{hyperref}
\usepackage{orcidlink}

\usepackage{graphicx,amssymb,amsmath,amsfonts}
\usepackage{hyperref}
\usepackage{enumerate}
\usepackage{xcolor}

\numberwithin{equation}{section}

\usepackage[utf8]{inputenc}

\def\be{\begin{equation}}
\def\ee{\end{equation}}

\newcommand{\avg}[1]{\langle #1 \rangle}

\newmuskip\pFqmuskip

\newcommand*\pFq[6][8]{%
  \begingroup % only local assignments
  \pFqmuskip=#1mu\relax
  \mathchardef\normalcomma=\mathcode`,
  \mathcode`\,=\string"8000
  \begingroup\lccode`\~=`\,
  \lowercase{\endgroup\let~}\pFqcomma
  {}_{#2}F_{#3}{\left(\genfrac..{0pt}{}{#4}{#5}\Big | #6\right)}%
  \endgroup
}
\newcommand{\pFqcomma}{{\normalcomma}\mskip\pFqmuskip}
\newcommand{\pt}[1]{\left(#1\right)}

\newcommand{\pmat}{\begin{pmatrix}}
\newcommand{\fpmat}{\end{pmatrix}}
\newcommand{\eq}{\begin{equation}}
\newcommand{\feq}{\end{equation}}
\newcommand{\cas}{\begin{cases}}
\newcommand{\fcas}{\end{cases}}

\newcommand{\eqarray}{\begin{eqnarray}}
\newcommand{\feqarray}{\end{eqnarray}}
\newcommand{\Tr}[1]{\operatorname{Tr}\pt{#1}}

		\def\b{\beta}

\newcommand{\sff}{\mathrm{SFF}}

\newcommand{\vlam}{{\vec\lambda}}

\newcommand{\pri}{^{\prime}}

\title{Multi-dimensional chaos I: \\ Classical and quantum mechanics}

\author[a,b]{Massimo~Bianchi\,\orcidlink{0000-0002-7591-3870},}
\affiliation[a]{Dipartimento di Fisica, Universit\`a di Roma Tor Vergata,\\
Via della Ricerca Scientifica 1, 00133, Roma, Italy}

\affiliation[b]{INFN sezione di Roma Tor Vergata, \\
Via della Ricerca Scientifica 1, 00133 Roma, Italy}
\emailAdd{massimo.bianchi@roma2.infn.it}

\author[c]{Maurizio~Firrotta\,\orcidlink{0000-0003-0849-5995},}
\affiliation[c]{Crete Center for Theoretical Physics, Institute for Theoretical and Computational Physics,
Department of Physics, Voutes University Campus,
GR-70013, Vasilika Vouton, Heraklion, Greece} 
\emailAdd{mfirrotta@physics.uoc.gr}

\author[d]{Jacob~Sonnenschein\,\orcidlink{0000-0001-9283-1554},}
\affiliation[d]{The Raymond and Beverly Sackler School of Physics and Astronomy,\\
Tel Aviv University, Ramat Aviv 69978, Tel Aviv, Israel}
\emailAdd{cobi@tauex.tau.ac.il}

\author[e]{and Dorin~Weissman\,\orcidlink{0000-0001-9697-0252}}
\affiliation[e]{INFN Sezione di Napoli,\\Monte S. Angelo, Via Cintia, 80126 Naples, Italy}
\emailAdd{dorin.weissman@na.infn.it}

\abstract{We introduce the notion of multi-dimensional chaos that applies to processes described  by erratic functions of several dynamical variables.  We employ this concept in the interpretation of classical and quantum scattering off a pinball system.  In the former case it is illustrated by means of two-dimensional plots of the scattering angle and of the number of bounces.  We draw similar patterns for the quantum differential cross-section for various geometries of the disks. We find that the eigenvalues of the $S$-matrix are distributed according to the Circular Orthogonal Ensemble (COE) in random matrix theory (RMT), provided the setup be asymmetric and the wave-number be large enough. We then consider the electric potential associated with charges randomly located on a plane as a toy model that generalizes the scattering from a leaky torus. We propose several  methods to analyze the distribution of spacings between the extrema of such functions. We show that these follow a repulsive Gaussian $\beta$-ensemble distribution even for  Poisson-distributed positions of the charges. A generalization of the spectral form factor is introduced and determined.   We apply these methods to the cases of a chaotic $S$-matrix and of the quantum pinball scattering. The spacings between nearest neighbor extrema points and ratios between adjacent spacings follow a logistic and Beta distributions correspondingly. We conjecture about a potential relation with random tensor theory.
}

\begin{document}

\maketitle
\flushbottom
\clearpage
%%%

\section{Introduction}

Chaotic processes often admit a description in terms of  erratic functions of a certain continuous variable. Examples  of such a behavior are the scattering angle as a function of the incident angle in a  pinball experiment \cite{Gaspard:1989exq}, the leaky torus phase shift \cite{Gutzwiller:1983} as a function of the wave-number and the decay amplitude of highly excited string state (HES)  into two low-mass states or a scattering amplitude of a HES with three low-mass states as a function of an angle \cite{Gross:2021gsj,Rosenhaus:2021xhm}.

In \cite{Bianchi:2022mhs} and \cite{Bianchi:2023uby} we proposed a novel measure of chaotic behavior based on a map between the set of maxima of erratic  scattering amplitudes and the eigenvalues of random matrices. The spacings  between adjacent peaks  and their ratios were shown to admit a gaussian $\beta$-ensemble distribution in a similar manner to the Wigner-Dyson  map of the energy eigenvalues of chaotic complex nuclei and molecules \cite{Dyson:1962es}.

An obvious generalization  of  erratic functions of one variable is that of erratic functions of several continuous variables. In many systems it is a necessary generalization, since a scattering process can depend on many variables. For pinball scattering it takes the form of the dependence of the scattering angle on both the incident angle and the impact parameter, while for HES scatterings it takes the form of the dependence of the amplitude on several kinematical factors. 

It may look like a trivial generalization but in fact it is not. The extremum points of erratic functions in the ``one-dimensional'' case can be potentially replaced in the multi-dimensional case by a combination of extremum points, saddle points, curves, in particular ridges and valleys, etc.  Such ``topographical structures" exist also for non-erratic functions of two or more variables. Thus, the main question behind our investigation  is whether there are clear signs and indications of  multi-dimensional chaotic behavior that do not show up in integrable cases, and how to quantify them.

In the present paper we take the first step in answering this question by analyzing  the structure of the extremum points of two-dimensional patterns. If in the one-dimensional case we denote the locations of these points by $\lambda_i$, for a  multi-dimensional case  we will have a set of vectors $\vec \lambda_i$. The study of other structures, different from the extremum points, will be deferred to a future research work. The analysis of the peaks is richer in two-dimensional cases than in one-dimensional cases. Clearly, in a two-dimensional problem, we can fix one of the two kinematic variables and then switch back to the one-dimensional case. However, we can also define, for example, nearest neighbors (NN) pairs in the plane and study the distribution of the spacings between them. In this way we can determine for instance the ``repulsion" between the extremum points in two dimensions similar to the level-repulsion in one dimension as for the eigenvalues of random matrices. 

In the present work we analyze two-dimensional chaotic processes in the context of the two following frameworks: the classical \cite{Gaspard:1989cla} and the quantum \cite{Gaspard:1989exq} pinball scattering, and a toy model that generalizes the leaky torus scattering \cite{Gutzwiller:1983}. The latter takes the form of an electric potential produced by a set of charges located on a plane in three spatial dimensions.  

For the classical pinball scattering we determine the scattering angle and number of collisions as a function of the incident angle and the impact parameter which is expressed also in terms of an angle. In fact the emerging patterns were studied intensively in the past, see \cite{Sweet:1999} and references therein. To understand the patterns associated with chaotic processes, we compare the patterns of the three-disk chaotic cases to those of the non-chaotic scattering from two disks. In this way we identify the regions in the pattern that can be associated with the chaotic behavior.

It is a long-standing conjecture \cite{Blumel:1990zz} that the eigenvalues of the  $S$-matrices  of quantum chaotic processes could be mapped into those of random matrices, specifically those of the circular ensembles of random unitary matrices.\footnote{Chaotic $S$-matrix theory has been intensively investigated, see for instance  \cite{Beenakker:1996esw,Haake:2010fgh} and references therein.}
Surprisingly, we found that for the fully symmetric setup, in spite of the fact that classical scattering is chaotic, the eigenvalues of the quantum $S$-matrix follow a non-chaotic distribution, namely a Poisson one. On the other hand for asymmetric setups and for a large enough wave-number $k$ the distribution of the $S$-matrix eigenvalues  is the expected COE (circular orthogonal ensemble) distribution, associated with chaotic systems.\footnote{A similar observation was made in \cite{Balasubramanian:2024ghv} regarding the level spacing statistics in triangular billiards systems.}

We propose four different methods of analyzing the distributions of extremum points in two dimensions. In particular we suggest to study the distribution of nearest-neighbor spacings, as well as the distribution of consecutive spacing ratios along a path. To that end we propose a simple algorithm that defines the path that visits all the eigenvalues once, in a determined order, once an initial point is specified. We then apply this method to a toy model where the locations of the peaks are chosen randomly, to a general two-dimensional scattering amplitude driven by a random $S$-matrix, and to the pinball scattering.  
For the case of the toy model we found that the spacings and the ratios follow  certain Gaussian $\beta$-ensemble distributions even for the case where the locations of the charges are  Poisson distributed. On the other hand for the chaotic $S$-matrix and for the pinball scattering we found logistic distribution and Beta distributions for the spacings and ratios correspondingly.

Some similar measures in two dimensions has been analyzed in \cite{Sa:2020fpf} and related works, in the context of dissipative quantum chaotic systems. There one encounters eigenvalues of the Hamiltonian which lie on the complex plane and can compare with Laguerre ensembles of random complex matrices. Since these methods do not always generalize to higher dimensions, we will not utilize them directly in this work.

%There are several measures used to identify both classical and quantum chaotic processes. What is the interplay between them and the notion of multi-dimensional chaos? For instance classically the butterfly effect is described in terms of the Lyapunov exponent. It is clear that for evolutions that depend on initial conditions of different variables there should be generically the same number of Lyapunov exponent as the number of the variables. For instance in the pinball case one can make a variation of the initial condition both in the incident angle as well as in the impact parameter. Thus, in this case there are two distinct Lyapunov exponents. 
%For the quantum Lypunov exponent related to the %OTOC correlators see e.g. \cite{ Maldacena:2015waa} 

A commonly used quantity to identify chaos is the spectral form factor (SFF), see for instance \cite{Cotler:2016fpe}. In \cite{Bianchi:2024fsi} an analogous quantity, dubbed the scattering form factor (ScFF), was introduced. It is straightforward to generalize this quantity for the vectors $\vec \lambda_i,\ i=1,...N$ and correspondingly a vector of ``times" $\vec t$. 
\begin{equation}
    \mathrm{SFF}(\vec t) = \frac{1}{N^2}\sum_{i=1}^N\sum_{j=1}^N e^{i(\vlam_i-\vlam_j)\cdot \vec t}  
\end{equation}
We determine this generalized SFF for various cases of the toy model.

In the present investigation we focus on the multi-dimensional nature of chaotic scattering processes. It is pretty clear that one can use the methods developed in the present work also for other erratic functions that are not related at all to scattering both in the classical or in the quantum realm. A hint of this is the example of the electric potential of randomly located electric charges that we will discuss later on. 

It is plausible that similarly to the map between one-dimensional chaos and the RMT,  there should exist a map  between the multi-dimensional chaotic processes and random tensor theories. We will make few preliminary remarks about this possibility at the end.  

A cornerstone of quantum chaotic behavior is the ability to map certain discrete set of observables associated to a given system to  the set of eigenvalues of random matrices \cite{Bohigas:1983er}. This was done originally by Wigner and Dyson \cite{Dyson:1962b} and recently with the novel measure that we have proposed \cite{Bianchi:2022mhs}.  The Wigner-Dyson distribution of the (normalized) level spacings in the one-dimensional case is, explicitly:
\begin{equation} p_{\beta}(\delta) = {\cal C}_{\beta} \delta^\beta \exp(-\gamma_\beta \delta^2) \label{eq:beta_delta} \end{equation}
 where $\delta_n=\lambda_{n+1}-\lambda_n$. The parameter $\beta$ takes the value $\beta =1$ for GOE (Gaussian Orthogonal Ensemble), $\beta =2$ for GUE (Gaussian Unitary Ensemble) and $\beta =4$ for GSE (Gaussian Symplectic Ensemble). The distribution is well-defined for any $\beta >0$, such that these three ensembles can be said to be particular cases of the more general Gaussian $\beta$-ensemble (GBE). The normalization constants are given in general by:
\begin{equation} {\cal C}_\beta = 2\frac{[\Gamma(\frac{\beta+2}2)]^{\beta+1}}{[\Gamma(\frac{\beta+1}2)]^{\beta+2}}\,,\qquad \gamma_\beta = \left(\frac{\Gamma(\frac{\beta+2}2)}{\Gamma(\frac{\beta+1}2)}\right)^2 \quad .\end{equation}
The distribution of consecutive spacing ratios, $r_n \equiv \delta_{n+1}/\delta_n$, is \cite{Atas:2013dis}:
\begin{equation} f_\beta(r) = \frac{3^{\frac{3+3\beta}2}\Gamma(1+\frac\beta2)^2}{2\pi \Gamma(1+\beta)} \frac{(r+r^2)^\beta}{(1+r+r^2)^{1+\frac32\beta}} \label{eq:beta_r}\end{equation}
In contrast, in integrable systems one expects a Poisson distribution, which take the form
\begin{equation}
    p_{\text{P}}(\delta)=e^{-\delta}
\end{equation}
\begin{equation}
    f_{\text{P}}(r) = \frac{1}{(1+r)^2}
\end{equation}
for the level spacings and their ratios, respectively.

%%%%%%%%%%%%%%%%%%%%%%%%
The presentation is organized as follows. In the next section \ref{tdps} we review the classical scattering off the three-disk pinball system. We start with  a brief description of the system in \ref{tbs}. Then in \ref{ccs} we compute the classical  scattering angle and the number of collisions as a function of the two incident angles. These demonstrate the basic  features of classical  chaotic scattering, namely,  the  erratic behavior and   the self-similarity structure.  In \ref{tdd} the chaotic behavior is shown using two-dimensional plots of the scattering angle and number of collision in terms of the two angles. The ``topography" of these figures is analyzed and various structures are identified. The results are compared to the non-chaotic two-disk system.

Section \ref{tpqs} is devoted to the quantum scattering off the pinball system. We start in \ref{teqotps} with reviewing the determination of  the exact solution of the Schr\"{o}dinger equation associated with the pinball boundary conditions and the resulting $S$-matrix. In section \ref{tsmotp} we compute the $S$-matrix for the fully symmetric system and for asymmetric ones. For the former we find that the eigenvalues of the $S$-matrix admit a Poisson distribution. For the latter cases we show that for each asymmetry there is a large enough wave-number $k$ for which the distribution is chaotic, namely a COE distribution. In section \ref{tqsa} we compute the amplitude for various different wave-numbers and initial angles and then the cross-section for a given angle and the averaged one. The two-dimensional plots of the amplitudes are presented in subsection \ref{tddotqs}. We examine the dependence of the scattering amplitude on the angles and wave-number and compare the amplitude for chaotic and non-chaotic configurations.

Section \ref{mtsbposa} is devoted to the analysis of the spacings between peaks of the amplitudes that depend on two variables. In \ref{tpsotlt} we first briefly review the landmark case of the phase shifts for the leaky torus \cite{Gutzwiller:1983} . We then propose a  toy model that generalizes  it with two random matrices.  We consider a function of two random variables, which is the analog of the time delay of the leaky torus. This function can be also interpreted as the electric potential  produced by a set of charges randomly located on a plane in three space dimensions. Section \ref{tdeats} is devoted to the analysis of two-dimensional eigenvalues and their spacings.
%We start with the spectra of the eigenvalues (\ref{tdes}) first by fixing one variable and considering the peaks as a function of the other variable.
%We describe the unfolding procedure for that case.
We propose four different measures for spacings in two (or higher) dimensions, including a definition to form consecutive level spacings. These methods are applied to the toy model mentioned above in \ref{amwtirm}. We compute the corresponding distributions in two dimensions. We explain the resulting distribution by identifying the ``effective repulsion" in the toy model. This is done by performing analytic calculations.  Lastly we consider a case where there is no disorder in the eigenvalues, in which they are taken to be integers but the pairing between them is performed with a permutation matrix. In this case also a GOE distribution is found for the distribution of the spacings. We end the analysis of the toy model by considering the  two-dimensional scattering form factor in \ref{tdsff}.

In section \ref{taoteofsa} we apply the methods developed for the toy model to the quantum scattering in the pinball system, analyzing the spacings of peaks in two dimensions.  In section \ref{tsmaarm} we examine a model where the $S$-matrix of the pinball is taken to be a random COE matrix. Then, in section \ref{aopop} we compare this to the results of the quantum pinball system. In these cases we do not get a COE distribution for the spacings but rather a logistic distribution that is peaked around one. 

 % Section \ref{acoamtrtt} states a conjecture of a map between the multi-dimensional chaotic systems and random tensor theory.
 Section \ref{s} is a summary of the present investigation that includes several open questions, including some remarks on a possible connection with Random Tensor Theory.

 We also include two appendices. In appendix \ref{app:supplfig} we include several supplementary plots of the quantum pinball scattering amplitude. In appendix \ref{ecoasdfp}  we present the explicit computation of the two-dimensional spacing distributions for the Poisson distribution of eigenvalues. 
  %In  appendix \ref{beisv} we discuss  $\beta$ ensembles in several variables. 

%%%%%%%%%%%%%%%%%%%%%%%%%%%%%%%%%%%%%%%%%%%%%%%%%%%%%%%%%%%%%%%%%
\clearpage
\section{The three-disk pinball: Classical scattering}\label{tdps}
Rather than discussing multi-dimensional chaotic scattering in general, we focus in the next two sections on the scattering in a pinball system. It will serve us as an arena to review some of the  main properties of the multi-dimensional chaotic behavior. In the current section we discuss classical scattering and in the next one quantum scattering. We start with the basic structure of the pinball system. 

\subsection{The basic setup}\label{tbs}
In the pinball scattering experiment a point-like particle is scattered from a system of three hard disks. The system is characterized by the different radii and positions of the disks.  For simplicity, we can take the most symmetric configuration with three disks of equal size, whose centers are at the corners of an equilateral triangle.
For this fully symmetric case, we will consider disks of radius $R=1$, and place the three disks centered at the points:
\[ d_1 = \left(\frac{L}{\sqrt{3}},0\right)\,, \quad d_2 = \left(-\frac{L}{2\sqrt{3}},\frac L2\right)\,,\quad d_3 = \left(-\frac{L}{2\sqrt{3}},\frac L2\right) \]
such that the system is centered at the origin and the distance between (the centers of) the disks is $L > 2R$.
%\[ d_1 = (-\frac32,-\frac{\sqrt3}{2})\,, \quad d_2 = (\frac32,-\frac{\sqrt3}{2})\,,\quad d_3 = (0,\sqrt3)\]
%such that the system is centered at the origin and the distance between (the centers of) the disks is $L = 3$.

\begin{figure}[t!]
    \centering    \includegraphics[width=0.3\linewidth]{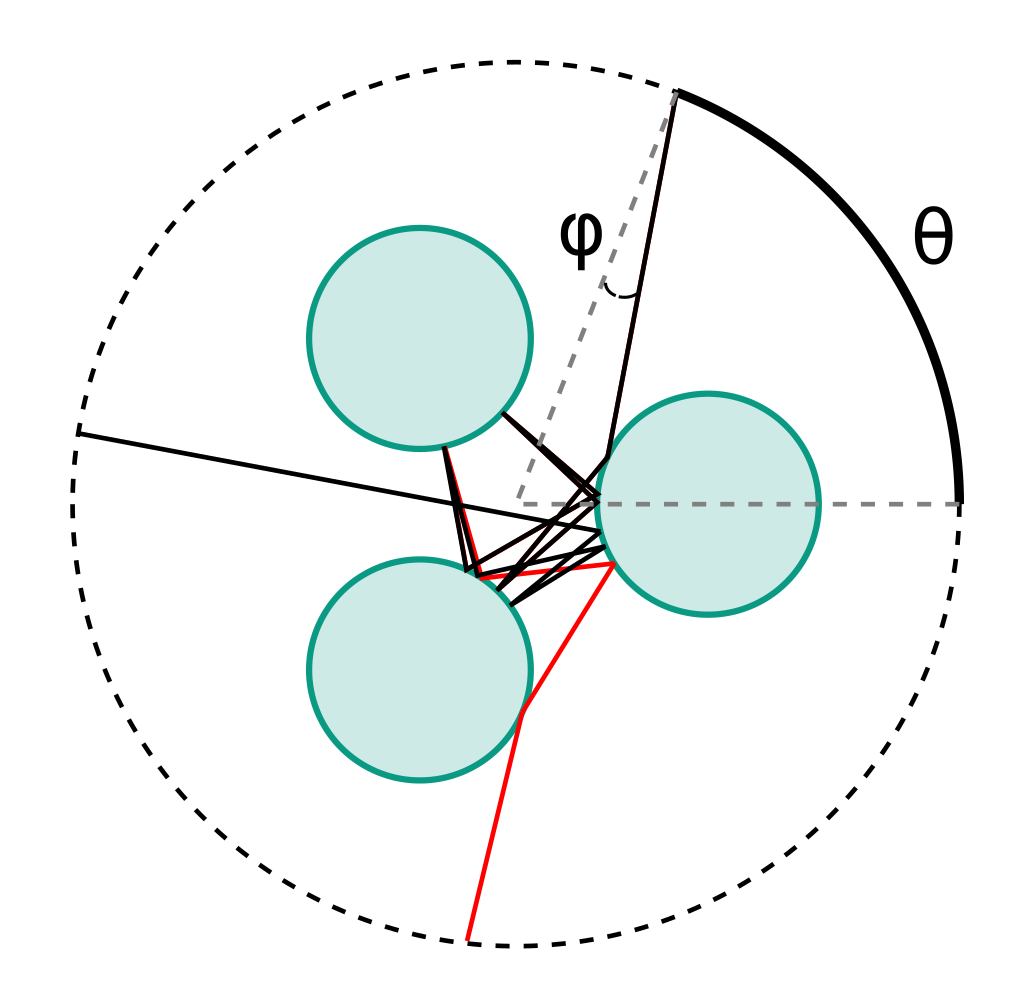}
    \caption{Symmetric setup for the three-disk pinball system using two angles to parameterize the initial condition. We plot two trajectories at fixed $\theta$, differing by $\delta\phi < 10^{-7}$.}
    \label{fig:pinball_setup}
\end{figure}

We can define the relevant parameters of the incoming particle using two angles. See figure \ref{fig:pinball_setup}.
\begin{itemize} 
\item The first angle $\theta$ parameterizes the \emph{initial position} of the particle. It is taken to be on a circle centered at the origin and having some fixed radius $R_0$, large enough to surround the entire system. For the present analysis we chose $R_0=4$.
\item The second angle $\phi$ is the direction of the \emph{initial velocity}. We define it relative to $\theta$ such that $\phi = 0$ always points to the origin. In absolute terms the direction of the velocity is then given by $\alpha_{in} = \pi + \theta + \phi$.
\end{itemize}
This parametrization covers all the ``phase space'' of the incoming particle in a compact way. Another common parametrization is in terms of an impact parameter. For example, considering  a particle coming in horizontally from the left and taking its position along the vertical axis to be the impact parameter $b$ means setting the angles according to:
\begin{equation}
 b = R_0 \sin\theta,\quad \alpha_{in} = \pi + \theta+ \phi = 0    
\end{equation}

The fully symmetric system is invariant under rotation by ${2\pi}/{3}$ around the axis perpendicular to the plane of the disks that is placed at the origin, as well as the three reflections around the lines from the centers of the disks to the origin. The absolute value of the velocity is conserved, being related to the kinetic energy, but plays a marginal role in that it can be reabsorbed into a rescaling of the time variable or of all the lengths.
Later, we will consider a general system, and not only the fully symmetric one. It will be parameterized by the three radii $R_1, R_2$ and $R_3$ and by the positions of the centers of the disks $d_1$, $d_2$, and $d_3$. 
%In fig (\ref{fig:asysetup}) we illustrate the asymmetric case with two examples one with unequal radii of the disks and the other one with their centers at a vertices of a non-equilateral triangle. 
Obviously, in the most general asymmetric setup the system is not invariant under any rotation. In particular cases the system may be invariant under certain symmetry transformations. 
%%%%%%%%%%%%%%%%%%%%%%%%%%%%
% \begin{figure}[h!]
%     \centering
%     \includegraphics[width=0.48\linewidth]{setup051015.pdf}
%     \includegraphics[width=0.48\linewidth]{setup02515075.pdf } \\ 
%     \caption{ Asymmetric setups: 
% Left: three different radii $R_1=0.5, R_2=1.0,R_3=1.5$ and same distances L=3. 
%     Right: same radii and asymmetric locations $d_1 = \left(\frac{L}{\sqrt{3}} , \frac L4\right) \quad d_2 = \left(-\frac{L}{2\sqrt{3}},\frac {3 L}{2}\right)\,,\quad d_3 = \left(-\frac{L}{2\sqrt{3}},\frac {3 L}{4}\right)$ }
%     \label{fig:asysetup}
% \end{figure}
%\clearpage
%%%%%%%%%%%%%%%%%%%%%%%%
\subsection{Classical chaotic scattering}\label{ccs}
The classical  chaotic behavior of the pinball scattering has been investigated thoroughly \cite{Gaspard:1989cla}.  Several observables, such as the Lyapunov exponent(s), the Kolmogorov-Sinai entropy and others have been used as a measure of chaos in this context. In the present analysis we focus on two observables. One is the scattering angle, defined as the difference between the incoming and outgoing angles of the velocity. The second is the number of collisions $n_c$, closely related to the time the particle spends inside the system before escaping. We are interested in the dependence of these  two observables on the two angles $\theta$ and $\phi$ defined above.

\begin{figure}[p!]
    \centering
    \includegraphics[width=0.40\linewidth]{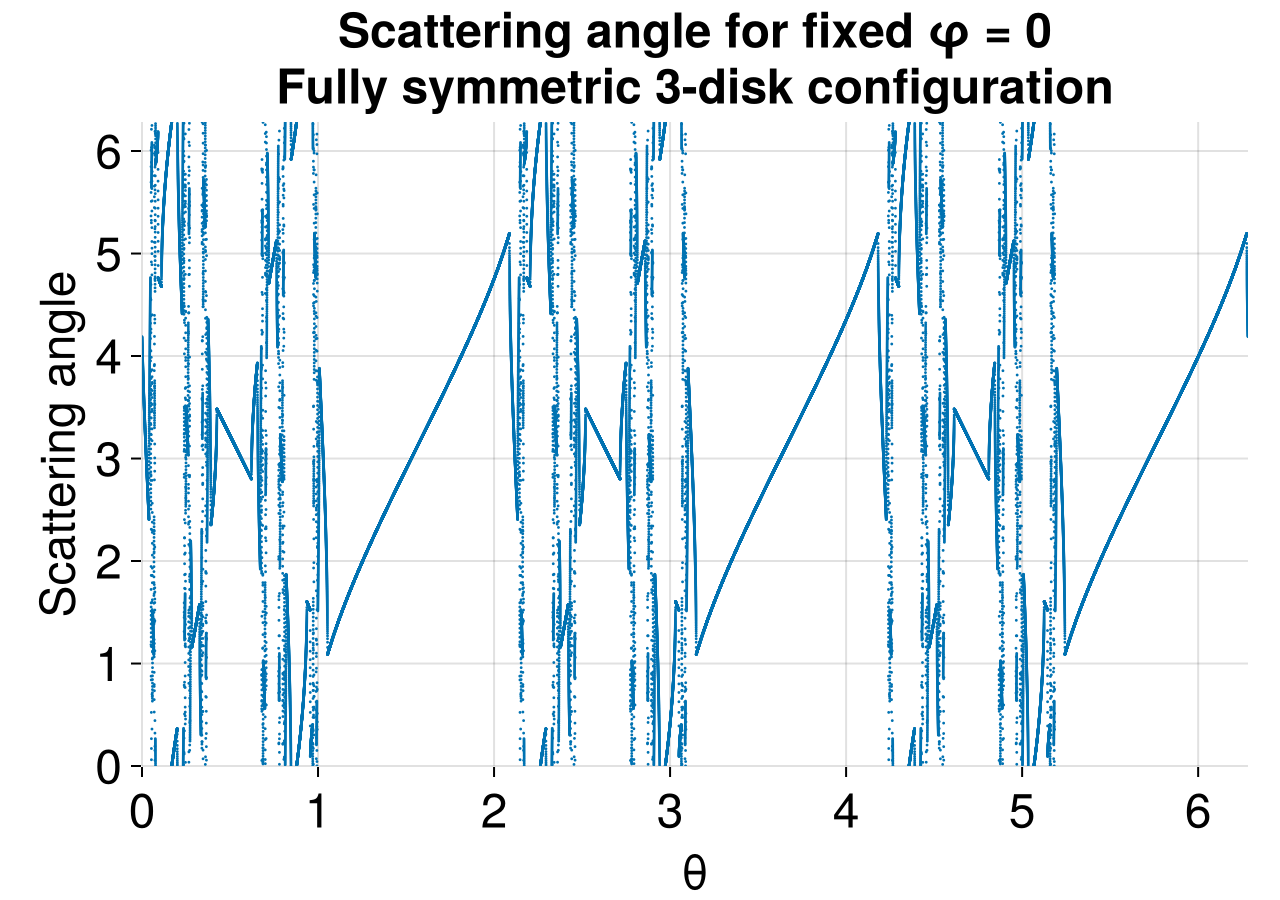}
    \includegraphics[width=0.40\linewidth]{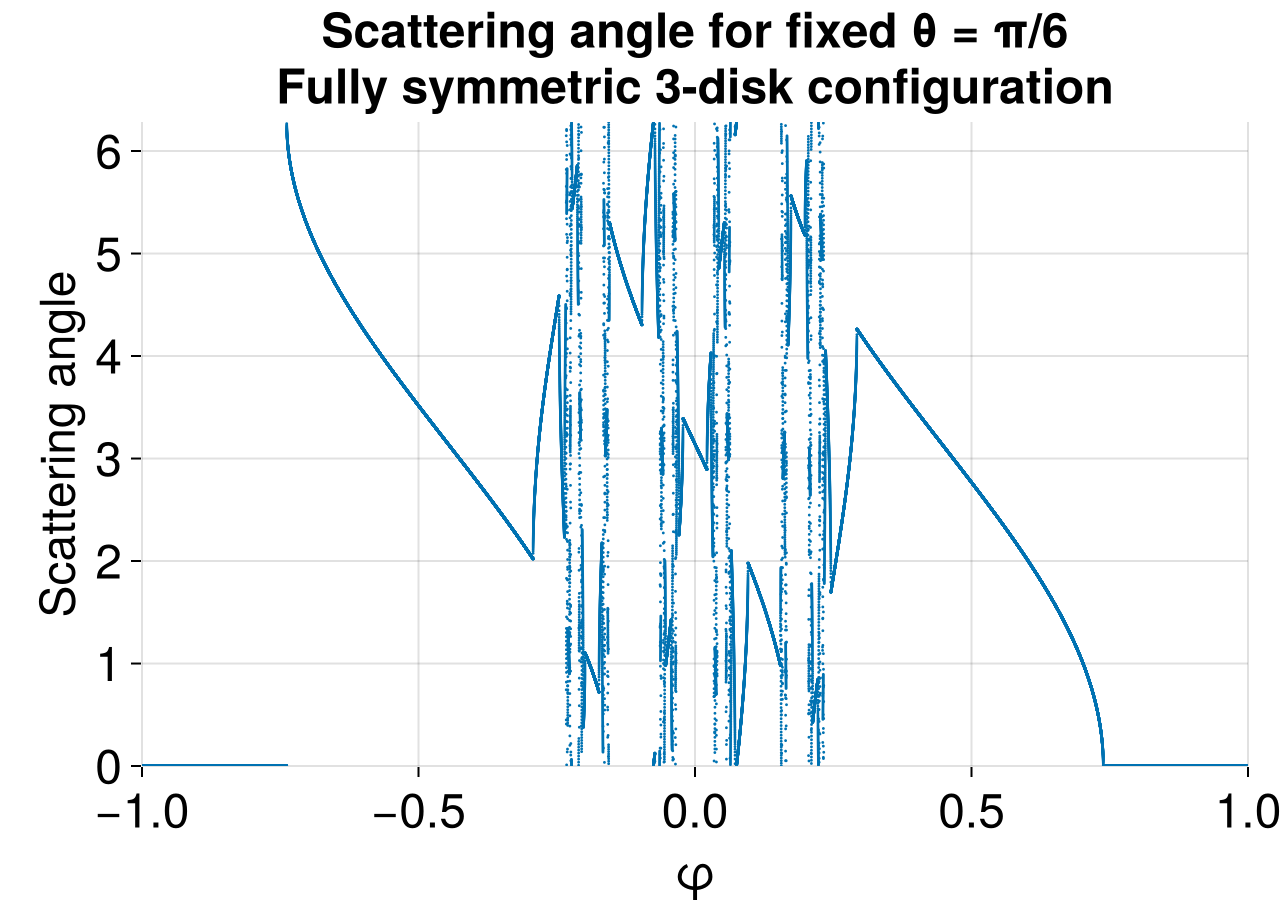} 
    \includegraphics[width=0.40\linewidth]{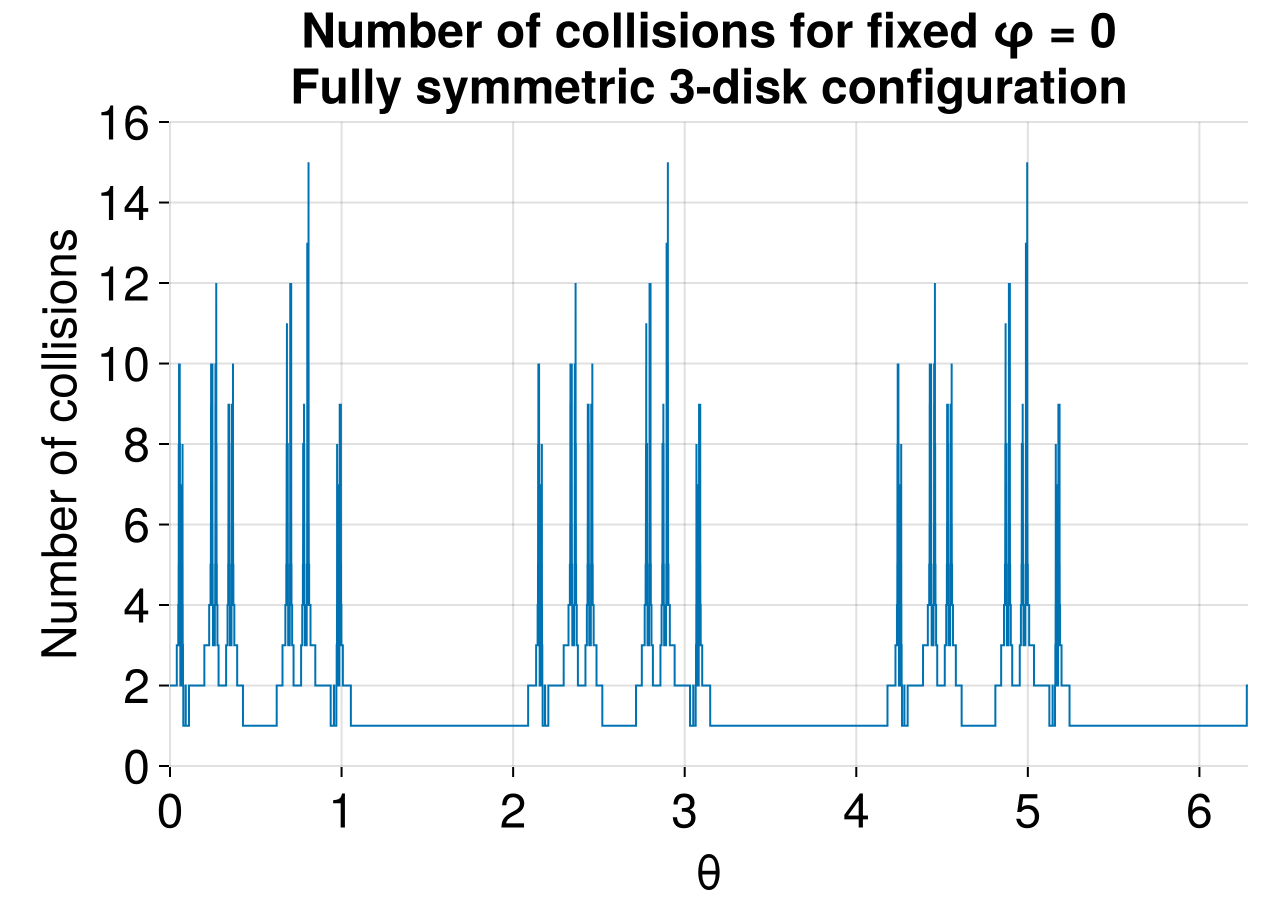}
    \includegraphics[width=0.40\linewidth]{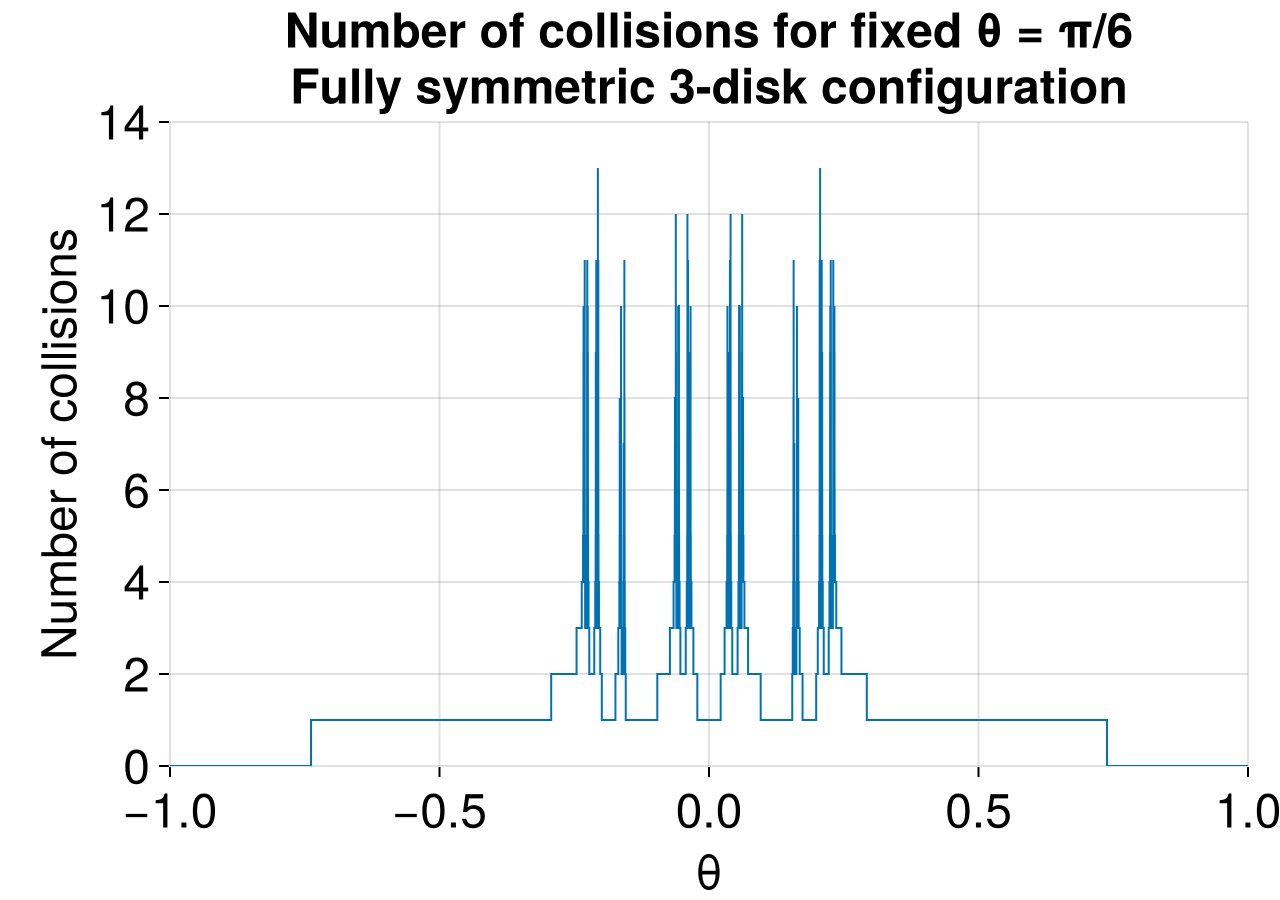}
    \caption{The scattering angle (top) is an erratic function of both angles, $\theta$ (left), and $\phi$ (right), when the other is kept fixed. The regions where the function is erratic correspond to regions where the number of collisions (bottom) is large. }
    \label{fig:pinball_angle_dependence}
\end{figure}

\begin{figure}[p!]
    \centering
    \includegraphics[width=0.40\linewidth]{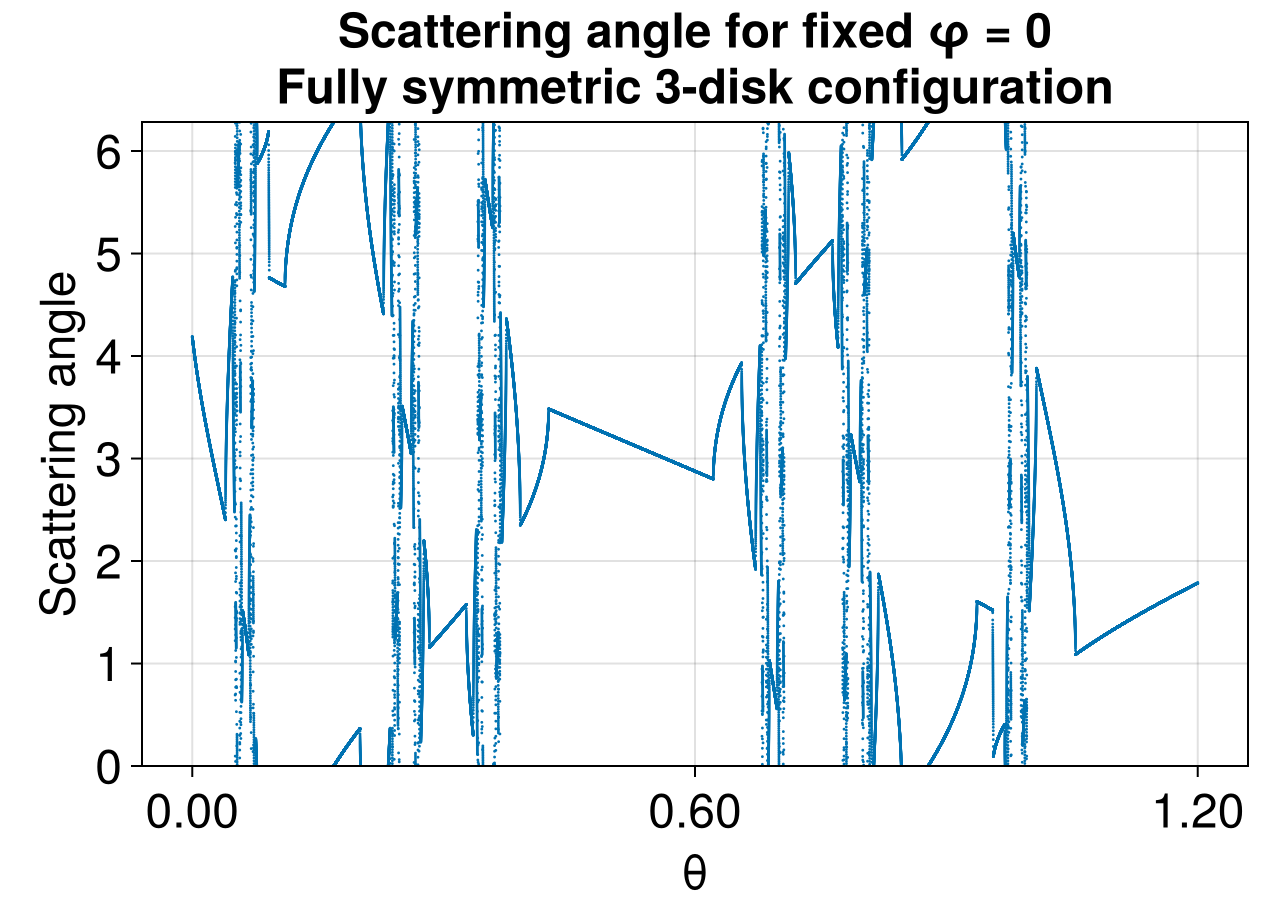}
    \includegraphics[width=0.40\linewidth]{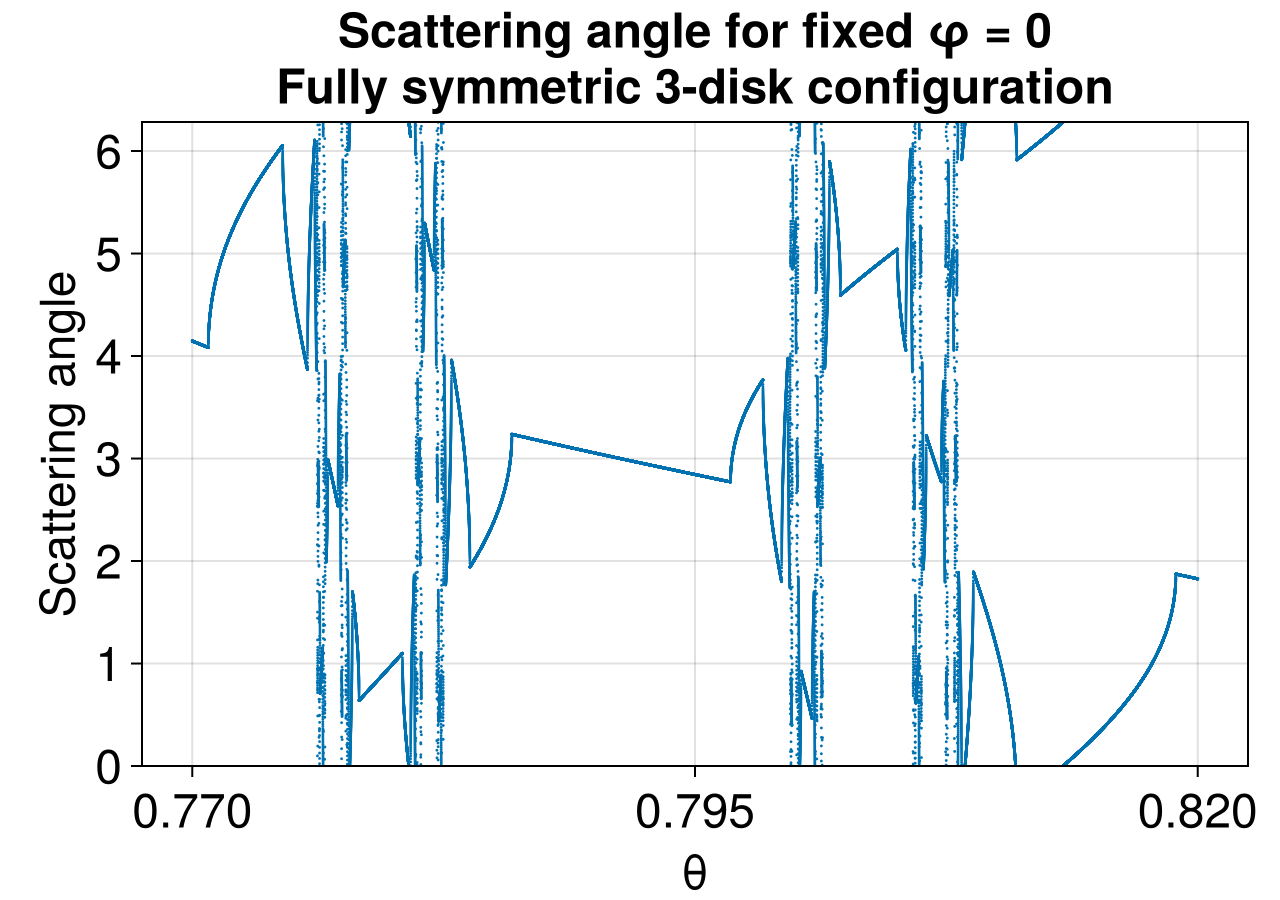} \\ 
    \includegraphics[width=0.40\linewidth]{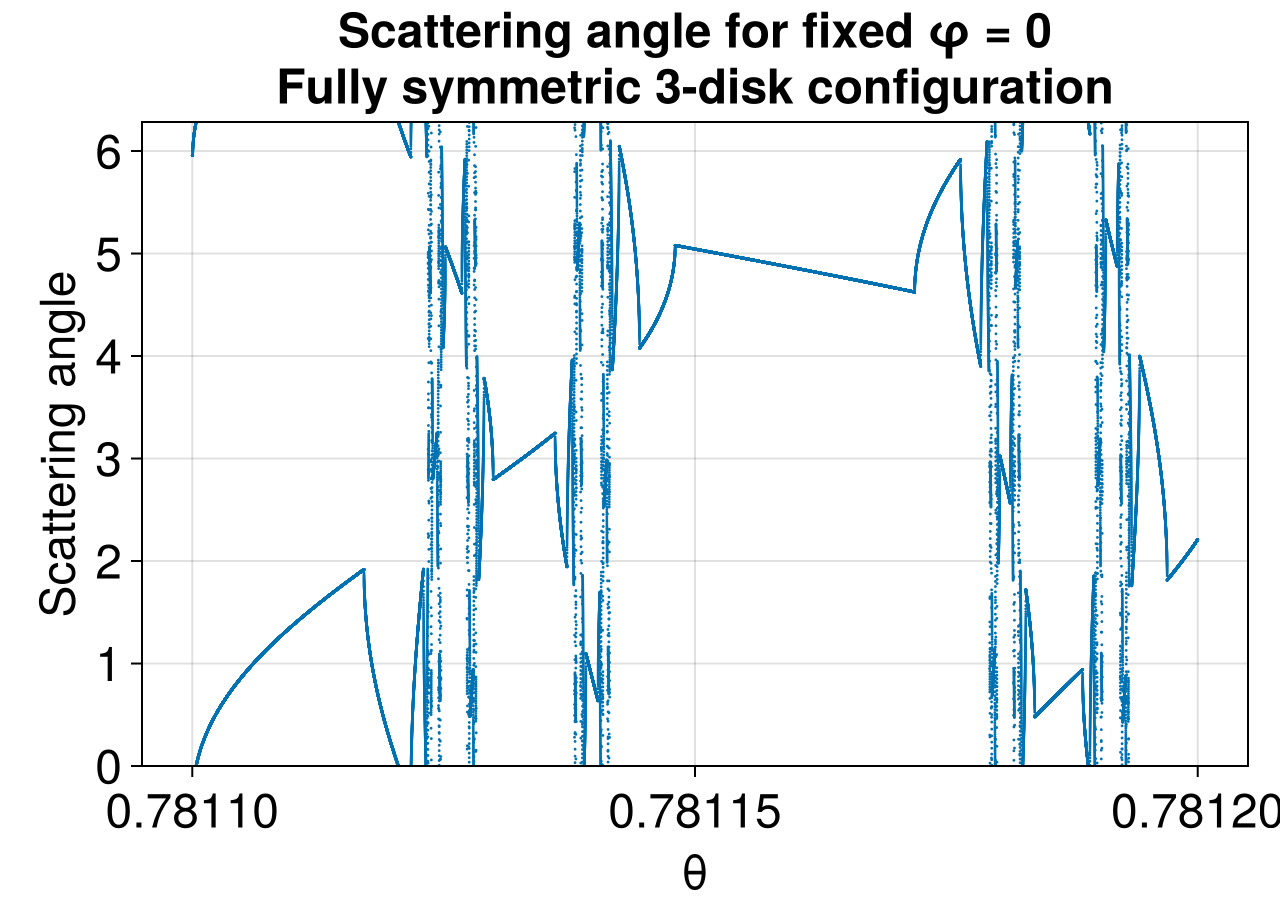}
    \includegraphics[width=0.40\linewidth]{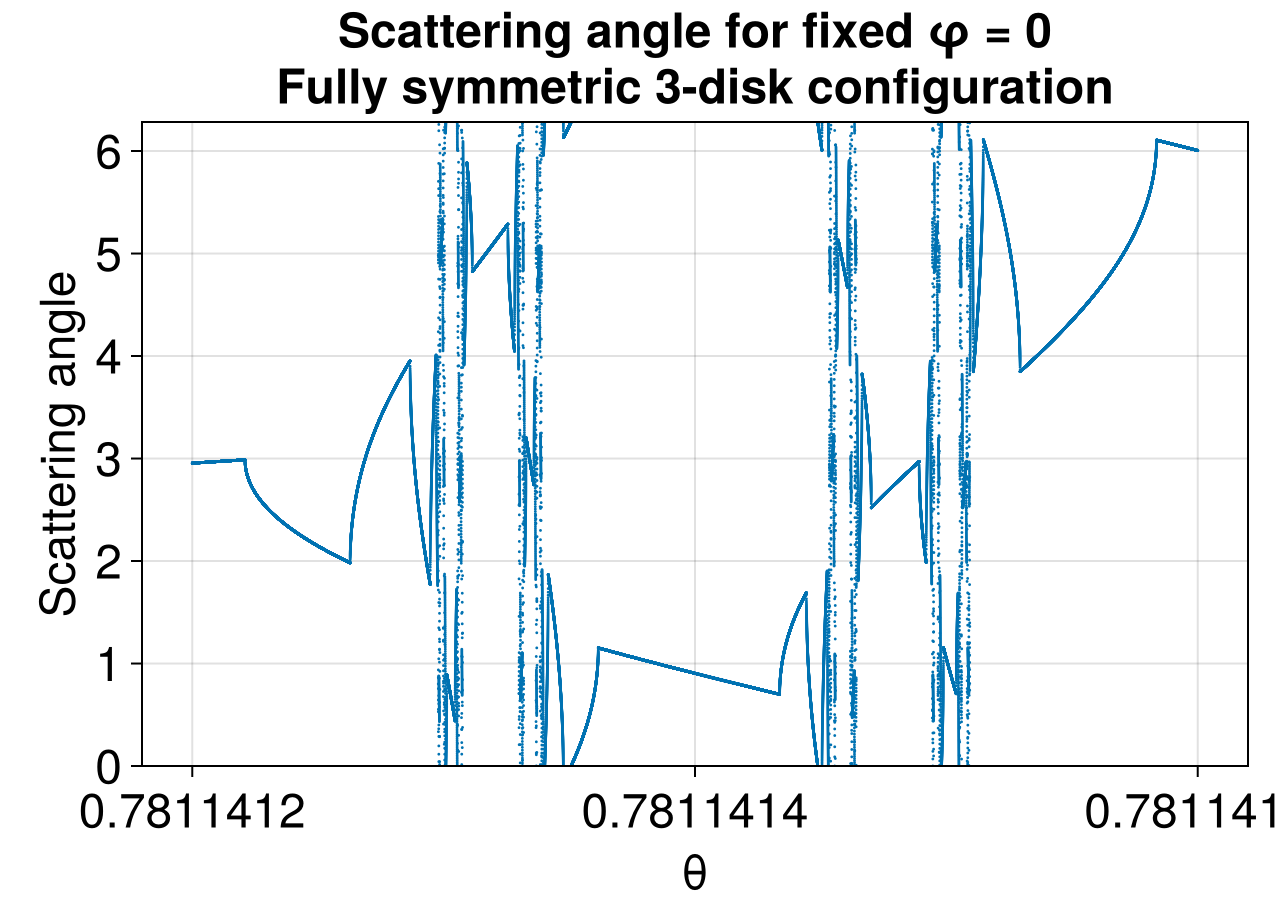}
    \caption{Successively zooming in on erratic regions in the plot reveals a self-similar structure.}
    \label{fig:pinball_fractal}
\end{figure}

Classical chaotic behavior is characterized  by the phenomena of erratic dependence on the basic input variables (in our case $\theta$ and $\phi$) and of self-similarity leading to fractal structures.  
We now study these two properties in the  two observables we chose.

For certain values of the parameters $(\theta,\phi)$, the particle will spend a long time inside the system, in such a way that the number of collisions becomes large. Then, there are regions of the parameter space in which the scattering angle will be an erratic function of each of the two variables. See figure \ref{fig:pinball_angle_dependence}.

A crucial feature of these functions is that they display a fractal, self-similar nature. Namely, when one zooms in on one of the chaotic regions, one always sees the same pattern of regions where the function is relatively smooth and points where the number of collisions is very large and the function is erratic. This pattern repeats indefinitely \cite{Gaspard:1989cla}. See figure \ref{fig:pinball_fractal}. In the figure we zoom in on an erratic region in the plot of the scattering angle as a function of $\theta$, seeing the same structure repeating at all scales. One can see the same behavior in the scattering angle as a function of $\phi$.

Self-similarity can be quantified by calculating the fractal dimension of the curve. A convenient numerical strategy is to employ a box-counting algorithm. The algorithm is as follows: divide the plane into boxes (squares in $2D$) of size $\epsilon\times\epsilon$, count how many boxes $N_b$ the curve passes through, and measure how this number scales with the size of the boxes. The fractal dimension is then given by the slope of a plot of $\log N_b(\epsilon)$ as a function of $\log\frac1\epsilon$. For an ordinary one-dimensional curve it would be one. For a fractal it is a number between 1 and 2. For the scattering angle as a function of one of the angles, plotted in figures \ref{fig:pinball_angle_dependence}, we have found this box-counting dimension to be in the range 1.5--1.6.

% \clearpage
\subsection{Two-dimensional descriptions }\label{tdd}
\begin{figure}[ht!]
    \centering
    \includegraphics[width=0.44\linewidth]{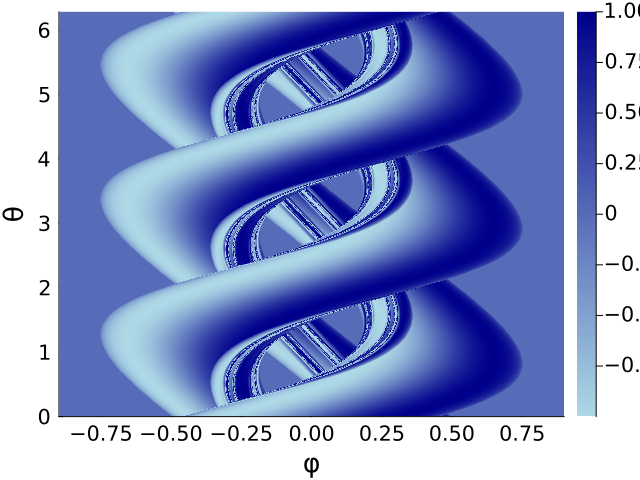}
    \includegraphics[width=0.44\linewidth]{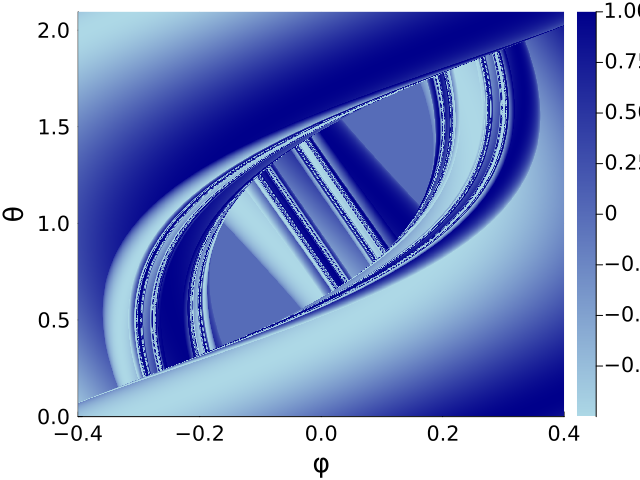} \\
    \includegraphics[width=0.44\linewidth]{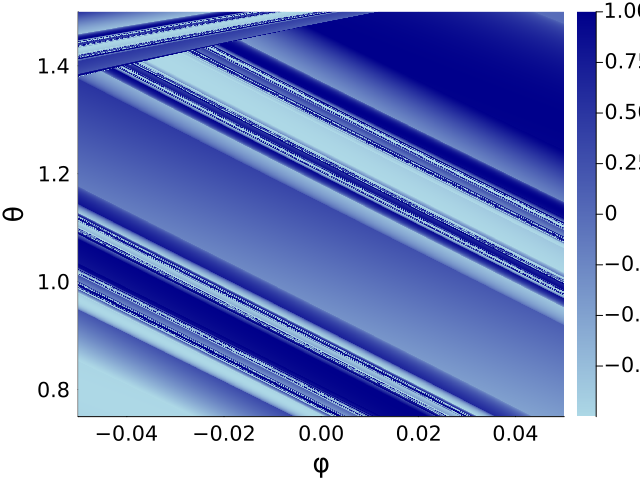}
    \includegraphics[width=0.44\linewidth]{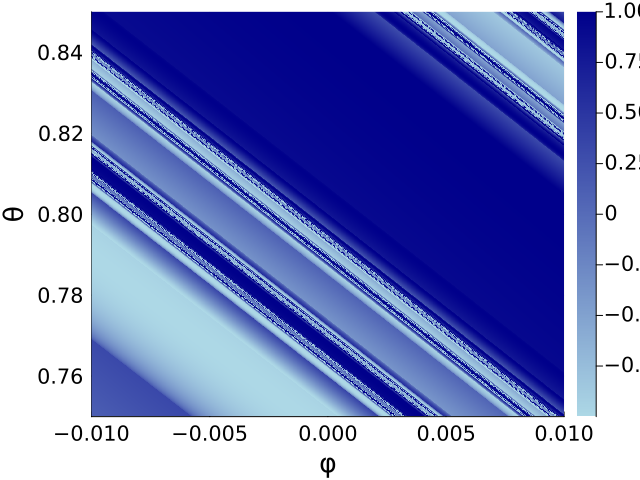}
    \caption{Sine of scattering angle as a function of   $\theta$ and $\phi$.}
    \label{fig:pinball_2d_plot}
\end{figure}

\begin{figure}[t!]
    \centering
    \includegraphics[width=0.44\linewidth]{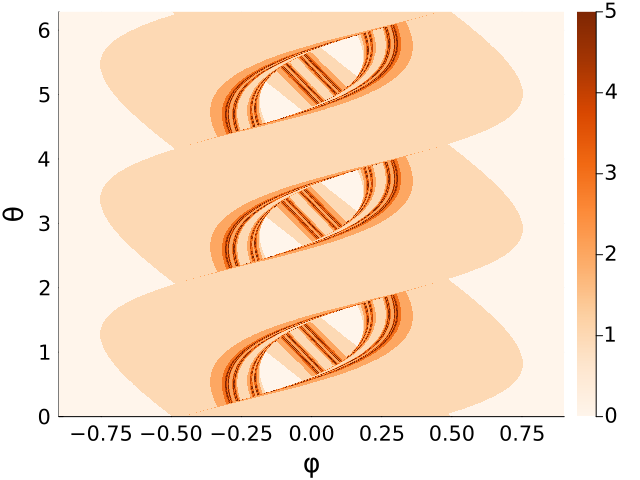}
    \includegraphics[width=0.44\linewidth]{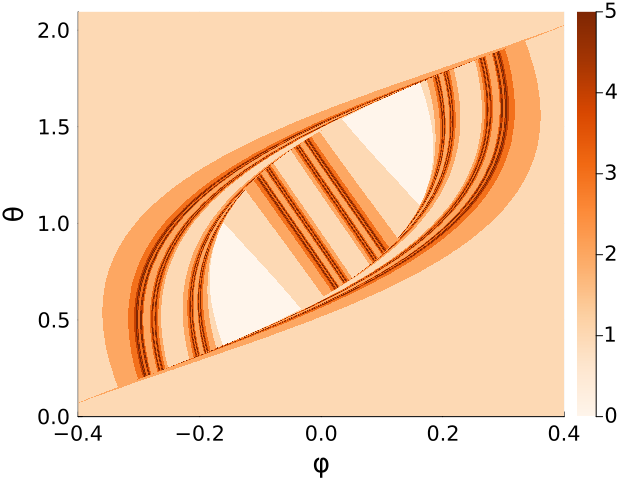} \\
    \includegraphics[width=0.44\linewidth]{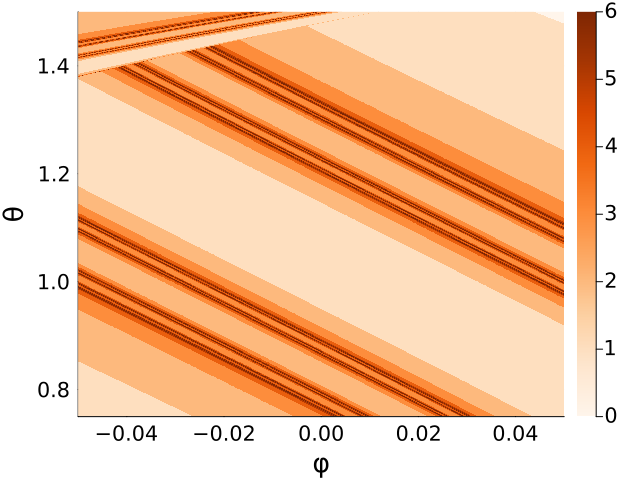}
    \includegraphics[width=0.44\linewidth]{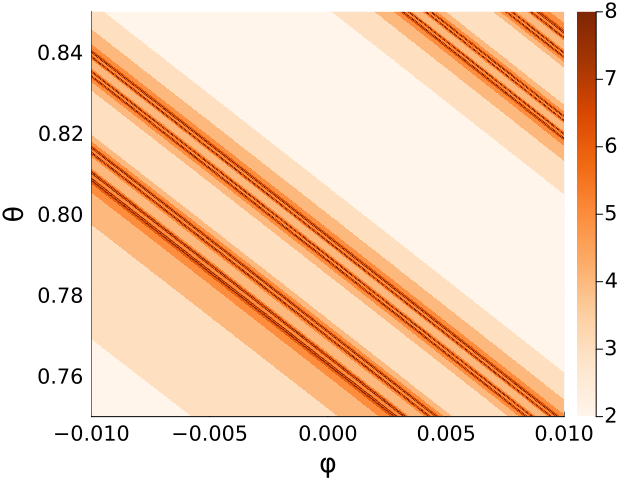}
    \caption{Number of collisions as a function of both angles. The regions where the number of collisions is large and there is erratic behavior follows line patterns.}
    \label{fig:pinball_2d_plot_nc}
\end{figure}

A main point of the present investigation is the extraction of information from the multi-dimensional (in the pinball case two-dimensional) nature of the scattering.
For this purpose we start by plotting the scattering angle  as a function of both variables.\footnote{We plot only the sine of the scattering angle, to avoid jumps in the plot when the angle goes from $2\pi$ to $0$.} The results are drawn in figure \ref{fig:pinball_2d_plot}. In figure \ref{fig:pinball_2d_plot_nc} we plot the number of collisions. Most of the features of the plots of the scattering angle repeat themselves also for the number of collisions.

The two-dimensional patterns and their self-similar structure have been intensively studied in the past, see for instance \cite{Sweet:1999} and references therein. In these references the scattering patterns were analyzed using the notion of Wada basins. As with the one-dimensional plots, one can also employ the box counting algorithm, in a straightforward generalization to two-dimensions, and find that the surfaces plotted in the figures below have dimension higher than 2.

Let us make some simple observations motivated by the plots, that will be relevant for our analysis:
\begin{itemize}
\item 
There are three-``eye-like" structures associated  with the three disks. In fact, there are three large bands where the number of collisions is 1, one band for each disk, with the three ``eyes'' located in the gaps between the disks, where the particle can enter inside the system.
\item
In each eye there is a group of curves for which the sine of the scattering angle is $1$. These curves do not intersect and in between any adjacent ones there is a curve for which the sine of the scattering angle is $-1$. The phenomenon of ``repelling"  curves  shows up in other multi-dimensional descriptions of chaotic behavior. The self-similarity property can be seen in the zoomed-in pictures.
\item 
There are two ``parallel"  curved stripes of maximal and minimal values of the sine of the angle and also such stripes ``perpendicular" to the former ones. 
\item
The three eye-like structures resides in a region of zero scattering angle.
\item Inside the eyes there are parallel lines of large number of collisions separate by lines, or stripes of zero number of collisions. Once again the lines are non-intersecting.
% \item 
% In the plot of the number of collisions, unlike for the scattering angle,  the lines are connecting  the two sides of each eye.
\end{itemize}

%%%%%%%%%%%%%%%%%%%%%%%%%%%%%%%%%%%%%%%%%%%%

\begin{figure}[p!]
    \centering
    \includegraphics[width=0.40\linewidth]{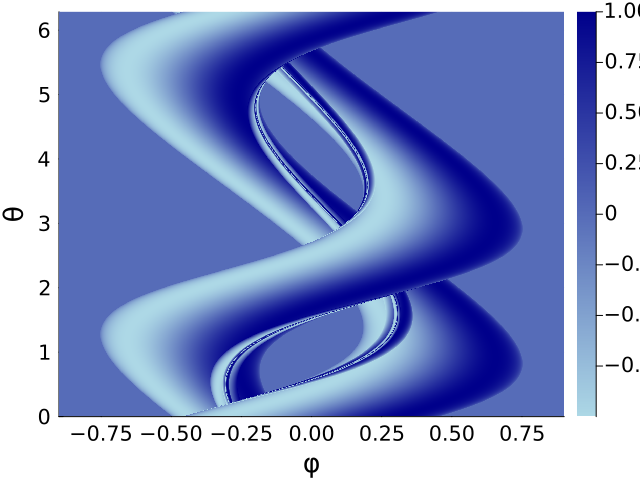}
    \includegraphics[width=0.40\linewidth]{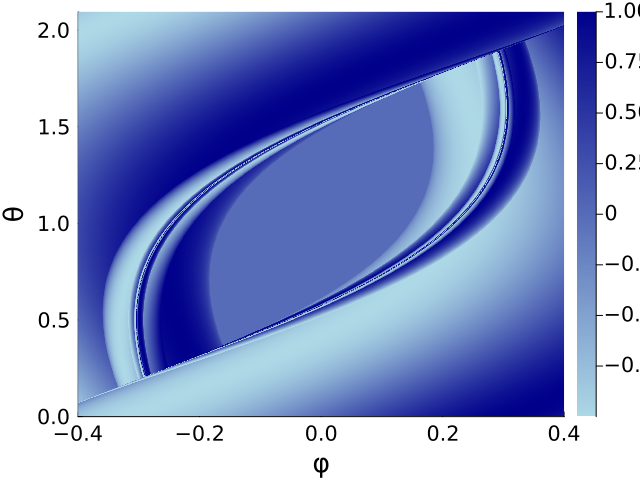} \\
    \includegraphics[width=0.40\linewidth]{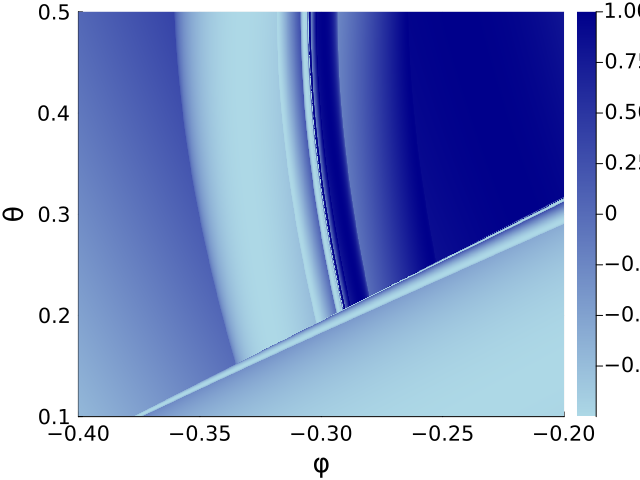}
    \includegraphics[width=0.40\linewidth]{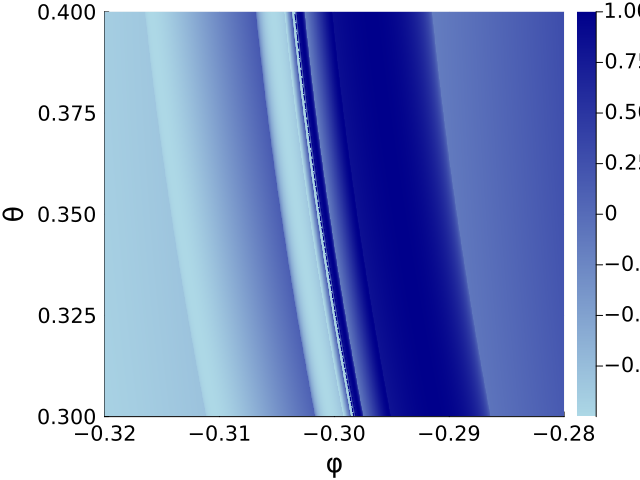}
%     \caption{Scattering angle as a function of both angles for the non-chaotic two-disk system.}
%     \label{fig:pinball_2d_plot_twodisk}
% \end{figure}

% \begin{figure}[h!]
%     \centering
    \includegraphics[width=0.40\linewidth]{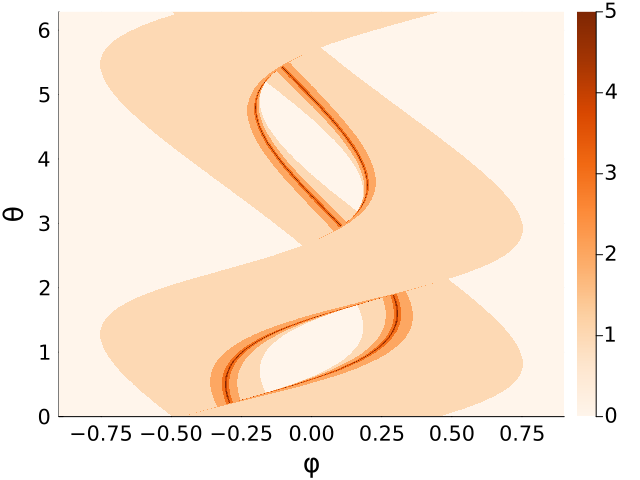}
    \includegraphics[width=0.40\linewidth]{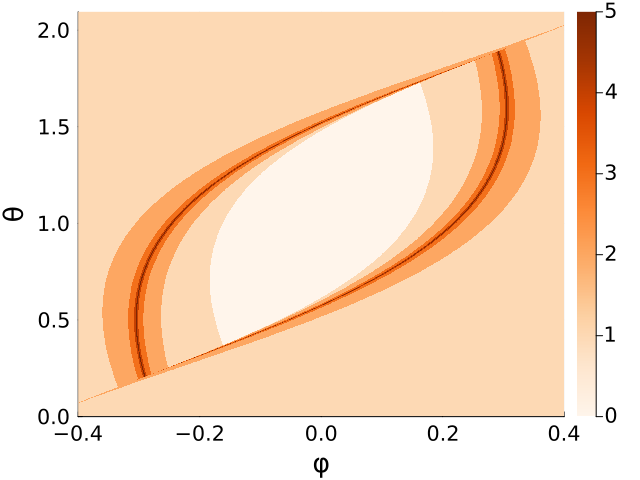} \\
    \includegraphics[width=0.40\linewidth]{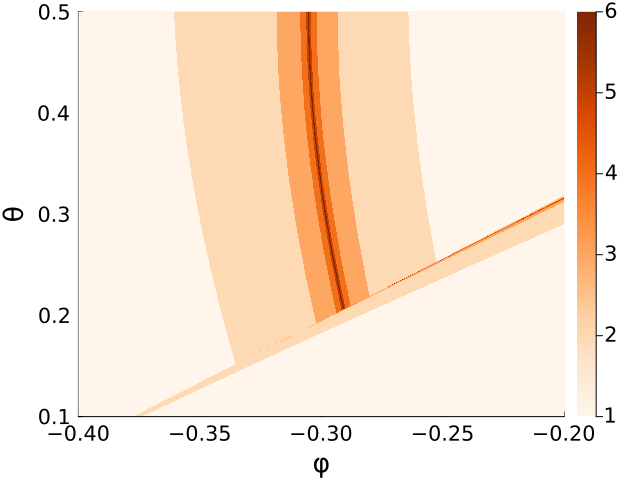}
    \includegraphics[width=0.40\linewidth]{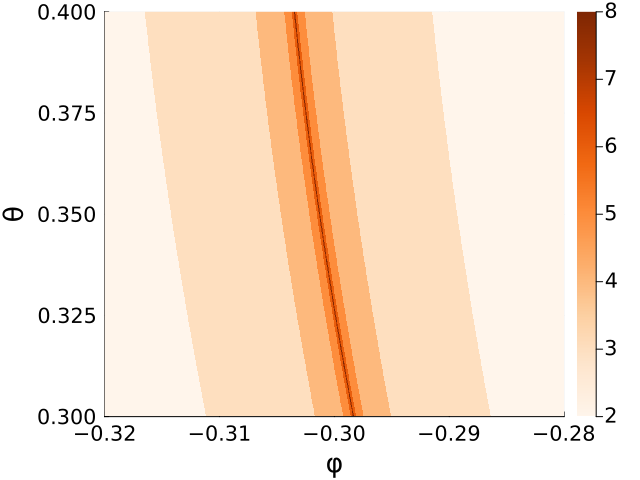}
%    \caption{Number of collisions as a function of both angles for the non-chaotic two-disk system.}
\caption{The sine of the scattering angle (top four plots) and number of collisions (bottom four) as a function of both angles for the non-chaotic two-disk system.}
    \label{fig:pinball_2d_plot_nc_twodisk}
\end{figure}

\paragraph{Two-dimensional scattering from two disks:} To identify the chaotic sub-patterns of both the plots of the scattering angle and of the number of collisions, we compare the results of figures \ref{fig:pinball_2d_plot} and \ref{fig:pinball_2d_plot_nc}  with the corresponding figures associated with scattering from two disks. Recall that the latter is non-chaotic.  In figure \ref{fig:pinball_2d_plot_nc_twodisk} we plot the sine of the scattering angle and the number of collisions  for the  two-disk case including zoomed-in pictures. 

There are now two bands with two ``eyes'' between them. It is clear that inside the two eyes there is nothing, zero collisions and the scattering angle is always zero. This follows from the fact that  the eyes are located where the particle passes in between the disks. If there is no third disk the particle will just pass through without any collisions.

For two disks there are only two simple curves where the number of collisions is large. If one zooms in on one of the two, one observes  that it is  really a simple line of
zero width. Contrary to the chaotic three-disk case, for the two disks the lines do not open up in a fractal way to reveal more lines between them. This corresponds to the fact that there is only one possibility to trap the particle between the two disks, and only one possible trajectory in which it bounces between the disks infinitely many times, when the particle is traveling (near to, but never exactly on in the scattering experiment) the line between the centers of the two disks.

% \clearpage
\section{Quantum scattering amplitude of the pinball system}\label{tpqs}
\subsection{The exact quantization of the three-disk pinball system}\label{teqotps}
The quantum version of the three-disk pinball scattering system was solved exactly by Gaspard and Rice \cite{Gaspard:1989exq}. The problem is simple to define. One has to solve the free wave-equation in the plane,
\begin{equation}
    (\nabla^2 + k^2)\psi_k(x,y) = 0 \label{eq:freewaveeq}\,,
\end{equation}
subject to the boundary conditions that the wave-function vanishes on the boundaries of the three disks.

In \cite{Gaspard:1989exq} the authors focus on the most symmetric case of three disks of equal radii centered on the corners of an equilateral triangle, the same as the classical setup we explored above. But this is not necessary, as the system can be solved for any general set-up, including $n$ disks of different radii and generic positions. We follow here the detailed review of \cite{Wirzba:1997hy}, which contains the general solution. The analogous system of scattering from $n$ spheres in three dimensions has also been solved exactly using the same techniques \cite{Henseler:1997}.

We will restrict our attention to the three-disk system, but keep the setup general. The solution is expressed in the basis of spherical waves of fixed angular momentum $l$:
\begin{equation}
    \psi^{(0)}_{k,l}(r,\phi) = J_l(kr)e^{il\phi}\,.
\end{equation}
where $J_l$ is a Bessel function of the first kind.\footnote{Note that $l\in {\bf Z}$  in two dimension and, while $J_{-l}=(-1)^l J_l$, the angular dependence is different for positive and negative $l$.} We can define the $S$-matrix by writing an asymptotic solution at large $r$, far away from the scattering system, as a sum of incoming and outgoing waves as
\begin{equation}
    \psi_{k,l}(r,\phi) \approx \frac{1}{\sqrt{2\pi k r}}\sum_{l\pri=-\infty}^\infty \left(\delta_{ll\pri} e^{-i (kr -l\frac\pi2-\frac\pi4)}+S_{ll\pri}e^{i (kr -l\frac\pi2-\frac\pi4)}\right)e^{il\phi}\,.
\end{equation}
The scattering amplitude is then defined as
\begin{equation}\label{scatamp}
    f(k\,;\phi,\phi^\prime) = \frac{e^{-i\pi/4}}{\sqrt{2\pi k}} \sum_{l, l^\prime}^{-\infty,+\infty}   e^{-i l (\phi-\frac\pi2)} \left(S_{ll^\prime}-\delta_{ll^\prime}\right) e^{i l^\prime(\phi^\prime-\frac\pi2)}\,,
\end{equation}
where $\phi$ is the angle for the incoming wave, and $\phi\pri$ the outgoing angle. The differential cross-section is
\begin{equation}\label{difcros}
    \frac{d\sigma}{d\phi^\prime}(k;\phi,\phi^\prime) = |f(k;\phi,\phi^\prime)|^2\,.
\end{equation}
The total cross-section for fixed incoming angle $\phi$ is given by integrating over $\phi^\prime$, with the result
\begin{equation}
    \sigma(k;\phi) = \frac{1}{k} \sum_{l, l^\prime}^{-\infty,+\infty} e^{-i l(\phi-\frac\pi2)} \big(T T^\dagger)_{l,l^\prime} e^{il^\prime(\phi-\frac\pi2)}
\end{equation}
where we defined the $T$-matrix as usual as $S \equiv 1+ i\, T$. We can average also over the incoming angle $\phi$ to get the average total cross-section as a function of the energy $E=k^2$, given by
\begin{equation}
    {\bar \sigma}(k) = \frac{1}{k} \Tr{T T^\dagger}
\end{equation}
The $S$-matrix can be computed exactly using Green's theorem to solve the equations for the wave-function with the correct boundary conditions. Here, we will not repeat the derivation and only write the final results. The details can be found in \cite{Gaspard:1989exq,Wirzba:1997hy}.

\begin{figure}[t!]
    \centering
    \includegraphics[width=0.36\textwidth]{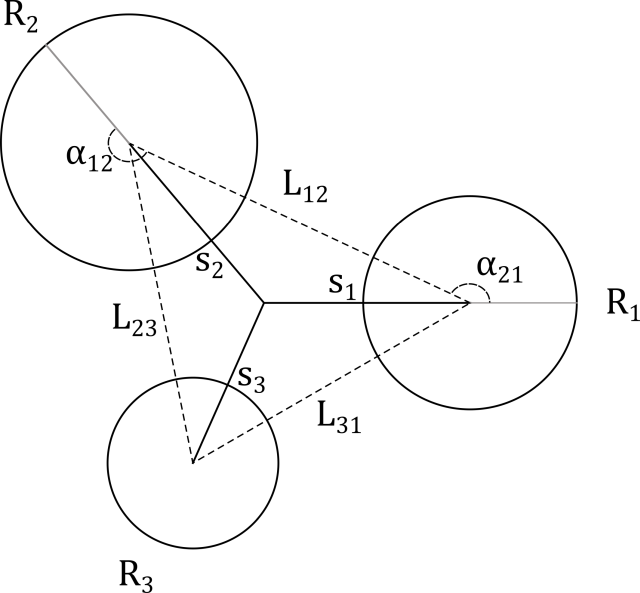}
    \caption{A generic set-up for the three-disk pinball with definitions of the various parameters.}
    \label{fig:genericpinball}
\end{figure}

The set-up is the following. We consider three disks of radii $R_{a}$ with ${a}=1,2,3$. They are centered at the points $\vec r_{a} \equiv s_{a}(\cos\chi_{a},\sin\chi_{a})$, such that the distance between the origin and each disk is indicated by $s_{a}$ and the angle in polar coordinates is denoted as $\chi_{a}$.

We denote the distance between a pair of disks as $L_{{a}{b}} = |\vec{r}_{a}-\vec{r}_{b}|$. We will also need to define the angle variable $\alpha_{{b}{a}}$ for ${a}\neq {b}$ which is the angle between the vector to the center of the  ${a}$-th disk $\vec{r}_{a}$ and the vector pointing from the center of ${a}$-th to the center of the ${b}$-th disk, $(\vec r_{b}-\vec r_{a})$, see the diagram in figure \ref{fig:genericpinball}. Note that $\alpha_{{a}{b}}\neq\alpha_{{b}{a}}$.

The answer for the $S$-matrix is given by
\begin{equation}
 S_{ll^\prime}(k) \equiv \delta_{ll^\prime} + i T_{ll^\prime}(k) = \delta_{ll^\prime} + i\, C_{l}^{a m} (M^{-1})_{am}{}^{a\pri m\pri} D_{l^\prime,a\pri m\pri}
\end{equation}
or $S = 1 + i C M^{-1} D$.  The matrices $C$ and $D$ are related to the free propagation before and after scattering from the disks. Before inversion, the matrix $M_{{a},m}{}^{{a}^\prime m^\prime}$ describes propagation between the disks ${a}$ and ${a}^\prime$. The inverse matrix $M^{-1}$ represents the multiple reflections on the various disks.

The values of $k$ for which $\det M = 0$ and the matrix is singular are identified with the resonances of the system. Resonances are generally found for complex values of $k$ and correspond to classical trajectories with a large number of bounces.

The matrices $C$, $M$, and $D$ all admit closed-form expressions, which are as follows:
\begin{equation} \label{eq:Mamam}
M_{{a}m}{}^{{a}^\prime m^\prime}  = \delta_{a}{}^{{a}^\prime}\delta_{m}{}^{m^\prime} + (1-\delta_{a}{}^{{a}^\prime})\frac{R_{a}}{R_{{a}^\prime}}\frac{J_m(k R_{a})}{H_{m\pri}^{(1)}(k R_{{a}^\prime})}H_{m-m\pri}^{(1)}(k L_{{a}{a}^\prime})e^{i[m\alpha_{{a}{a}^\prime}-m\pri(\alpha_{{a}^\prime {a}}-\pi)]}
\end{equation}
\begin{equation} \label{eq:Clam}
    C_{l}^{{a}m} = \frac{2i}{\pi R_{a}} e^{il\chi_{a}} \frac{J_{l-m}(k s_{a})}{H_m^{(1)}(kR_{a})}
\end{equation}
\begin{equation} \label{eq:Dlam}
    D_{l,am} = \pi R_{a}  J_{l-m}(k s_{a}) J_m(k R_{a}) e^{-il\chi_{a}}
\end{equation}
where $H^{(1)}_l$ is a Hankel function of the first kind, describing outgoing spherical waves at infinity.

Suppressing the internal angular momentum indices $m$ and $m^\prime$, and making explicit the labels of the disks $a$ and $b$, which become labels for blocks of $C$, $M$, and $D$, we can write the $S$-matrix also as:
\begin{equation} S_{l,l\pri} = \delta_{l,l\pri} + i \left(C_l^1\: C_l^2\:C_l^3\right) \begin{pmatrix}
    1 & M_1{}^2 & M_1{}^3 \\  M_2{}^1 & 1 & M_2{}^3 \\  M_3{}^1 &  M_3{}^2 & 1 \\ 
\end{pmatrix}^{-1} \begin{pmatrix} D_{l\pri,1} \\ D_{l\pri,2} \\ D_{l\pri,3} \end{pmatrix}
\end{equation}

In principle, the angular momenta can take any integer value and the matrices are infinite dimensional. On the other hand, one can see that for a given configuration at finite $k$, there is a natural cutoff, that one can deduce from the behavior of the Bessel functions, when expanded for large order. For positive large $l$ and $x$ fixed \cite{NIST:DLMF},
\begin{equation} J_l(x) \approx \frac{1}{\sqrt{2\pi l}}\left(\frac{ex}{2l}\right)^l\,,\qquad H^{(1)}_l(x) \approx \frac{-i}{\sqrt{2\pi l}}\left(\frac{ex}{2l}\right)^{-l}\end{equation}
Since the arguments of the Bessel functions in our case are always $k$ times one of the characteristic lengths $L_{char}$ in the system ($R_a$, $L_{ab}$) this means that at fixed $k$ there is an effective finite dimension of the matrix $N \sim {\cal O}(kL_{char})$. In \cite{Gaspard:1989exq} it is pointed out that the only condition for the combination of Bessel functions appearing in $M_{am}{}^{a\pri m\pri}$ to be small at large enough $m$ or $m\pri$ is that the disks do not overlap. This is always implicitly understood to be true in the pinball system.

Similarly, all components $T_{ll\pri}$ of the $T$-matrix are highly suppressed when $l$ is much larger than all of the $k R_j$ and $k L_{ij}$. Then, when $l$ is very large, the $T$-matrix is effectively zero and $S_{ll\pri} \approx \delta_{ll\pri}$. We can compute the non-trivial part of $S(k)$ to high accuracy by applying a finite cutoff to the size of the matrices, and that cutoff will scale with $k$. This sets a computational limit on how large  $k$  can be taken.

\subsection{The \texorpdfstring{$S$}{S}-matrix of the pinball system}\label{tsmotp}
We now examine the $S$-matrix to search for a correspondence with random matrix theory. Due to time-reversal symmetry of the system, one would expect the $S$-matrix to follow the statistics of the circular orthogonal ensemble (COE).

Though this was long conjectured, we have not found in the literature any explicit confirmation, as we will present in the following. One reason for this might be the surprising fact that the correspondence with COE does not hold unless one considers a sufficiently asymmetric configuration of the three disks. The symmetric configuration instead exhibits what appears to be a Poisson distribution, even though classically the system is chaotic regardless of the symmetry. 

As an illustrative example, we choose the configuration where the centers of the disks are at the corners of an equilateral triangle whose sides are of length $L = 3$. If the disks are all of equal radii, $R_1 = R_2 = R_3 = 1$ in our case, the system is symmetric under the discrete symmetry group $C_{3v}$ of rotations and reflections of the equilateral triangle. We can break the symmetry either by changing the positions or the radii of the disks. We will compare the fully symmetric case to the case where we break the symmetry by setting \begin{equation}
    R_1 = 1\,,\quad R_2 = 1-\epsilon\,,\quad R_3 = 1+\epsilon\,.
\end{equation}
with $0\leq \epsilon < 1$. Our main result is that even when the symmetry breaking parameter $\epsilon$ is small, the eigenvalues of the $S$-matrix are distributed as in the COE if we calculate it at large enough energy $k$. This is compatible with the classical behavior that is always chaotic and largely independent of $k$ or the energy. Indeed, the eikonal approximation is reliable for large $kL_{char}$, where $L_{char}$ is any of the physical scales relevant in the scattering process. Since we have set the scale of lengths in the system to be of order 1, for small $\epsilon$, one expects that $k$ needs to be large and of order $k\approx 1/\epsilon$.

We have computed the distributions as follows. After setting the parameters of the three-disk system, at any given value of $k$, the $S$-matrix is determined from equations \eqref{eq:Mamam}--\eqref{eq:Dlam}. After the numerical computations, the matrix is unitary as expected, thus its eigenvalues are all on the unit circle, $\lambda_n = e^{i\alpha_n}$. We remove the eigenvalues near $\lambda=1$, which come from the trivial part of the $S$-matrix at large $l$, and examine as usual the distribution of spacings of the eigenphases,
\begin{equation} \delta_n \equiv \alpha_{n+1}-\alpha_n \end{equation}
of the remaining eigenvalues.

We performed the analysis by sampling the system at multiple values of $k$ and checking the average distribution of spacings when all are combined together. After removing the eigenvalues very near $\lambda=1$,\footnote{More precisely, we selected only eigenvalues with $\lambda = e^{i\phi}$ with $0.1 < \phi < 2\pi-0.1$. The threshold of 0.1 was chosen by examining the eigenvalue density and choosing a region where it is approximately constant, to eliminate any ``edge'' effects from the special point $0$/$2\pi$.} we find that the average density of the remaining eigenvalues is approximately constant. Then, we can find agreement with the Wigner--Dyson distribution of the GOE, equal to the expected distribution of spacings of eigenphases in the COE:
\begin{equation} f(\delta) = \frac{\pi}{2}\, \delta\, e^{-\frac\pi4\delta^2} \,,\end{equation}
and the corresponding distribution of the normalized spacing ratios, $\tilde r_n = \min\left(\frac{\delta_{n+1}}{\delta_n},\frac{\delta_n}{\delta_{n+1}}\right)$ \cite{Atas:2013dis},
\begin{equation}
    \tilde f(\tilde r) = \frac{27}{4} \frac{\tilde r+\tilde r^2}{(1+\tilde r+\tilde r^2)^\frac52}\,, \qquad 0\leq\tilde r \leq 1\,.
\end{equation}
However, in order to see this one must consider a sufficiently asymmetric configuration of the three disks, as noted before.

\begin{figure}[t!]
    \centering
     \includegraphics[width=0.48\textwidth]{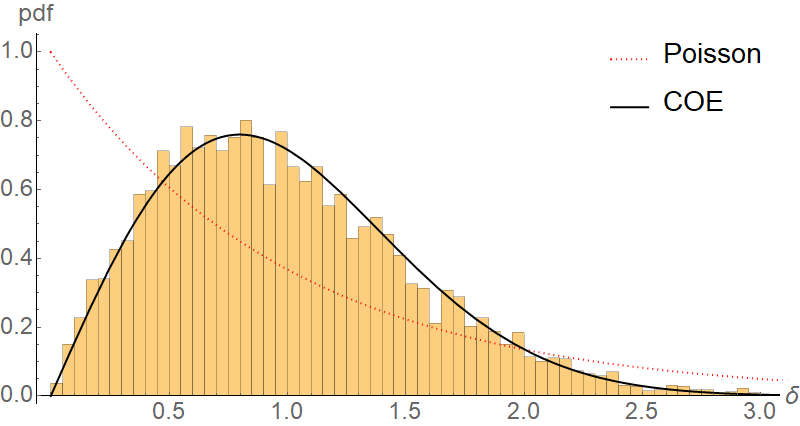}
    \includegraphics[width=0.48\textwidth]{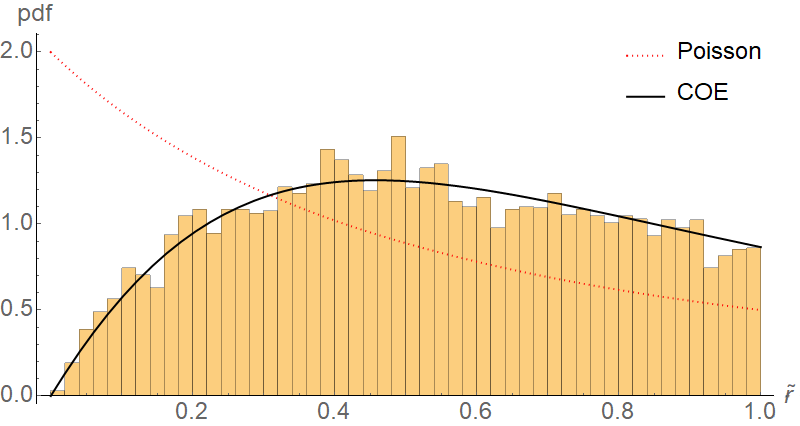}
    \caption{The spacings and spacing ratios of eigenvalues of the $S$-matrix for the asymmetric pinball system, with disks of radii $R_1 = 1$, $R_2 = 0.8$, $R_3 = 1.2$ placed on the vertices of an equilateral triangle of side length $L=3$.  The distribution is the accumulated distribution for eigenvalues of $S(k)$ with $k = (10,10.5,\ldots,40)$. We find excellent agreement with the Wigner-Dyson distribution of the COE.} 
    \label{fig:quantum_pinball_COE}
\end{figure}
\begin{figure}[t!]
    \centering
    \includegraphics[width=0.48\textwidth]{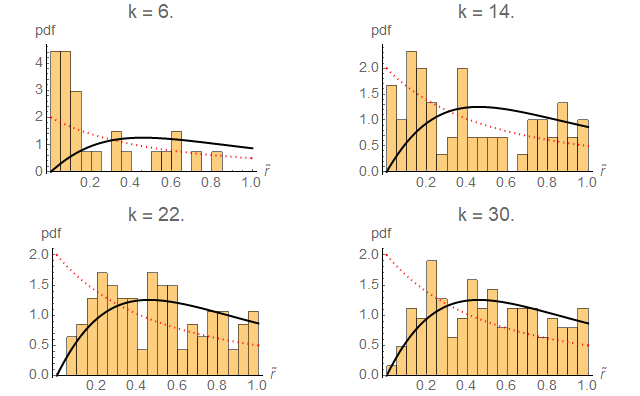}
\includegraphics[width=0.48\textwidth]{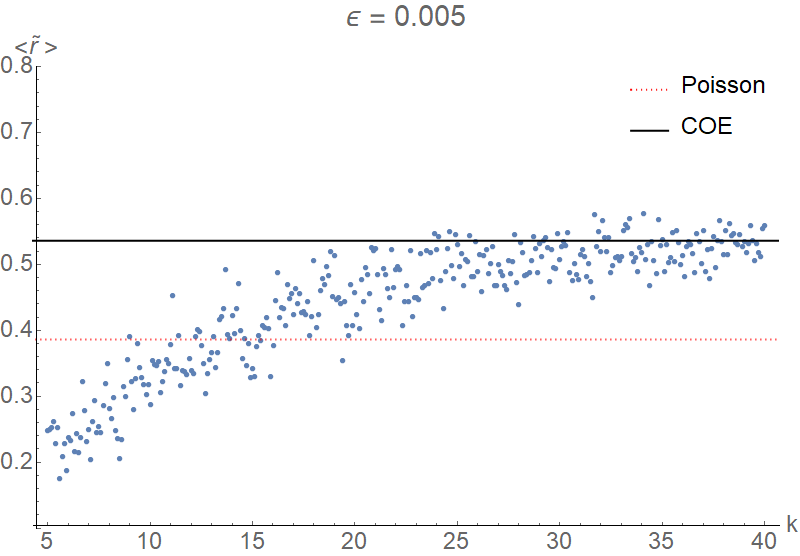}
    \caption{For a small symmetry breaking parameter $\epsilon=0.005$, the system exhibits a transition from a Poisson-like distribution (dashed red line) to COE (solid black). We see this either from looking directly at the distributions of eigenvalue spacings for different values of $k$ (left), or by plotting $\avg{\tilde r}$ as a function of $k$.}
    \label{fig:quantum_pinball_asym_transition}
\end{figure}

\begin{figure}[t!]
    \centering
    \includegraphics[width=0.48\textwidth]{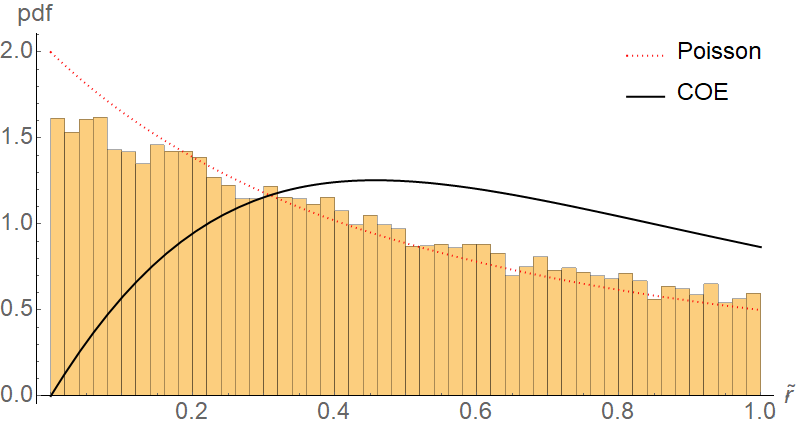}
\includegraphics[width=0.48\textwidth]{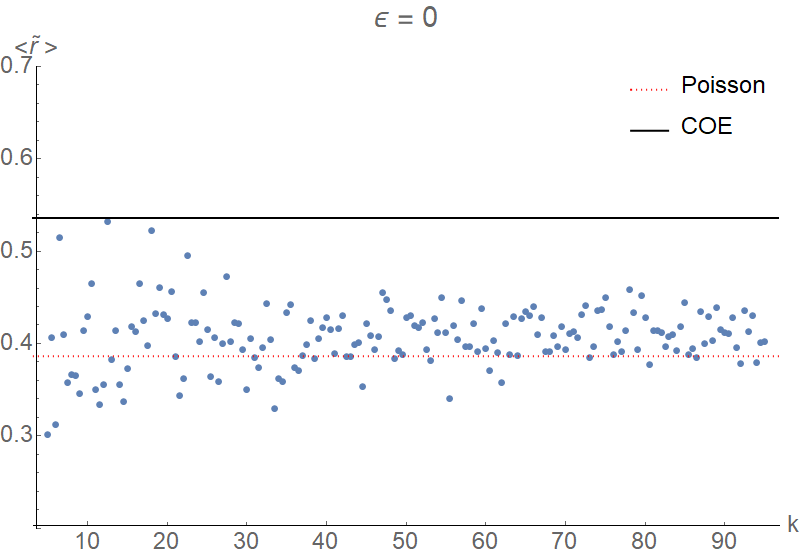}
    \caption{For the symmetric three-disk system, we observe a Poisson-like distribution (left) of the spacings of the unique eigenvalues of the $S$-matrix. Plotting $\avg{\tilde r}$ as a function of $k$ shows no increase from the Poisson value in that range (right). The data was collected for the range $k = (5,5.5,\ldots95)$.}
    \label{fig:quantum_pinball_sym_rn}
\end{figure}

Moreover, contrary to the classical case, one finds that such a physical observable as the distribution of the eigenphases depends on the energy $k$. If the symmetry breaking parameter is small, one can see a transition in the distribution of the eigenvalues of $S(k)$, changing from a Poisson-like distribution at small $k$, to the COE distribution at large $k$.  In figure \ref{fig:quantum_pinball_COE} we plot the distributions of $\delta_n$ and $r_n$ when $\epsilon = 0.2$. The eigenvalues were collected by accumulating the eigenvalues of $S(k)$ for $k = (10,10.5,\ldots,40)$ and drawing the average, combined distribution. The results show good agreement with the COE distribution.
 
We have observed, numerically, that the smaller the symmetry breaking parameter is, the larger $k$ has to be to reach the COE. If we plot the average value of $\tilde r_n$ as a function of $k$, taking $\epsilon = 0.005$ as an example, one can see that at small $k$ it increases, taking the Poisson value of $\langle \tilde r\rangle_{\text{Poisson}}\approx 0.39$ at around $k = 15$, and reaches the COE value of $\langle\tilde r\rangle_{\text{COE}} \approx 0.54$ from $k \geq 25$. This can be seen in figure \ref{fig:quantum_pinball_asym_transition}. Note that this transition occurs around $k \epsilon \approx 0.12$, which is not quite ${\cal O}(1)$. We observed that the transition occurs at smaller $k$ for larger values of $\epsilon$, but did not verify explicitly whether the expected dependence of $k \sim 1/\epsilon$ holds.

We compare it with the symmetric case $\epsilon = 0$. In this case, there is a degeneracy in many of the eigenvalues due to the symmetries, and we consider the spacings only between unique non-degenerate eigenvalues, removing the point at $\delta=0$ from the distribution. The result is a Poisson-like distribution - though it does not match Poisson exactly, it is peaked at zero and the average value $\avg{\tilde r}$ is near the expectation value for the Poisson case. Plotting $\avg{\tilde r}$ as a function of $k$ (figure \ref{fig:quantum_pinball_sym_rn}) reveals that it does not increase with $k$, at least not in the range where we have performed the numerical computations of up to $k \sim {\cal O}(100)$.\footnote{Note that this and all other computations in this work were performed on an ordinary personal computer. With more computing power or time, one can examine much larger values of $k$, but it is unnecessary for our purposes.}

Based on these results, one can conjecture that for finite $\epsilon$, one can always find a large enough $k$ for which the distribution becomes COE, while in the $\epsilon\to 0$ limit, the transition occurs at $k\to\infty$, corresponding to taking a classical eikonal limit, as discussed above.

One can repeat the same analysis for the case where the symmetry is broken not by taking disks of different sizes, but by placing three disks of radius $R=1$ on the corners of an asymmetric, scalene triangle. The results are the same. One finds the COE distribution if one goes to large enough $k$ relative to the symmetry breaking scale.

%\clearpage
\subsection{The quantum scattering amplitude}\label{tqsa}
In \cite{Bianchi:2022mhs,Bianchi:2023uby} we showed that for string scattering amplitudes involving highly-excited string states, one can find RMT distributions when looking at the spacings of consecutive peaks in the angular dependence of the amplitude. We would like to see if similar distributions emerge when looking at the angular dependence of the scattering amplitude in the pinball system.

We will consider the differential cross-section, which is a function of the two angles, incoming and outgoing, as well as the energy, related to $k$. At fixed $k$, the differential cross-section given by $|f(\phi,\phi\pri)|^2$ is a complicated fluctuating function of the two angles, which we will examine in detail in the remainder of this section. Even though in this system we have already observed that there is a correspondence of the $S$-matrix to COE, we would like to examine if one could see this from examining directly the scattering amplitude, as we did for string amplitudes in \cite{Bianchi:2022mhs,Bianchi:2023uby}.

\begin{figure}[t!]
    \centering
    \includegraphics[width=0.48\linewidth]{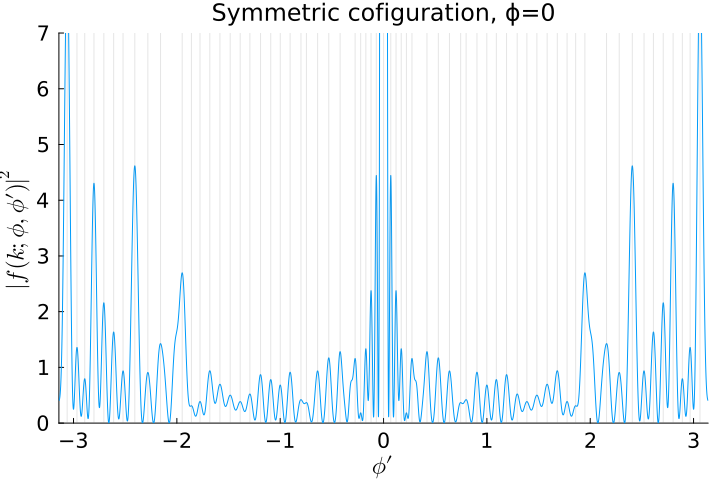}
    \includegraphics[width=0.48\linewidth]{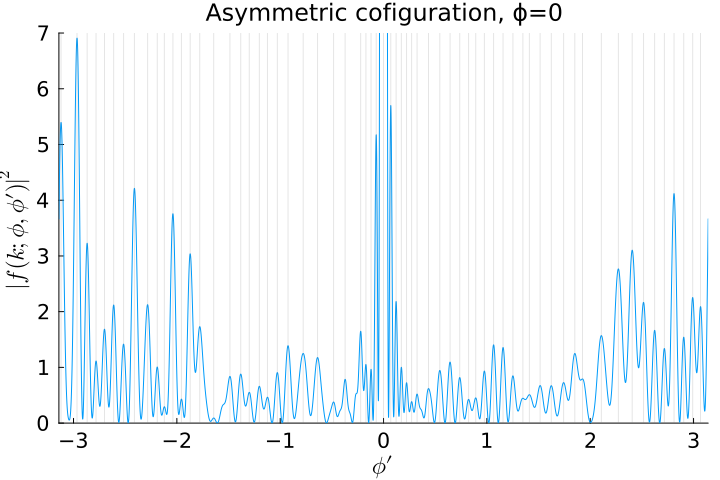}
    \caption{The differential cross-section at fixed $k = 25$ and incoming angle $\phi = 0$, as a function of the outgoing angle. for the pinball system with $L=3$ for the symmetric system with $R_i = 1$, and the asymmetric system $R_i = (1,1.2,0.8)$. The value of the functions at the large peak at $\phi=\phi\pri=0$ is around 70 in this case.}
    \label{fig:quantum_pinball_1d}
\end{figure}

We can begin by taking the function at fixed $k$ and incoming angle $\phi$, and plot it as a function only of the outgoing angle $\phi\pri$. We can see that function exhibits a large peak at $\phi=\phi\pri$, corresponding to forward scattering, and many smaller peaks besides. We plot the function in figure \ref{fig:quantum_pinball_1d}, for two configurations where the disks are on the equilateral triangle with $L=3$, taking once the fully symmetric system with $R_1=R_2=R_3 = 1$, and once the asymmetric system with $R_1=1$, $R_2 = 1.2$, $R_3 = 0.8$. When we plot the function at $\phi = 0$, we see for the former the reflection symmetry $\phi\pri\to-\phi\pri$.

The peaks in the differential cross-section as a function of the angle are marked as vertical lines in the plot. They appear to be almost regularly spaced. We will analyze the distribution of these spacings in later sections, considering the full two-dimensional picture.

\begin{figure}[t!]
    \centering
    \includegraphics[width=0.48\textwidth]{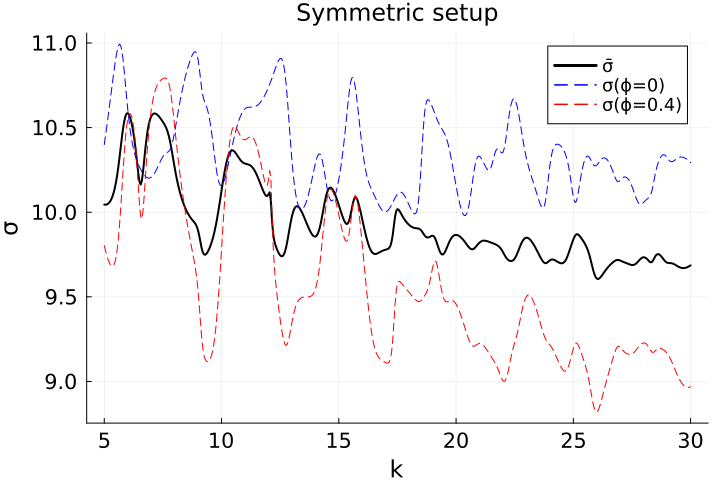}
    \includegraphics[width=0.48\textwidth]{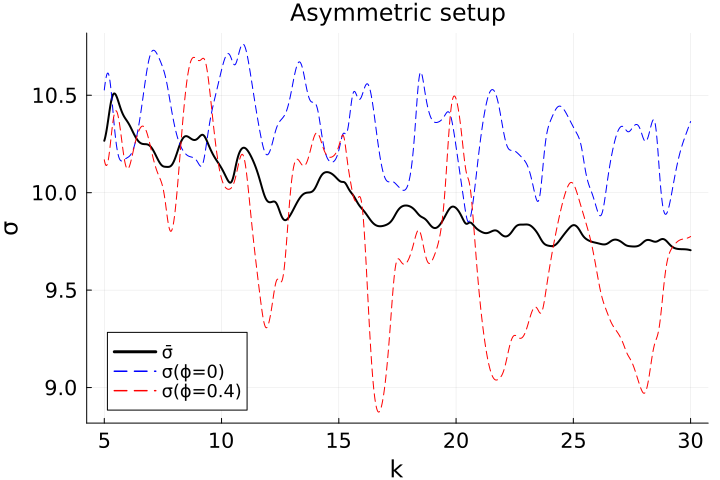}
    \caption{The total cross-section as function of $k$ for the pinball system with $L=3$ for the symmetric system with $R_i = 1$, and the asymmetric system $R_i = (1,1.2,0.8)$. We plot the cross-section at fixed incoming angle $\sigma(k;\phi)$ for two different values of $\phi$, as well as the total average cross-section $\bar\sigma(k)$, averaged over all incoming angles.}
    \label{fig:sigmaofk}
\end{figure}

In the classical case, there was no non-trivial dependence on the velocity, but in the quantum system the dependence of $k$ is significant. If we plot the total cross-section as a function of $k$, we can see a structure of resonances. However, the positions of peaks in the plot depend on whether one looks at a specific incoming angle or the average total cross-section, as can be seen in figure \ref{fig:sigmaofk}. The actual positions of the resonances (which have also an imaginary part) is most accurately determined from the locations of the singularities of the matrix $M$ (eq. \eqref{eq:Mamam}), and these also are quite regularly spaced, as noted already in \cite{Gaspard:1989exq}. Some long-lived resonances, \textit{i.e.} with small imaginary part, can be observed as a small kink in the plot of the total cross-section, for example around $k = 12$ for the symmetric system plotted in figure \ref{fig:sigmaofk}. It depends on the configuration of the system whether such long-lived resonances appear. 

Other than the larger cross-section, there appears to be no qualitative difference in the behavior of the scattering amplitude as a function of the angle depending on whether the chosen value of $k$ is near a resonance or not. In particular, we do not observe any correlation between the total cross-section and the parameter $\langle\tilde r_n\rangle$ of the distribution of $S$-matrix eigenvalues.

%%%%%%%%%%%%%%%%%%%%%%%%%%%%%%%%%%%%%%%%%%%%%%%%%%%%%%%%%%%%%%%%%

%\clearpage
%%%%%%%%%%%%%%%%%
\subsection{Two-dimensional description of the quantum differential cross-section}\label{tddotqs}
The differential cross-section given by eqs. \eqref{scatamp}-\eqref{difcros} is a function of the boundary conditions, namely the radii and locations of the disks as well as the incident and outgoing angles and the wave-number $k$.

In our analysis of the two-dimensional plots we would like to: 
\begin{enumerate}[(i)]
\item Determine the dependence on the wave-number $k$ and in particular whether the pattern  is different in the vicinity of a resonance. 
\item 
Identify ``topographic structures" like isolated peaks and minimum points, saddles points, ridges, valleys, etc. 
\item 
Determine the symmetries of the patterns. 
\item 
Examine whether the plots admit a self-similar structure  and whether  one can find the corresponding  fractal dimension like in the one-dimensional analysis of figure(\ref{fig:pinball_angle_dependence}).
\item 
Analyze the distances between peaks. For this we will need to introduce the two-dimensional measures in section \ref{tdes}. The analysis will then be carried out in section \ref{aopop}.
\end{enumerate}

These goals, in particular the last two, should enable us to define a measure for multi-dimensional chaoticity that can distinguish between chaotic and integrable systems.

To this end, it is most useful to compare the systems that we have found are chaotic in the sense that its $S$-matrix eigenvalues admit the COE distribution, to systems in which we find a Poisson distribution of the same eigenvalues.

Our basic configuration will again be the one in which we choose the centers of the three disks to be on the corners of the equilateral triangle with $L = 3$, and we choose the radii of the three disks to be $R_1 = 1$, $R_2 = 1.2$ and $R_3 = 0.8$, breaking the symmetry of the system.

We can plot the differential cross-section as a function of both angles for several configurations and values of $k$.

\begin{figure}[t!]
    \centering
    \includegraphics[width=0.48\linewidth]{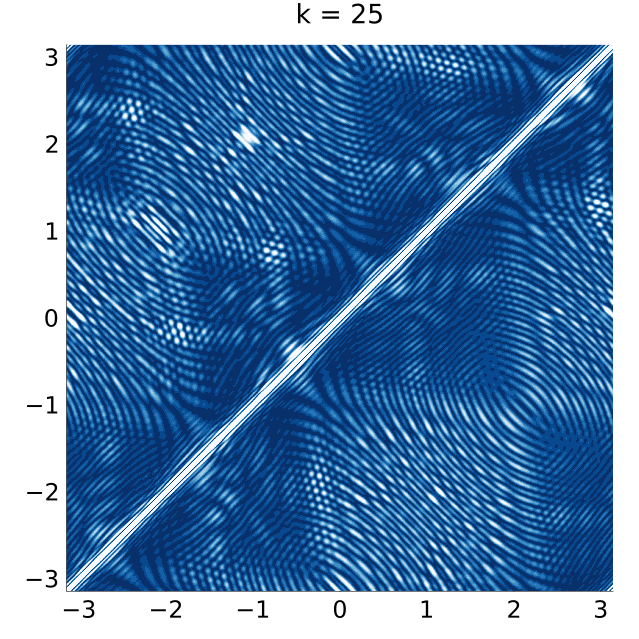} \\
    \includegraphics[width=0.48\linewidth]{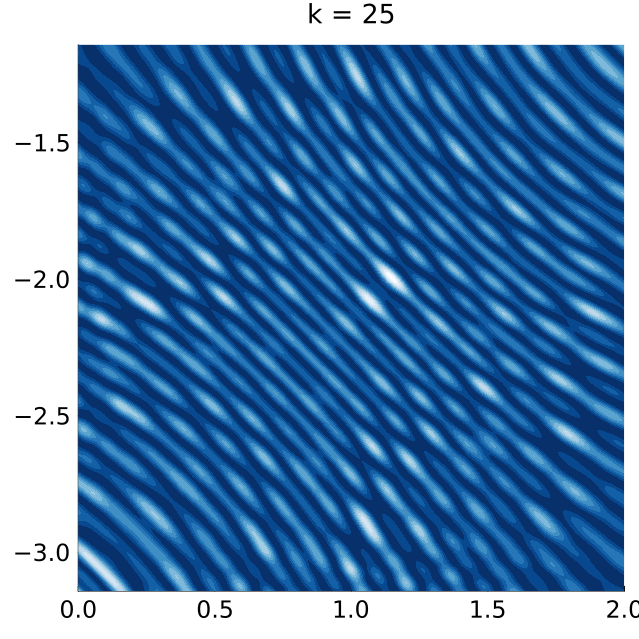}
    \includegraphics[width=0.48\linewidth]{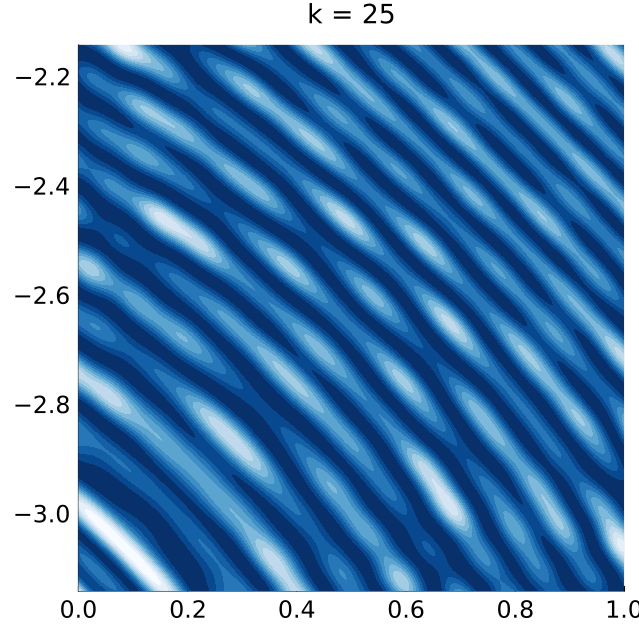}
    \caption{The differential cross-section for the chaotic system with $R_i= (1,1.2,0.8)$ and $L=3$ as a function of both angles, for $k = 25$. Lighter areas represent higher values. The plots on the bottom row are zoomed in versions of the top one.}
    \label{fig:quantum_pinball_zoom}
\end{figure}

The three plots in figure \ref{fig:quantum_pinball_zoom} show the asymmetric system at a fixed value of $k=25$, first the full range of the angles, and then zooming in. We can make the following observations, which are common to all the two-dimensional plots in the following:
\begin{itemize}
    \item There is \textbf{no self-similarity}. This is not surprising since there  are other quantum chaotic systems that are not self similar, for instance the decays and scattering of highly excited string states \cite{Bianchi:2023uby,Hashimoto:2022bll}.
    \item When we zoom in on the function, we see that it has many \textbf{isolated peaks}.
    \item As for the \textbf{symmetries} of the patterns, all the two-dimensional plots are invariant under $Z_2^{\rm par}\times Z_2^{\rm per}$ transformation, namely reflection around the $\phi=\phi\pri$ line and then reflection around the $\phi=2\pi-\phi\pri$, taking:
\begin{equation}
(\phi,\phi\pri)\rightarrow (\phi-\pi,\phi\pri-\pi)
\end{equation}
This is a consequence of the time-reversal invariance, being an exchange of the incoming and outgoing angles.
\item As expected there is a line of maxima along the forward scattering direction $\phi=\phi\pri$. 
\end{itemize}

In appendix \ref{app:supplfig} we include several additional figures inspecting the dependence on $k$ and the parameters of the three-disk system. We summarize here the conclusions:

\begin{itemize}
\item Increasing $k$ has an effect of zooming in to get ``higher-resolution'' images, with more peaks.
\item There is no qualitative difference in the angular dependence at the near-resonance value of $k=12.34$.

\item At all $k$ there persist some large ``macroscopic'' structures, which are made more distinct at larger $k$. For instance there are hexagonal-like cells along the line $\phi=\phi\pri$. 
\end{itemize}

In additional to the dependence on $k$, we can plot the figures as we change the configuration from the asymmetric system of $L = 3$, $R_i = (1,1.2,0.8)$, where the distribution of $S$-matrix eigenvalues was COE, to other systems where the same distribution was Poisson. In all cases there is a continuous parameter that we can use to smoothly change from one to the other.

\begin{figure}[p!]
    \centering
    \includegraphics[width=0.48\linewidth]{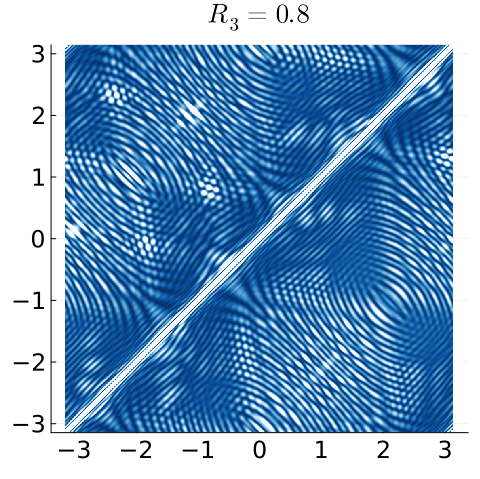}
    \includegraphics[width=0.48\linewidth]{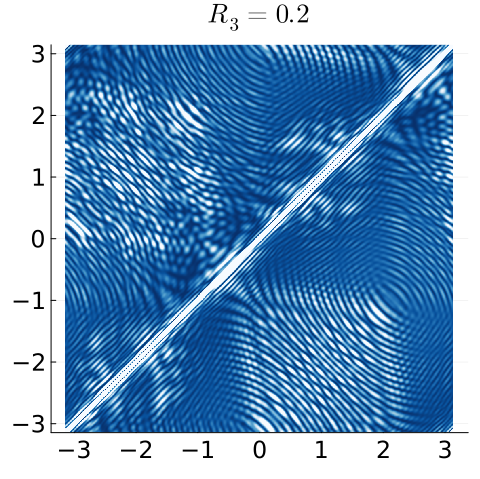}
    \includegraphics[width=0.48\linewidth]{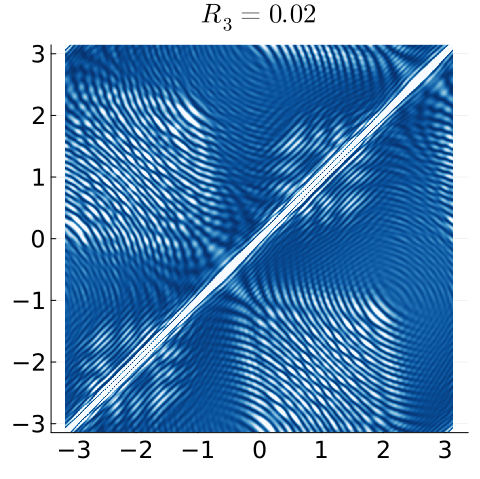}
    \includegraphics[width=0.48\linewidth]{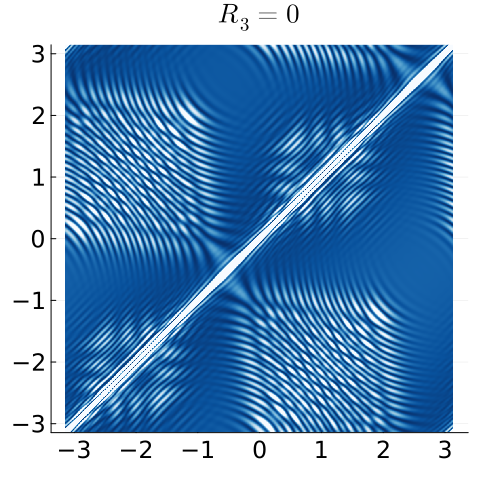} 
    \caption{The differential cross-section for the system with $R_1=1$, $R_2=1.2$ and $L=3$ where the radius of third disk is shrunk from $R_3=0.8$ (chaotic three-disk system) to $0$ (non-chaotic two-disk) in several steps. We set $k=25$ for all plots.}
    \label{ShrinkingR3}
\end{figure}

There are at least three ways to do that:
\begin{itemize}
\item 
Shrinking the radius $R_3$ gradually to move from the three-disk to the non-chaotic two-disk system.  
\item 
Gradually removing the asymmetry in the radii by taking $R_1=1$, $R_2=1+\epsilon $ and $R_3=1-\epsilon$, and decreasing $\epsilon$ from 0.2 to 0, the fully symmetric system.

\item Starting from a triangular configuration, then moving the disks continuously to bring the system to a configuration where the centers of the three disks all lie on a straight line.

\end{itemize}
The transition from three to two disks is plotted here in figure \ref{ShrinkingR3}. The remaining two cases are shown in the appendix.\footnote{In addition, we have prepared a few animated versions of these figures to visualize the transitions. They are available at \cite{stringchaosgithub:2510}.}

Our purpose in this section has been to search for qualitative measures or identifications of two-dimensional chaotic behavior. The patterns associated with the symmetric setups admit certain symmetries discussed above and contain certain symmetric regions like hexagons and rhombi. These structures are likely not signs of chaos. They appear also in the case of the two disks, which is known to be integrable.

It does not seem possible from simple visual comparison to distinguish between the cross-sections associated with chaotic systems, and those of integrable ones. To do that we will need to develop quantitative tools, which will be the focus of the following sections.

%The symmetric setups do not admit an $S$-matrix that can be associated with any $\beta$ ensemble as was shown in section \ref{tsmotp}. For the asymmetric cases  the patterns are naturally less symmetric.

Finally, we note that, although we will not discuss any generalization to pinball systems with more than three disks, they are solvable both classically and quantum mechanically and display chaotic behaviors in the same observables as we have considered for three-disk system. The code associated with this paper \cite{stringchaosgithub:2510} can be also used to solve generic $n$-disk systems.

% {\bf Referee 2, Comment 1:}

% {\it The paper is unfocused and overly long relative to its core results. A large number of plots are presented — many showing negative or null results (e.g., no self-similarity in the quantum case, no qualitative difference near resonances, similar patterns for chaotic and non-chaotic configurations in the 2D cross-section plots). While documenting these is not without value, the sheer volume dilutes the impact of the genuinely new findings. The paper would benefit substantially from a more disciplined presentation that foregrounds the key positive results and relegates exploratory/negative observations to supplementary material.}

\clearpage
\section{Modeling the spacings between peaks in two dimensions}\label{mtsbposa}
For the pinball system both the plots of the classical scattering angles and those of the quantum differential cross-section as a function of the two angles admit a rich structure that includes isolated peaks, minimum points, saddle points, ridges, valleys, etc. In our previous work \cite{Bianchi:2022mhs,Bianchi:2023uby,Bianchi:2024fsi}, we emphasized the important role that the spacings between adjacent peaks of the scattering amplitudes play in chaotic scattering. We would like now to analyze this topic in the context of two-dimensional scattering. 

For this purpose we introduce four measures of the distribution of spacings useful in two- or higher-dimensional systems: the spacings of all pairs, the spacings of nearest neighbors, consecutive spacing ratios on a path, and spacings of projections on an axis.

To provide a concrete example of the behavior of these measures, we apply them to a simple toy model inspired by the system of the ``leaky torus'' \cite{Gutzwiller:1983}. We introduce and study a toy model for erratic functions with various types of ensembles of randomly-positioned peaks. The model is based on a generalization of the phase shift in the leaky torus system, which we review in the following subsection. We rely on random matrices for the locations of the peaks.

We use this toy model to illustrate the behavior of the spacings of peaks in two dimensions. We end with the computation of the generalization of the scattering form factor (ScFF) introduced in \cite{Bianchi:2024fsi} for the toy model, and show how it reveals additional information about preferred directions in the two-dimensional system.
 
\subsection{The phase shifts of the leaky torus and the inspired toy model}\label{tpsotlt}
A landmark model that displays chaotic behavior is scattering on the leaky torus. Originally proposed by Gutzwiller \cite{Gutzwiller:1983}, the leaky torus geometry is constructed by taking the two-dimensional hyperbolic plane with the metric 
\begin{equation} ds^2 = \frac{dx^2+dy^2}{y^2} \label{eq:leaky_metric} \end{equation}
where we set the radius of curvature to unity. One looks at the region, in the upper half plane \(y>0\), between the geodesics (i) \(x=-1\), (ii) \(x = 1\), (iii) \((x-\frac12)^2+y^2 = (\frac12)^2\), and (iv) \((x+\frac12)^2+y^2 = (\frac12)^2\).
Then, identifying boundary (i) with (iii) and (ii) with (iv), the result is a torus with a cusp point at infinity.

Scattering in this setting involves sending an incoming free wave from the cusp point \(y=\infty\) and measuring the phase shift of the outgoing wave at some finite \(y = y_0 > 0\). The $S$-matrix and phase shift are found to be exactly\footnote{The standard convention for partial waves is $S_\ell(E) = e^{2i\delta_\ell(E)}$. For consistency with the original references we absorb the factor of 2 into $\delta(k)$.}
\begin{equation} S(k) \equiv e^{i\delta(k)} = \frac{\pi^{-ik}\Gamma(\frac12+ik)\zeta(1+2ik)}{\pi^{+ik}\Gamma(\frac12-ik)\zeta(1-2ik)} \end{equation}
where $k = \sqrt{2E}$ is the momentum of the incoming wave. 

The Wigner time delay function is given in general by the determinant of the logarithmic derivative of the $S$-matrix, which here reduces to\footnote{We mostly follow chapter 8 of \cite{Hurt:1997quantum}.}
\begin{equation} {\cal T}(k) \equiv \frac{d\delta(E)}{dE} = \frac{1}{k}\frac{d\delta(k)}{dk}\end{equation}
It is given explicitly by ${\cal T}(k) \equiv \frac{1}{k}\tau(k)$ with
\begin{align}
     \tau(k) = 4\left(\log(2\pi)-1-\frac{\gamma}{2}\right)-\frac{1}{\frac14+k^2} + \sum_{k_n > 0}\frac{1}{(\frac14)^2+k_n^2} + \nonumber\\
     + \sum_{k_n > 0}\left(\frac{1}{(\frac14)^2+(k+k_n)^2}+\frac{1}{(\frac14)^2+(k-k_n)^2}\right) 
\end{align}
where $\gamma = - \Gamma'(1)$ is Euler-Mascheroni constant and the sum runs over the non-trivial zeroes of the Riemann zeta function located at $z_n = \frac12 + 2i k_n$.\footnote{Riemann conjectured that all the non-trivial zeros lie on the critical line ${\rm Re}(z)=1/2$.} The term in the second line comes from the fluctuating part of $S(k)$ (involving only the zeta function), and it is given as a series of resonances located at (half the imaginary parts of) the zeta function zeros $k = \pm k_n$, all having the same residue and width.

The chaotic nature of this function comes from the fact that the spacings of the non-trivial zeros of the Riemann zeta function follow the Wigner--Dyson distribution of the GUE \cite{Odlyzko:1987}. It follows that $\tau(k)$ is a function with randomly spaced peaks \cite{Rosenhaus:2020tmv}, the same notion of chaos that we proposed for scattering amplitudes in \cite{Bianchi:2022mhs,Bianchi:2023uby}.

We can isolate the fluctuating part of $\tau(k)$ that has this property. It has the form:
\begin{equation} \label{eq:taufl}
\tau_{\text{fl}}(k) = \sum_{n}\frac{1}{(k-\lambda_n)^2 + \Gamma_n^2} \end{equation}
where $\lambda_n$ can be identified with the eigenvalues of a random matrix, and $\Gamma_n$ is the width of each resonance. In the leaky torus case the correspondence is to GUE eigenvalues, and the widths are all equal.

%\subsubsection{The toy model}\label{ttm}
One possible construction of a similar function to $\tau_{\text{fl}}(k)$ in two dimensions is
\begin{equation}
 {\cal F}(x,y) = \sum_{n=1}^N \frac{1}{\sqrt{(x-\lambda^{(1)}_n)^2+(y-\lambda^{(2)}_n)^2+z_0^2}} \label{eq:f2d}   
\end{equation}
where $\lambda^{(1)}$ and $\lambda^{(2)}$ are the eigenvalues of two independent random matrices, $M^{(1)}$ and $M^{(2)}$, drawn from some ensemble and $z_0$ is a non-zero real constant.

We have defined the function with a square root in the denominator to allow for a simple physical interpretation: ${\cal F}(x,y) = V(x,y,z=z_0)$ is the electric potential in three-dimensional space, generated by $N$ identical point charges located on the 2D plane $z=0$, as measured at $z=z_0$. The charges are assumed to be randomly distributed, and are located at the points $(x_n=\lambda^{(1)}_n,y_n=\lambda^{(2)}_n)$ with $0\le \lambda^{(1)}_n,\lambda^{(2)}_n\le N$.
Note that whereas the phase shift of the leaky torus is a quantum property, the electric potential just mentioned is classical. It is thus an example of an erratic function that has both classical and quantum interpretations.

% {\bf Referee 2, Comment 3:}

% {\it 
% The toy model's relevance to actual scattering is unclear. In the two-matrix toy model (Section 4.3), the 2D "eigenvalues" are constructed from two independent sets of random-matrix eigenvalues paired via a random permutation. In physical scattering, however, the kinematic variables (angles, momenta) are not independent — they are constrained by unitarity, crossing symmetry, and the specific dynamics encoded in the S-matrix. The authors acknowledge that the eigenvectors of the physical S-matrix are not random (unlike in the COE model of Section 5.1), but do not address the more fundamental issue that the factorised structure $\lambda_n = (\lambda_n^{(1)}, \lambda_n^{(1)})$
%  has no obvious physical justification for the pinball or string amplitudes. This weakens the interpretive power of the toy model results considerably.}

\begin{figure}[ht!]
    \centering
    \includegraphics[width=0.33\textwidth]{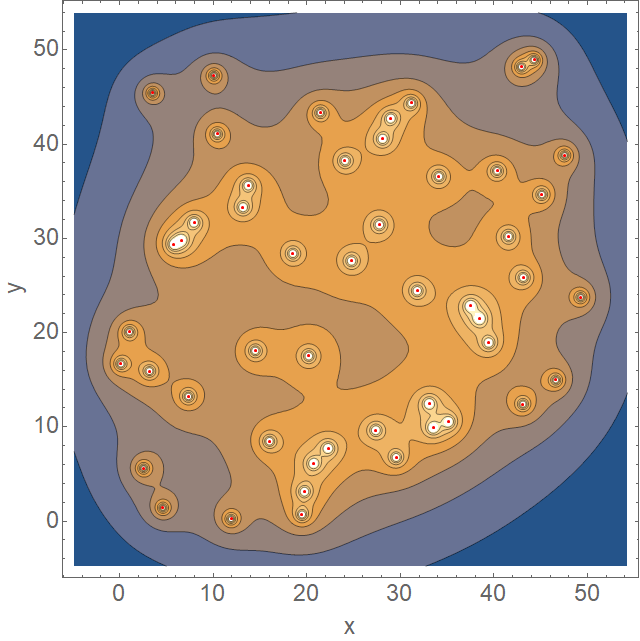}
    \includegraphics[width=0.40\textwidth]{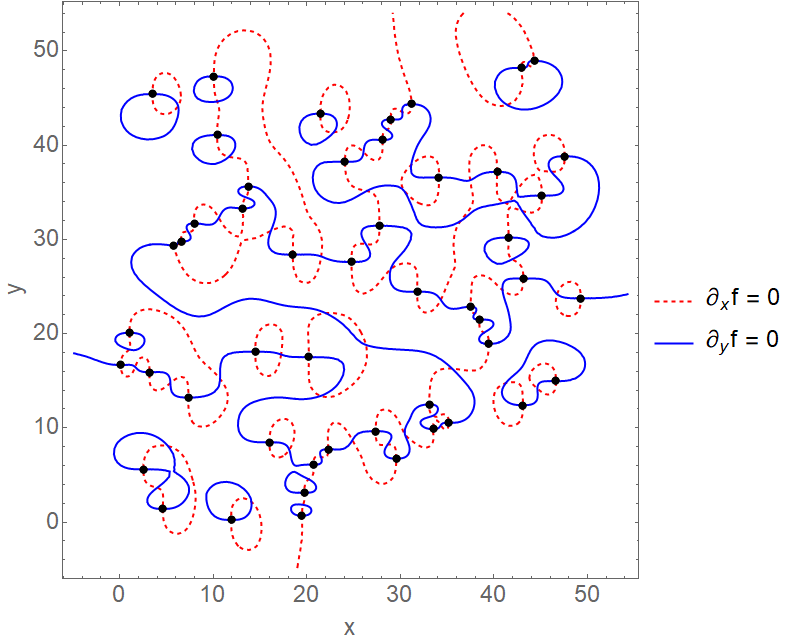}
    \caption{An example of the function \eqref{eq:f2d} with eigenvalues $\{\lambda_n^{(1)}\}$ and $\{\lambda_n^{(2)}\}$ corresponding to unfolded GOE spectra. Here $N=50$ and $z_0 = \frac14$. Left: Contour plot of the function. Lighter areas represent higher values, and the positions of peaks are marked with red points. Right: Plot of lines where the partial derivatives of $f$ are zero, with peaks marked in black. }
    \label{fig:2dpeaks}
\end{figure}

This function has randomly located peaks on the plane. If we choose $z_0$ such that it is smaller than the typical spacing between peaks, they will typically not overlap and we can see them all distinctly. For illustration we plot in figure \ref{fig:2dpeaks} an instance of this function with $\{\lambda_n^{(1)}\}$ and $\{\lambda_n^{(2)}\}$ drawn independently from the GOE. We also plot the lines along which the partial derivatives ${\partial}{\cal F}/{\partial x} $ and ${\partial}{\cal F}/{\partial y}$ are zero. The peaks are located at the intersections of these lines.

For the rest of this section we will ask the question of how, given such a function, we can perform an analysis of the positions of its peaks to uncover the underlying distribution of eigenvalues.

\subsection{Two-dimensional eigenvalues and their spacings}\label{tdeats}
\subsubsection{Two-dimensional eigenvalue spectra}\label{tdes}
Our purpose is to analyze the spacings between the peaks as a function of two variables. We want to ask the following question. Let us assume that we have analyzed some physical system, and obtained a function of two variables with a ``spectrum'' of peaks given by ${\vec\lambda}_n$. How would one diagnose whether this system is chaotic or not?

One option is to fix one of the variables, and analyze the spacings between the peaks as a function of the other variable. It is clear that for a function such as \eqref{eq:f2d} this loses a lot of information, but it is a useful indicator nonetheless.

After analyzing  such a function  we have a set of two-dimensional eigenvalues,
\begin{equation}
{\vec\lambda}_n = (x_n, y_n)\,,\qquad n = 1,\ldots N \end{equation}
We can unfold the spectrum by generalizing the standard procedure. After measuring the 2D density function $\rho(x,y)$, the unfolding procedure should map the eigenvalues to new variables 
\begin{equation}
    \vlam_n = (x_n,y_n) \to {\vec s}_n=\left(u(x_n,y_n),v(x_n,y_n)\right) \label{eq:unfold2d}
\end{equation}
in terms of which the density would be constant,
\begin{equation} du\, dv = \rho(x,y) dx\, dy \end{equation}
so the problem is to find a transformation for which the Jacobian determinant is one.

One solution is to use cumulative distribution functions. In terms of
\begin{equation}
    \rho_x(x) = \int_{-\infty}^\infty \!\!dy \,\rho(x,y)
\end{equation}
the new coordinates are
\begin{align}
    u(x,y) &= \int_{-\infty}^{x} dx\pri \,\rho_x(x\pri) \\
    v(x,y) &= \frac{1}{\rho_x(x)}\int_{-\infty}^y \!\! dy\pri \,\rho(x,y)
\end{align}
In the case where $x$ and $y$ are independent variables and $\rho(x,y) = \rho_x(x)\rho_y(y)$, this reduces to performing the standard 1D unfolding using the CDF on each of the variables separately.

\subsubsection{Two-dimensional spacings of eigenvalues}\label{tdsoe}
In two or higher dimensions, the eigenvalues cannot be ordered, so we cannot directly define the consecutive level spacings equivalent to 1D spectra.

There are several ways to define the spacings. We will examine the following:

\paragraph{I. Spacings for all possible pairs:}

We can look at the distribution of all $N(N-1)/2$ spacings defined by:
\begin{equation}
\delta_{mn} \equiv |{\vec\lambda}_m-{\vec\lambda_n}|\,, \qquad m < n     
\end{equation} 
This has the advantage that we do not need to define an ordering of the eigenvalues before using it. However, since they are not ordered, we cannot define the spacing ratios, and therefore we will need to unfold the spectra to uncover the universal behavior.

\paragraph{II. Spacings of nearest neighbors (NN):}

This would be the distribution of
\begin{equation}
\delta^{(\text{NN})}_{n} \equiv \min_{m\neq n}\left(|{\vec\lambda}_n-{\vec\lambda_m}|\right)    
\end{equation} 
Namely for each point we take the distance between it and its nearest neighbor. This means looking at only a small subset of all the spacings $\delta_{mn}$. Note that each point has exactly one nearest neighbor by definition, but a given point can be the nearest neighbor of more than one of the other points.

% A related measure was proposed in \cite{Sa:2020fpf} there is a definition of
% \begin{equation}
% r_n^{(\text{NN})} = \frac{|\vlam_n-\vlam_{\text{NN}(n)}|}{|\vlam_n-\vlam_{\text{NNN}(n)}|}
% \end{equation}
% being the ratio between the distance to the nearest neighbor and the distance to the next-to-nearest neighbor of any point. Like the ordinary spacing ratio it has the advantage that it does not require any unfolding or normalization. We will propose another notion of spacing ratios as follows.

\paragraph{III. Consecutive spacings along a path:}

We can use the following algorithm to define an ordering and the notion of successive spacings:
\begin{enumerate}
    \item Begin by choosing an initial point, ${\vec \lambda}_1$.
    \item The next point, ${\vec \lambda}_2$, is the nearest point to ${\vec \lambda}_1$ in the sense that $|{\vec\lambda}_1-\vec{\lambda}_2|$ is minimized.
    \item Repeat: ${\vec \lambda}_{k+1}$ is defined as the point nearest to ${\vec \lambda}_k$, excluding points that have previously been chosen.
\end{enumerate}
This algorithm defines a path that starts from a chosen initial point, and visits all other points once, in a determined order. There are $N$ possible paths, one per each initial point. There is some degree of overlap between these paths, but there can be considerable differences between one path and the other. There are several possible ways to address this issue: one can always choose a special reference point such as the point nearest to the origin, the point with the smallest value of $x$, the point nearest to the center of the distribution, and so forth. Conversely, one can choose an initial point at random, or perform an average over all possible paths.

The spacings defined as
\begin{equation} \delta_n = |{\vec\lambda}_{n+1}-{\vec\lambda}_n| \end{equation}
will be analogous to successive level spacings in 1D. This algorithm also allows us to define the usual spacing ratios as \begin{equation} r_n = \delta_{n+1}/\delta_n \end{equation}
One disadvantage of this definition is that it is difficult to perform any analytic computations to predict the distribution of $\delta_n$ and $r_n$.

\paragraph{IV. Spacings of projections on an axis:}

Given $\vlam_n = (x_n,y_n)$, after we have chosen the $x$ and $y$ axes, we can consider the distribution of spacings of the projections on the two axes. If one orders the eigenvalues such that $x_1 \leq x_2 \leq ...$, then one can analyze the distribution of the (positive) variable
\begin{equation} \delta^{(1)}_n = x_{n+1}-x_n  \: .\end{equation}
Similarly one can consider the distribution of $\delta^{(2)}_n=y_{n+1}-y_n$ after ordering the eigenvalues by their $y$-value.

This method has the disadvantage that, if one does not choose the correct axes, one can fail to see the level repulsion characteristic of chaotic spectra, as we will show in the following.

\begin{figure}
    \centering
    \includegraphics[width=0.40\textwidth]{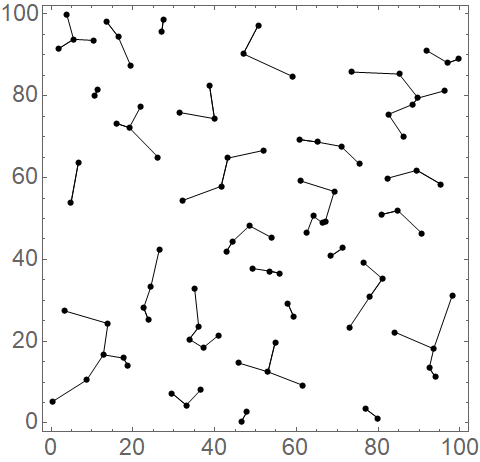}
    \includegraphics[width=0.40\textwidth]{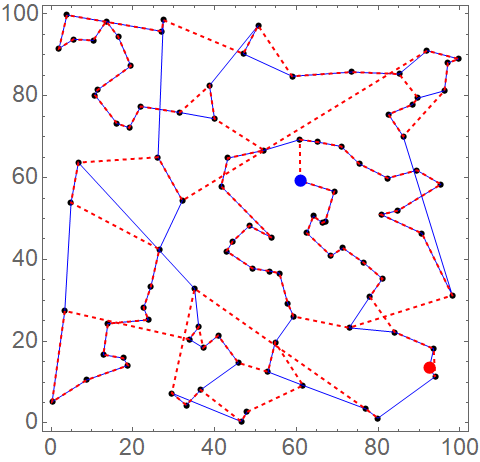}
    \caption{A spectrum of $N=100$ eigenvalues of in $x$ and $y$, both drawn independently from GOE, and their spacings. On the left, each point is connected to its nearest neighbor. Right: two choices of paths connecting all points in a specific order, with different starting points, marked in red and blue.}
    \label{fig:2dspacings}
\end{figure}

We illustrate choices \textbf{II} and \textbf{III} in figure \ref{fig:2dspacings}.

\subsection{A model with two independent random matrices}\label{amwtirm}
\subsubsection{A two-dimensional spectrum from random matrices}\label{atdsfrm}
A simple way to construct a two-dimensional spectrum of erratically spaced points is using two independent sets of eigenvalues, $\{x_n=\lambda_n^{(1)} \}$ and $\{y_n=\lambda_n^{(2)}\}$, each drawn from a separate distribution. We will take as examples the cases where $x_n$ and $y_n$ are either the eigenvalues of random matrices from the GOE or GUE, 
%{\bf Wigner's semicircle law  with $\rho(\lambda) \approx {8\over \pi N^2} \sqrt{\lambda (N-\lambda)}$ ... with $0\le \lambda \le N$
%$p_\beta(\delta) = ...{\cal C}_\beta\delta^\beta e^{-\gamma_\beta \delta^2}$ for the spacings with ... and $\beta=1$ for GOE and $\beta=2$ for GOE and $f_\beta(r) = ....$ for the ratio's and  $f_\beta(\tilde{r}) = ....$ for the normalized ratio's}
or taken from a uniform distribution $\rho(\lambda) = 1/N$. In the last case, a uniform distribution of eigenvalues leads to a Poisson distribution for their spacings (in one dimension) $p_P(\delta) = e^{-\delta}$, with $f(r) = 1/(1+r)^2$ for the ratios and $f(\tilde{r}) = 2/(1+\tilde{r})^2$ for the normalized ratios, so we will refer to it as the Poisson case in the following.

We will consider all six possible pairs: Poisson-Poisson, Poisson-GOE, Poisson-GUE, GOE-GOE, GOE-GUE, and GUE-GUE.

We assume the sets to be ordered within themselves, \textit{i.e.} 
\begin{equation} 0\le x_1 \leq x_2 \leq \ldots \leq x_N\le N\,,\qquad  0\le y_1 \leq y_2 \leq \ldots \leq y_N \le N
\end{equation}
We construct the two-dimensional eigenvalues by packing $\{x_n\}$ and $\{y_n\}$ together, defining:
\begin{equation} \vlam_n = (x_n, y_{\sigma(n)})\,, \end{equation}
where $\sigma$ is a randomly chosen permutation of $\{1,2,\ldots,N\}$. The use of a random permutation ensures that we do not introduce correlations between $x_n$ and $y_n$, such that the eigenvalues do not lie on one particular curve in the 2D plane.\footnote{Equivalently, we could have declared from the beginning that the $x_n$ and $y_n$ are unordered, but it is convenient for later to make this point explicit by introducing the permutation $\sigma$.} This construction sets the correlation between the two variables to zero, which for realistic physical systems would be too strong an assumption. Nevertheless, we proceed with the analysis of this as the simplest model of two independent (sets of) variables.

We can also assume that the 2D spectrum of $\{\vlam_n\}$ was unfolded as in eq. \eqref{eq:unfold2d}, and work from the beginning only with unfolded 1D spectra of $\{x_n\}$ and $\{y_n\}$.

In this model, the way to see the underlying RMT spacing distribution is to look independently at the $x$- and $y$-positions of the peaks. This is measure \textbf{IV} of the previous section.

However, if we did not know in advance which are the natural axes in the problem on which to project the ${\vec\lambda}_n$, this can in fact hide the chaotic nature, even if $x_n$ and $y_n$ are exactly RMT spectra. That is, if instead of $x$ and $y$ one looks at the spacings of the positions of the peaks in some rotated coordinates,
\begin{equation}
x\pri_n(\theta) = x_n\cos\theta + y_n\sin\theta\,, \qquad y\pri_n(\theta) = y_n\cos\theta-x_n\sin\theta \,, \label{eq:rotatingxy} \end{equation}  and the consecutive spacings on, say, the $z$-axis:
\begin{equation}\delta_{n}(\theta) = x\pri_{n+1}(\theta)-x\pri_n(\theta) \end{equation}
the distribution of $\delta_n(\theta)$ is Poisson, as long as the angle $\theta$ is larger than $\sim 1/N$. This holds for all six possibilities of choosing the distributions for $x$ and $y$. When $N$ is large even very small rotations can take us from a Wigner-Dyson distribution in $(x,y)$ to Poisson distributions in $(x\pri,y\pri)$.

This is a puzzling feature. A part of the explanation is that, while the spectrum exhibits eigenvalue repulsion in the $x$- and $y$-directions, this will not hold when projecting to the new $x\pri$-axis: points that are apart in $x$ and $y$ can still have the same value of $x\pri$ or $y\pri$.

Because of this it is preferable to use a measure that does not depend on knowing in advance the exact variables corresponding to RMT in the system, such as one of the measures \textbf{I}--\textbf{III} introduced above.

These three measures will all exhibit the eigenvalue repulsion in two dimensions. However, as we will see, the two-dimensional toy model provides another source of repulsion that is largely independent of the underlying distributions of $\{x_n\}$ and $\{y_n\}$. 

\subsubsection{Distributions for spacings in two dimensions for the toy model}
As outlined above, now we will consider the distributions of \textbf{I.} all 2D spacings, \textbf{II.} nearest neighbor spacings, and \textbf{III.} consecutive spacings on a path, for the two-matrix toy model.

It turns out that in this model, where the two sets of eigenvalues $\{x_n\}$ and $\{y_n\}$ are taken to be completely independent of each other, there is an effective linear repulsion of eigenvalues, which is not sensitive to the underlying distributions of $x$ and $y$.

Note that the following discussion is valid for large matrices only, and there are qualitative differences between small and large $N$ which we will discuss towards the end.

\paragraph{I. Distribution of all spacings:}

For Poisson-Poisson the distribution of all spacings can be computed exactly (see details in section \ref{ac}). It turns out to be\footnote{Given $0\le x_n,y_n\le N$, then $0\le \delta \le \sqrt{2} N$.}
\begin{align}\label{PPdistr}
    p_{2D}(0\leq\delta\leq N) &= \frac{1}{N}\left[2\pi\left(\frac{\delta}{N}\right) - 8 \left(\frac{\delta}{N}\right)^2 + 2\left(\frac{\delta}{N}\right)^3\right] \\
    p_{2D}(N\leq\delta\leq \sqrt{2}N) &= \frac{1}{N}\left[(2\pi-4)\left(\frac{\delta}{N}\right)-2\left(\frac{\delta}{N}\right)^3+8\frac{\delta}{N}\frac{\sqrt{\delta^2-N^2}}{N} - 8 \frac{\delta}{N}\arctan\frac{\sqrt{\delta^2-N^2}}{N}\right] \end{align}
The distribution functions also seem to agree with the other cases with GOE, GUE - except in the region where $\delta$ is small (relative to $N$), and then in the RMT cases $f(\delta)$ goes faster to zero. This is probably because at long distances, the interaction of eigenvalues is very similar in all cases, especially after we have unfolded the spectra. We can then focus on other measures that probe only the spacings between the neighboring eigenvalues.

\paragraph{II. Distribution of nearest neighbor spacings:} To focus on the region of small spacings, we examine the distribution of nearest neighbor spacings. We find that they can be fitted to the $\beta$-ensemble distribution for spacings, eq. \eqref{eq:beta_delta}, with $\beta $ in the range $1.1$--$1.5$, depending on the chosen ensembles, and $\approx0.9$ for the Poisson-Poisson case. We average over 1,000 spectra with $N = 100$, and normalize the spacings such that $\langle \delta\rangle = 1$, simply by dividing by the average value before normalization. Taking $N$ to be larger (100 spectra of $N = 1000$), the values of $\beta$ are all closer to $\beta = 1$, as seen in table \ref{tab:betavaluesnn}.

Note that values in this and the following tables are indicative only as they are based on a single large sample. We have not made a systematic attempt to evaluate the errors, but the best-fit values of $\beta$ were observed to change by at most a few percent when taking different samples from the same ensembles.

\begin{table}[h!]
\centering
\begin{tabular}{c|ccc|} 
     $N = 100$    &    Poisson&GOE&GUE\\  \hline
         Poisson& 
      0.86 & 1.12 & 1.19 \\
    GOE& 1.12& 1.45 & 1.49\\
    GUE& 1.19& 1.49 &1.53 \\\end{tabular}
\qquad
    \begin{tabular}{c|ccc|} 
     $N = 1000$    &    Poisson&GOE&GUE\\  \hline
    Poisson & 0.95 & 1.08 & 1.07 \\
    GOE     & 1.08 & 1.15 & 1.19 \\
    GUE     & 1.07 & 1.19 & 1.15 \\
    \end{tabular}
    \caption{\label{tab:betavaluesnn} Best fit for $\beta$ of distribution of nearest-neighbor spacings, for different choices of ensembles for $x$ and $y$. Average over 1000 spectra of size $N=100$ (left), or 100 spectra with $N=1000$ (right).}
\end{table}

\paragraph{III. Distribution of consecutive spacings on a path:} Here, since the eigenvalues are ordered, we can work directly with the ratios of consecutive spacings $r_n$, without need of unfolding or normalizing the spacings beforehand.

For each of the choices of ensembles, we can fit the distribution of spacing ratios on the path to the $\beta$-ensemble distribution, eq. \eqref{eq:beta_r}. We find that the best fit has $\beta \approx 1.3$ when both ensembles are one of GOE and GUE, $\beta\approx 1$ for the choices Poisson-GOE and Poisson-GUE. When we take both distributions to be Poisson, the distributions of spacings still have $\beta$ close to one, $\beta \approx 0.9$. Again, larger matrices bring the values of $\beta$ closer to one. We summarize our findings in table \ref{tab:betavaluesrn}.

\begin{table}[h!]
    \centering
\begin{tabular}{c|ccc|} 
    $N = 100 $     &    Poisson&GOE&GUE\\  \hline
         Poisson& 
      0.92&1.01 &1.05\\
    GOE& 1.01 & 1.23 &1.34\\
    GUE& 1.05 & 1.34& 1.35  \\
    \multicolumn{3}{c}{\vspace{0.20cm}} \\
    $N = 1000 $     &    Poisson&GOE&GUE\\  \hline
         Poisson& 
      0.99 & 1.08 & 1.06 \\
    GOE& 1.08 & 1.12 & 1.13 \\
    GUE& 1.06 & 1.13 & 1.13  \\
    \end{tabular}
    \quad
    \begin{tabular}{c|ccc|} 
       $N = 100 $     &    Poisson&GOE&GUE\\  \hline
         Poisson& 
      0.532&0.538&0.541\\
 GOE& 0.538& 0.555 &0.561\\
 GUE& 0.541& 0.561 & 0.562 \\
\multicolumn{3}{c}{\vspace{0.20cm}} \\
$N = 1000 $     &    Poisson&GOE&GUE\\  \hline
Poisson& 0.541 & 0.549 & 0.547 \\
    GOE& 0.549 & 0.550 & 0.550 \\
    GUE& 0.547 & 0.550 & 0.552  \\
 \end{tabular}
 \caption{\label{tab:betavaluesrn} Fitted $\beta$ (left) and the average value $\avg{\tilde r}$ (right) for the different choices of ensembles. Average over 1000 spectra of size $N=100$ (top row), or 100 spectra with $N=1000$ (bottom).}
\end{table}

\begin{figure}[ht!]
    \centering
    \includegraphics[width=0.35\textwidth]{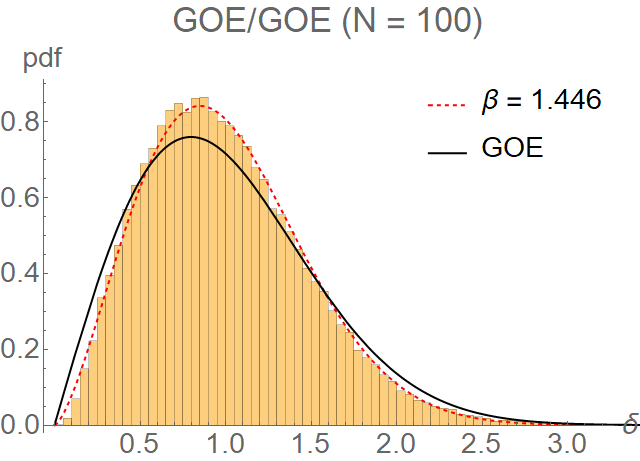}\hspace{1cm}
    \includegraphics[width=0.35\textwidth]{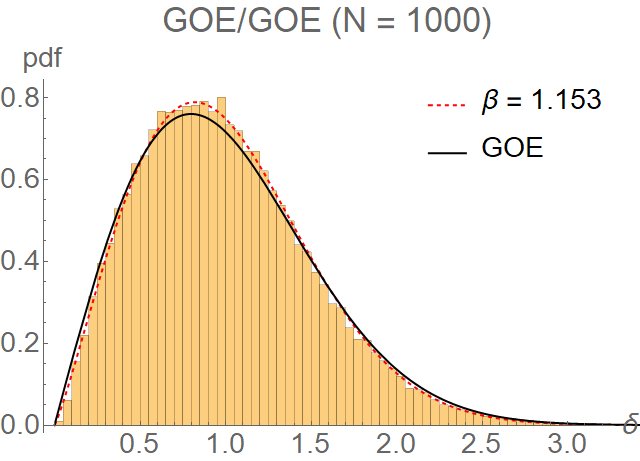} \\
    \includegraphics[width=0.35\textwidth]{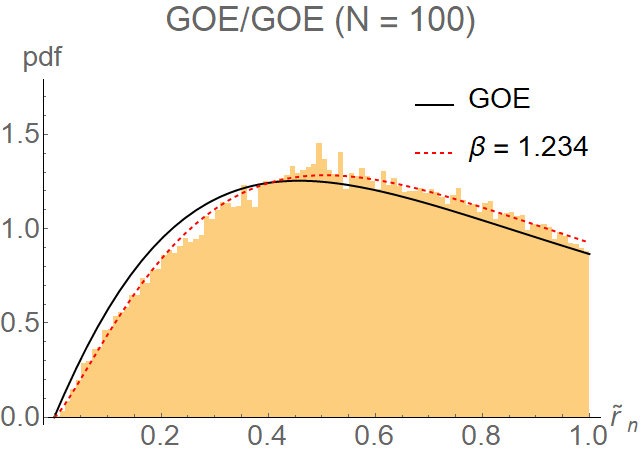}\hspace{1cm}
    \includegraphics[width=0.35\textwidth]{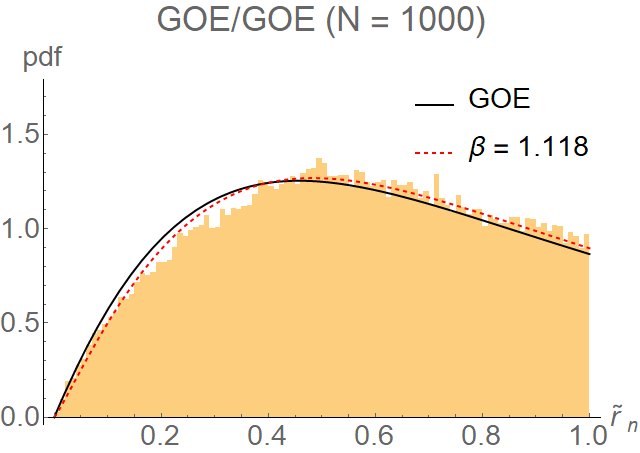}
    \caption{The distributions of $\delta_n$ (top) and $r_n$ (bottom) for the GOE-GOE case, for $N = 100$ (left) and $1000$ (right).}
    \label{fig:toymodelspacings}
\end{figure}

In figure \ref{fig:toymodelspacings} we plot the resulting distributions for the GOE-GOE case, for both the nearest neighbor spacings $\delta_n$ and the spacing ratios on a path $r_n$, with the fit to the $\beta$-ensemble distribution. The other choices of ensembles lead to very similar plots, with different values of $\beta$ as summarized in the tables.

The three measures above are all consistent in that they find an effective repulsion of eigenvalues in the plane, that is $p(\delta) \sim \delta^\beta$ for small $\delta$, with $\beta$ being approximately in the range $1.3$--$1.5$ for GOE-GOE, GOE-GUE, and GUE-GUE, and $0.9$--$1.1$ when one of the distributions is Poisson. The fact that we get repulsion even when both distributions are Poisson, and that the value of $\beta$ does not depend strongly on the initial ensemble for the RMT cases, suggests that the mechanism for this repulsion is different from the one-dimensional repulsion governing the spectra of $x_n$ and $y_n$.

\subsubsection{Effective repulsion in the toy model: analytic calculations}\label{ac}
We can see the source of the effective repulsion if we compute the probability distribution function of all spacings:
\begin{equation}
\delta_{ij} \equiv |{\vec\lambda}_i-{\vec\lambda}_j|     
\end{equation} 
in our model. We can focus on small $\delta$ to see the repulsion.

Since we do not put the eigenvalues in any particular order, we can fix w.l.o.g. $i = 1$, and then average over $j = 2,\ldots,N$.

We assume that the two-dimensional eigenvalues ${\vec\lambda_i} = (x_i,y_i)$, have independent components, \textit{i.e.} the PDF of $\{\vlam_i\}$ factorizes as:
\begin{equation}
f(\vlam_1,\ldots,\vlam_N) = f_X(x_1,\ldots,x_N) f_Y(y_1,\ldots,y_N) \end{equation}
We can compute the cumulative distribution function (CDF) of $\delta_{ij}$ by computing the probability that, for $\vlam_1=(x_1,y_1)$, there is another eigenvalue in the circle of radius $\delta$ around the point.
Let us denote by ${\cal C}_\delta(\vlam)$ the region:
\begin{equation}{\vec v} = (x,y) \in {\cal C}_\delta(\vlam) \quad \Leftrightarrow \quad |\vlam-{\vec v}| \leq \delta \end{equation}
and call ${\cal V}$ the full region where the probability $P(\vlam_i=\vlam)$ is non-zero (it can be the full 2D plane).

Then the CDF is given by
\begin{equation}F(\delta) = \int_{\cal V} d\vlam_1 \int_{{\cal C}_\delta(\vlam_1)} d\vlam_2 \int_{\cal V} d\vlam_3 \ldots \int_{\cal V} d\vlam_N f(\vlam_1,\ldots,\vlam_N)\,\, +\,\, \text{(permutations)} \end{equation}
In each term we pick one point to be inside the circle around $\vlam_1$, and integrate over all others. We can fix here $j=2$ w.l.o.g., since the contributions from other $j$ will be the same, so
\begin{equation} F(\delta) = \label{eq:CDFintegral} (N-1)\int_{\cal V} d\vlam_3 \ldots \int_{\cal V} d\vlam_N \int_{\cal V} d\vlam_1 \int_{{\cal C}_\delta(\vlam_1)} d\vlam_2  f(\vlam_1,\ldots,\vlam_N) \end{equation}

If there is already repulsion in the functions $f_X$ and $f_Y$, then clearly we do not need to do anything to get repulsion in the 2D variable. But we will try to get it without assuming anything about $f_X$ and $f_Y$.

We can write:
\begin{equation}F(\delta) = \int D\vlam_T \int dx_1 \int_{x_1-\delta}^{x_1+\delta} f_X(x_1,\ldots,x_N) \int dy_1 \int_{y_1-\Delta}^{y_1+\Delta} dy_2 f_Y(y_1,\ldots,y_N) \end{equation}
where
\begin{equation}\Delta \equiv \sqrt{\delta^2-(x_2-x_1)^2} \end{equation}
and $D\vlam_T$ denotes integration over the remaining eigenvalues $\vlam_j$ with $i=3\ldots N$. For the Poisson-Poisson case, the integral over these coordinates becomes trivial, and the integral over $\vlam_1$ and $\vlam_2$ becomes a geometric problem of computing the overlap of the circle $C_\delta(\vlam)$ and the square of size $N\times N$. The answer was given already in eqs. \eqref{PPdistr}, and the details of the derivation are in appendix \ref{ecoasdfp}.

Now change variables by defining
\begin{equation} x_2 - x_1 = \chi \delta\,, \qquad y_2-y_1 = \eta \Delta \end{equation}
which implies
\begin{equation}\Delta = \delta\sqrt{1-\chi^2} \end{equation}
\begin{equation} F(\delta) = \int D\vlam_T \int dx_1 \int_{-1}^{1}d\chi \, \delta f_X(x_1,x_1+\chi\delta,\ldots) \int dy_1 \int_{-1}^{1} d\eta \,\Delta f_Y(y_1,y_1+\eta\Delta,\ldots) \end{equation}

We can distinguish between three cases, depending on whether we have eigenvalue repulsion in the $1D$ distributions $f_X$ and $f_Y$ or not. In any case we can write:
\begin{equation} f_X(x_1,x_2+\chi\delta,\ldots) = (\chi\delta)^{\beta_X}\tilde f(x_1,x_3,\ldots,x_N) + \ldots \end{equation}
where $\beta_X = 0$ if there is no repulsion and can be identified with the Dyson index $\beta$ for the RMT distribution.

We can see that the CDF at $\delta\to0$ behaves as
\begin{equation} F(\delta) \sim \delta^{2+\beta_X+\beta_Y}
\end{equation}
which means that the PDF behaves as
\begin{equation} f(\delta) = F^\prime(\delta) \sim \delta^{1+\beta_X+\beta_Y}
\end{equation}
Repeating the same analysis in higher dimensions, we would see stronger repulsion, as $f(\delta)\sim \delta^{(d-1)+\beta_X+\beta_Y}$ in general.

This is not entirely consistent with what we observe, since we usually see behavior very close to $f(\delta)\sim \delta$ even when $\beta_X = \beta_Y = 2$. It is likely that this leading term behavior is valid only in a small region as $\delta\to0$.

There is one exception: when $N$ is taken to be small, for instance $N=3$, we can see in fact the repulsion with $\b_{\mathrm{eff}} = 1+\beta_X+\beta_Y$ when plotting the distributions of nearest neighbor spacings or path spacing ratios, as we did in the previous section. The effective linear repulsion occurs when $N$ is large, and is due to the way the 1D sets of eigenvalues are combined into 2D ones. Note that this is very different from the one dimensional case, where the distributions of spacings computed for small matrices, like the Wigner surmise, can be used without problem for large $N$ with very minor deviations.

Lastly, if the number of dimensions is $d$, we can observe that the effective repulsion in the toy model is not linear, but with $\beta{\text{eff}} = (d-1)$. This comes from taking the eigenvalues to be essentially uncorrelated as in the Poisson case and then $F(\delta) \sim \delta^d$ is just proportional to the volume of a sphere of radius $\delta$.

\subsubsection{The spacings of integer eigenvalues}\label{tsoie}
Lastly, we consider an example where there is no disorder in the eigenvalues themselves, but we get an RMT-like distribution due to the combination of the two sets $\{x_n\}$ and $\{y_n\}$ into 2D eigenvalues.

Let us assume that the sets of eigenvalues are simply taken to be
\begin{equation} x_n = n\,, \qquad y_n = n\,,\qquad n = 1,2,\ldots,N\,, \end{equation}
but they are combined in a random fashion, such that we can write:
\begin{equation} \vlam_n = (n,\sigma(n))\,. \end{equation}
for some permutation $\sigma$ of $N$.

One can think of this as choosing from an $N\times N$ lattice a set of $N$-points such that no two points are taken from the same row or from the same column in the lattice. This is like looking at the positions of a specific letter in the so-called Latin square. 

Now the only source of randomness is the permutation $\sigma$. We have $N!$ permutations, so there is still a large random ``ensemble'' from which we get our two-dimensional spectra.

Repeating the same analysis as before, calculating the distribution of spacing ratios on a path and averaging over 1,000 $N\times N$ spectra of $N = 100$, we find that they can be fitted to the $\beta$-ensemble formula as before and exhibit effective eigenvalue repulsion.

\begin{figure}[ht!]
    \centering
    \includegraphics[width=0.35\textwidth]{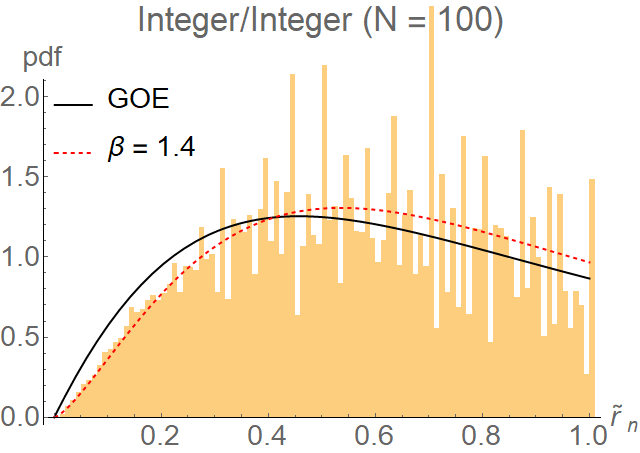} \hspace{1cm}
    \includegraphics[width=0.35\textwidth]{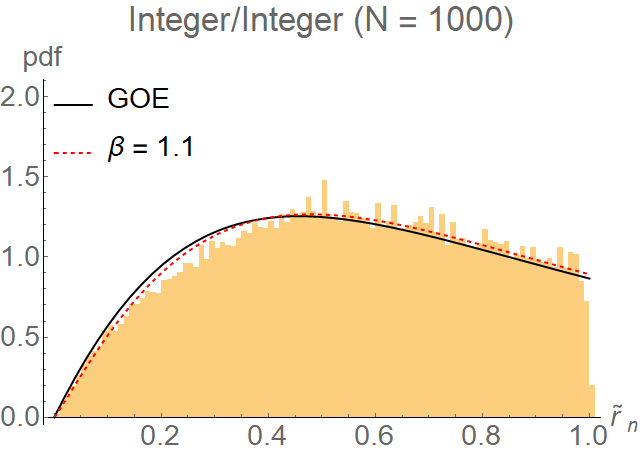}
    \caption{Distribution of path spacing ratios with only integer eigenvalues.}
    \label{fig:r_path_integers}
\end{figure}

The result does not fit the formula as neatly, partly due to the fact that here spacings $\delta_{ij} = |\vlam_i-\vlam_j|$ cannot take any value as they are always constrained to be square roots of integers. We find that $\avg{\tilde r} = 0.56$ and $\beta \approx 1.4$ for $N=100$, while $\avg{\tilde r} = 0.55$ and $\beta \approx 1.1$ for $N = 1000$, as we found for the combinations of RMT ensembles previously. The distributions are shown in figure \ref{fig:r_path_integers}.

\subsection{Two-dimensional spectral form factor}\label{tdsff}
The last tool that we will examine is the two-dimensional generalization of the spectral form factor, which in the context of scattering problem we also referred to as the ``scattering form factor'' \cite{Bianchi:2024fsi}. In the present case,  it is a function of two variables $\vec t = (t_1,t_2)$ that reads
\begin{equation}
    \mathrm{SFF}(\vec t) = \frac{1}{N^2}\sum_{i=1}^N\sum_{j=1}^N e^{i(\vlam_i-\vlam_j)\cdot \vec t} = \frac{1}{N^2}\sum_{i=1}^N\sum_{j=1}^N e^{i\left[(x_i-x_j)t_1 + (y_i-y_j)t_2\right]}
\end{equation}
On the lines $t_2 = 0$ and $t_1 = 0$ this reduces to the familiar one dimensional SFFs for $\{x_n\}$ and $\{y_n\}$, respectively.

The SFF can be decomposed into a disconnected and connected part:
\begin{equation}
\sff(\vec t) = {\cal G}_1^2(\vec t) + \frac{1}{N}(1 - {\cal G}_2(\vec t)) \label{eq:sff_rmt}     
\end{equation} 
The disconnected part can be computed as the Fourier transform of the average density function. In our case of independent $x_n$ and $y_n$ being unfolded eigenvalues with constant density from $0$ to $N$ it is
\begin{equation} {\cal G}_1(t_1,t_2) = \mathrm{sinc}\left(\frac12Nt_1\right)\,\mathrm{sinc}\left(\frac12 N t_2\right)
\end{equation}
where $\mathrm{sinc}(x) = \sin(x)/x=j_0(x)$, the lowest spherical Bessel function.

We have observed before that after rotating $(x_n,y_n)$ by an angle $\theta$ as in \eqref{eq:rotatingxy}, the result is a Poisson distribution of the spacings in the new coordinates $(x\pri_n,y\pri_n)$. In the SFF this manifests as the fact that the characteristic ramp associated with RMT spectra appears only in the directions in the $(t_1,t_2)$-plane conjugate to $x$ and $y$. In this way the SFF can identify the correct choice of axes to see the underlying RMT behavior of the 2D spectra. The SFF gives additional information about the spectrum than the distributions of spacings did not.

\begin{figure}[ht!]
    \centering
    \includegraphics[width=0.48\textwidth]{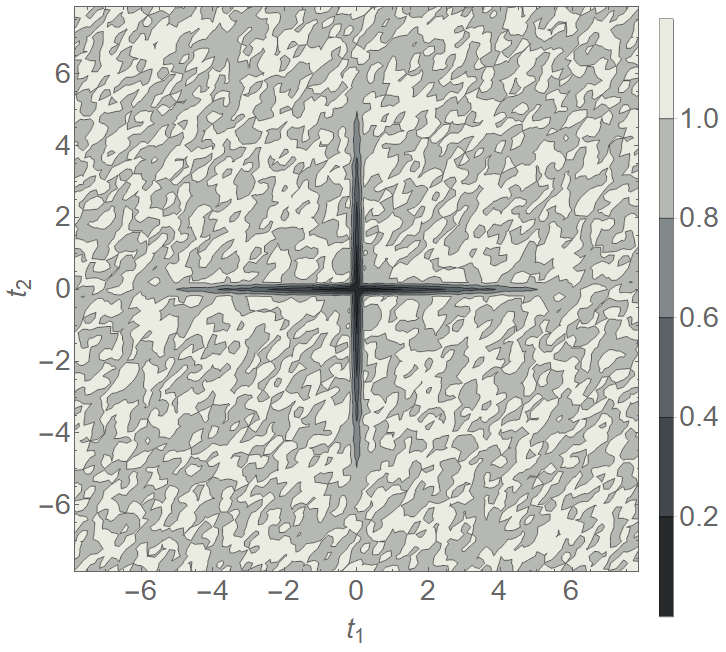}
    \includegraphics[width=0.48\textwidth]{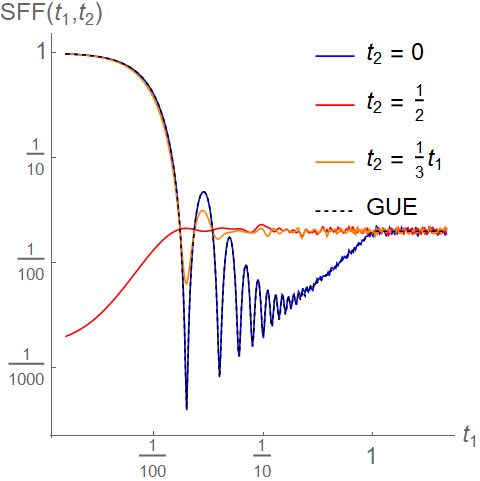} \\
    \includegraphics[width=0.48\textwidth]{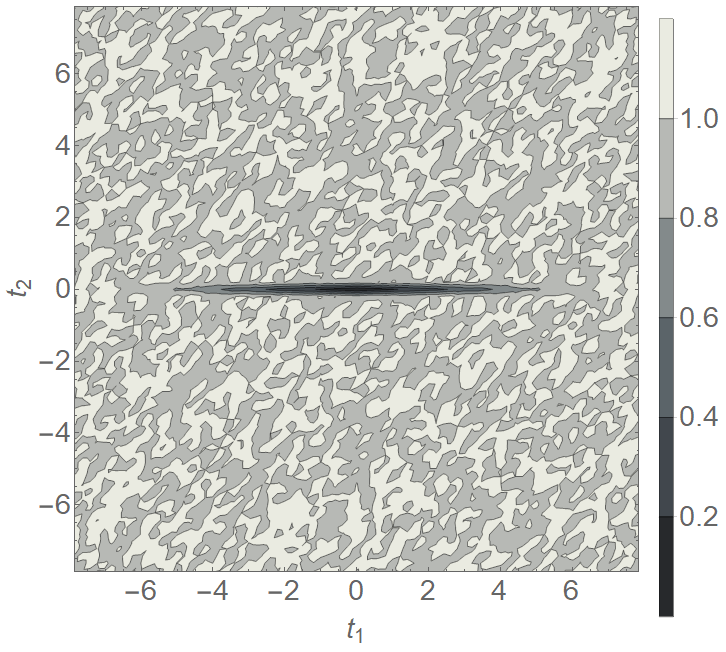}
     \includegraphics[width=0.48\textwidth]{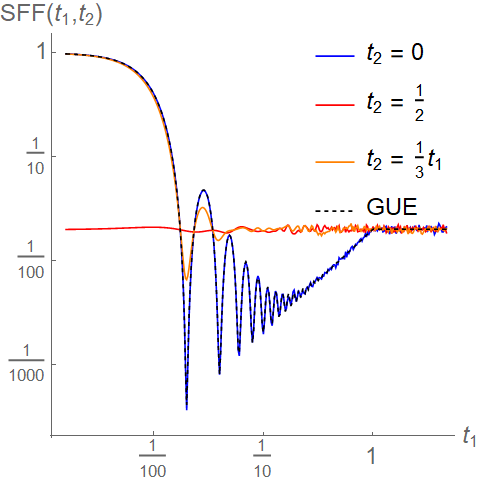} \\
    \caption{The two-dimensional SFF, averaged over 500 GUE-GUE (top row) or GUE-Poisson (bottom) spectra with $N=50$. Left: contour plots of connected part of the SFF, right: the full SFF on a logarithmic scale, computed on the lines $t_2 = 0$, $t_2 = \frac12$, and $t_2=\frac13t_1$.}
    \label{fig:sff_2D}
\end{figure}

In figure \ref{fig:sff_2D} we plot the SFF for the cases GUE-GUE and GUE-Poisson. In the two-dimensional figures, by plotting the connected part ($1-{\cal G}_2(t_1,t_2)$) of the SFF, we see clearly the existence of ramps in the specific directions corresponding to the GUE variables, two ramps for GUE-GUE, and a single ramp for GUE-Poisson. One-dimensional plots show the behavior along different lines: for $t_2 = 0$ we observe the usual GUE linear ramp, for $t_2 = \frac13 t_1$ (corresponding to a rotation of the original variables), we see a decline-to-plateau structure with no ramp, as in a Poisson spectrum. When plotting on the line $t_2 = \frac12$ there is a ``valley'' when the line crosses the ramp associated with $y_n$ in the GUE-GUE case, while for Poisson-GUE the SFF is nearly constant on that line.

\clearpage
\section{Two-dimensional spacing statistics for the quantum pinball scattering amplitude}\label{taoteofsa}
We will now use the measures of two-dimensional spacings proposed in section (\ref{tdeats}) on the scattering amplitude of the three-disk system. Before we do that, we will test a simple RMT model, where the $S$-matrix is taken to be a random unitary matrix from the circular ensembles, and find the distributions in that case for the nearest neighbor spacings, and the spacing ratios on a path. Then we will compare the two.

\subsection{The \texorpdfstring{$S$}{S}-matrix as a random matrix}\label{tsmaarm}
Recall that the scattering amplitude was given by:
\begin{equation}
    f(\phi,\phi^\prime) = \sum_{l=-\infty}^\infty \sum_{l\pri=-\infty}^\infty e^{-i l (\phi-\frac\pi2)} (S_{ll^\prime} - \delta_{ll^\prime})e^{i l\pri (\phi\pri-\frac\pi2)} \label{eq:amplitudeS}
\end{equation}
In the present case we want to model a system that is time-reversal invariant. In terms of the scattering amplitude and $S$-matrix the symmetry leads to the constraint:
\begin{equation} f(\phi,\phi^\prime) = f(\pi+\phi\pri,\pi+\phi)\qquad \Leftrightarrow \qquad S_{l,l\pri} = (-1)^{l+l\pri} S_{-l\pri,-l} \end{equation}
The circular unitary ensemble (COE) generates unitary matrices $M_{ij}$ that are symmetric in the sense that $M_{ij} = M_{ji}$, and we should modify it to enforce the correct symmetry of the $S$-matrix. If we define the matrix $P_{ij} = (-1)^i \delta_{ij}$ - i.e. the diagonal matrix with elements $(-1,+1,-1,\ldots)$ - then the new matrix $\tilde M = P M$ will satisfy $\tilde M_{ij} = (-1)^{i+j} \tilde M_{ji}.$, and we can use this to generate a random $S$-matrix compatible with time-reversal symmetry. This operation does not affect the probability distribution of the COE, since there is a one-to-one correspondence with symmetric matrices.

We will take the $S$-matrix to be a finite $N\times N$ size matrix with $N = 2\lambda + 1$, such that the angular momentum is $-\lambda\leq l\leq\lambda$, and the scattering amplitude is given by \eqref{eq:amplitudeS}. Since we draw the $S$-matrix from the COE, its eigenvalues will have the familiar distribution, but now we would like to see what the distributions of spacings of peaks of the amplitude \eqref{eq:amplitudeS} look like. An important difference between our approach here and the $S$-matrix one obtains from solving the equations for the quantum pinball system, is that here not only are the eigenvalues distributed as in the COE, but the eigenvectors will be completely random in the chosen basis.

\begin{figure}[ht!]
     \centering
     \includegraphics[width=0.48\textwidth]{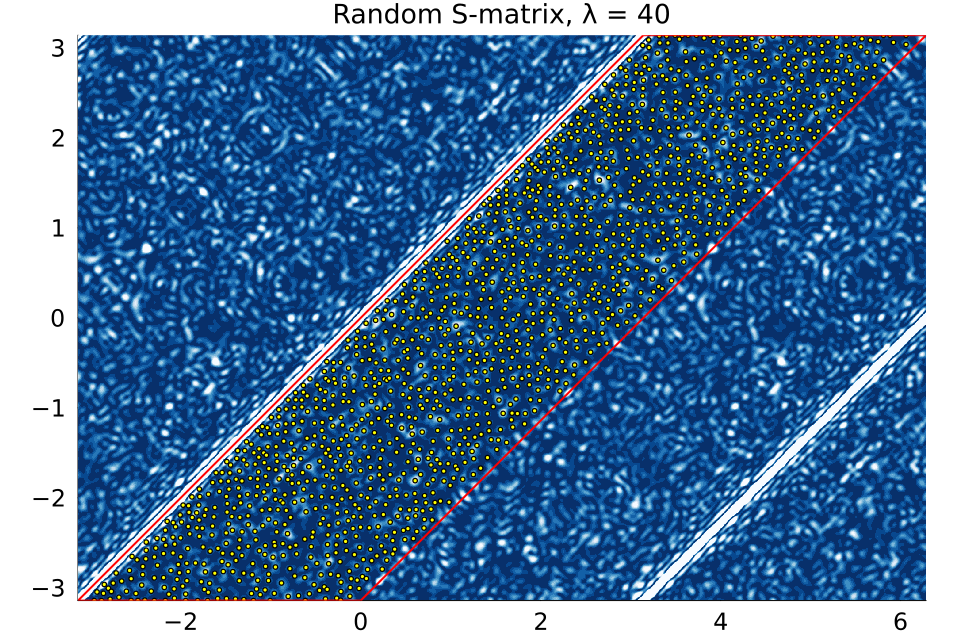}
     \includegraphics[width=0.48\textwidth]{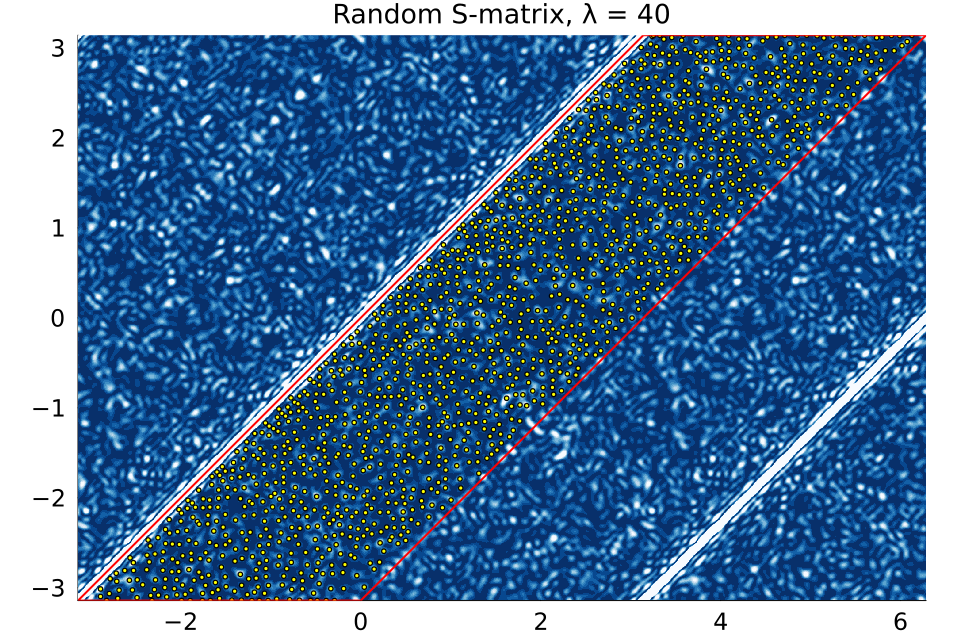}
     \\
     \includegraphics[width=0.48\textwidth]{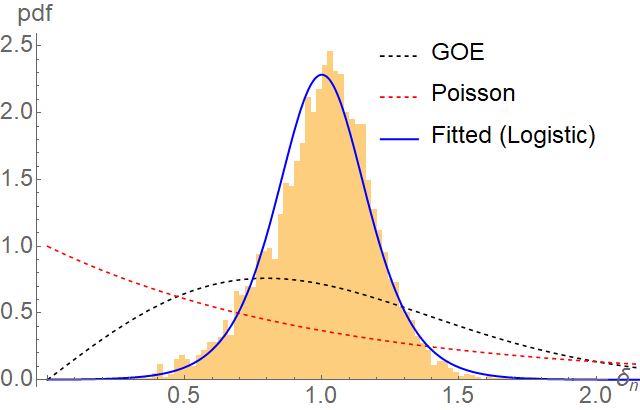}
     \includegraphics[width=0.48\textwidth]{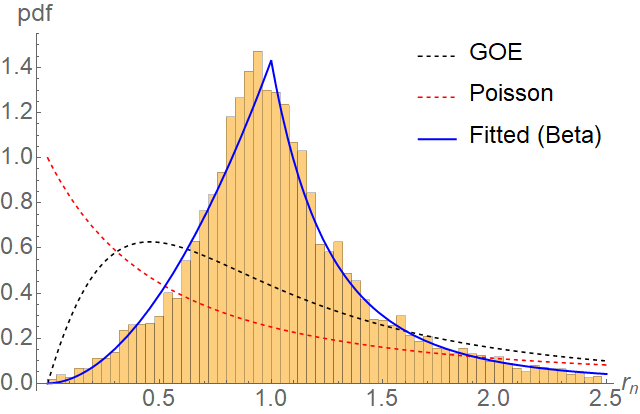}
     \caption{Top row: The differential cross-section for a random $S$-matrix and its peaks, for two instances of an random $S$-matrix with $\lambda=40$. Bottom: the distributions of spacings (left) and spacing ratios (right), averaged over five random matrices with $\lambda=40$.}
     \label{fig:randSmatrixspacings}
 \end{figure}

The amplitude defined in this way appears more irregular than that of the pinball, lacking completely the macroscopic structure of diffraction patterns we have observed in section (\ref{tqsa}). This can be seen in figure \ref{fig:randSmatrixspacings}, where we plot two instances of the function with a random $S$-matrix of size given by $\lambda=40$.

We look at the spacings of the peaks of this function using the two-dimensional measures of the distribution of spacings of nearest neighbors, and of spacing ratios on a path. We compute all the spacings in the region defined by $-\pi\leq \phi\pri <\pi $, and $\phi\pri +0.1 \leq \phi \leq \phi\pri+\pi-0.1$. Because of the parity symmetry, this gives the full range of the angles, except for the offset of 0.1 which was added to avoid the special lines $\phi=\phi\pri$ and  $\phi=\phi\pri+\pi$, considering only local maxima away from it.

The results are far from the one-dimensional distributions of COE spacings. The spacings of peaks of the amplitude are fairly regular, in the sense that in the distribution of spacings we find a large peak around the average value $\delta = 1$ for the nearest-neighbor spacings (measure \textbf{II}), or $r = 1$ for the spacing ratios on a path (measure \textbf{III}). These are also plotted in figure \ref{fig:randSmatrixspacings}.

The distribution of nearest-neighbor spacings can be modeled by fitting it to a logistic distribution, namely
\begin{equation}
    p_L(\delta) = \frac{1}{\sigma} \frac{e^{-(\delta-\mu)/\sigma}}{(1+e^{-(\delta-\mu)/\sigma})^2}
\end{equation}
where $\mu = 1$ (which is a choice of normalization) and $\sigma \approx 0.11$. Note that this cannot be the exact distribution since by definition $\delta$ is a positive variable and the logistic distribution is defined for all $\delta$, though with an exponential decay far from $\mu = 1$.

On the other hand, the distribution of $r_n$ along a path, which is again peaked at $r_n = 1$ unlike the COE distribution, fits a simple distribution with
\begin{equation}
    f_B(r) = \frac{1}{B(a,1)} \begin{cases} r^{a-1} & 0\leq r \leq 1 \\
    \frac{1}{r^{1+a}} & 1 \leq  r
    \end{cases}
\end{equation}
which is a special case of a Beta distribution $\mathrm{B}(a,b)$ with $b=1$,\footnote{The Beta distribution owes its name to Euler Beta function and should not be confused with the $\beta$-ensemble distribution of RMT.} extended to $r > 1$ by using the inversion symmetry $r\to1/r$, implying $f(r) = \frac1{r^2}f(\frac{1}{r})$. Alternatively, one can use the normalized ratios $\tilde r_n$ which are always between 0 and 1 and fit to an ordinary Beta distribution. In our fits the parameter $b$ is fixed to 1 while $a \approx 2.9$.

In the following, we consider these two distributions as the ``expected'' result for a chaotic system, and as such they will be compared with the measured distribution for the quantum pinball system.

\subsection{Analysis of positions of peaks for the quantum pinball}\label{aopop}

% {\bf Referee 2, Comment 4:}

% {\it The 2D spacing distributions for the pinball (Section 5.2) are not sharply diagnostic. The nearest-neighbour spacings follow a logistic distribution peaked at $\delta=1$ for both the chaotic (three-disk) and non-chaotic (two-disk) systems, differing only in the width parameter $\sigma$. Moreover, the chaotic system has the narrower distribution, which is counterintuitive and unexplained. The spacing ratios are even less discriminating, being sharply peaked at $r=1$ in all cases. The authors themselves note these difficulties but do not resolve them. As a result, the proposed 2D measures do not yet provide a reliable diagnostic of chaos from the scattering amplitude alone — which was stated as one of the main goals. }

To measure the distribution, we computed the location of the peaks of the scattering amplitude of the three-disk system, using the same configuration we used extensively in section \ref{tpqs}, where the centers of the disks are placed on an isosceles triangle with side $L = 3$, and the radii of the disks are $R_1 = 1$, $R_2 = 1.2$, and $R_3 = 0.8$.

We have computed the locations of the peaks for this system at $k = 25$ and $k = 50$, and again we look at the distributions of nearest-neighbor spacings $\delta_n$, and the spacing ratios $r_n$ along a path.

 The three-disk pinball system gives a distribution of the normalized $\delta_n$ which is very similar to the one we found in the RMT model, with a logistic distribution of width $\sigma \approx 0.1$. On the other hand, the distribution for $r_n$ is much more sharply peaked around 1 than the Beta distribution, and is not well fitted. There is some deviation in both distributions due to the presence of very different regions in the scattering amplitude, in which the density of peaks is higher or lower. This could explain, for instance, the second small peak in the distribution of $\delta$ around 1.3, coming from a region where the average spacing is larger than in others. We have not made an attempt to account for this effect. One could try to unfold the spectra to make the peaks uniformly distributed, or else to focus on the spacings only in specific regions.

 \begin{figure}[ht!]
     \centering
     \includegraphics[width=0.48\textwidth]{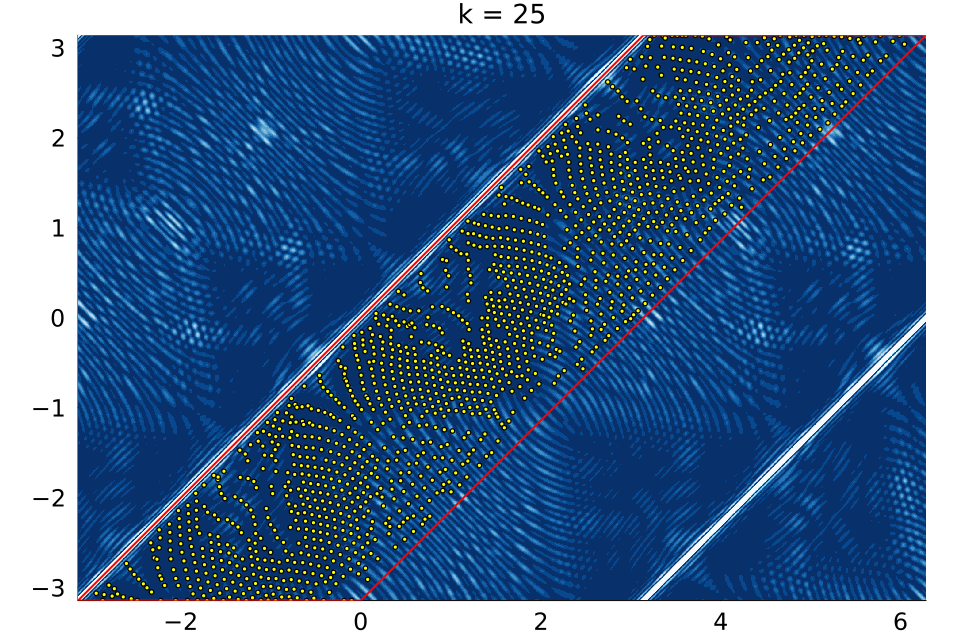}
     \includegraphics[width=0.48\textwidth]{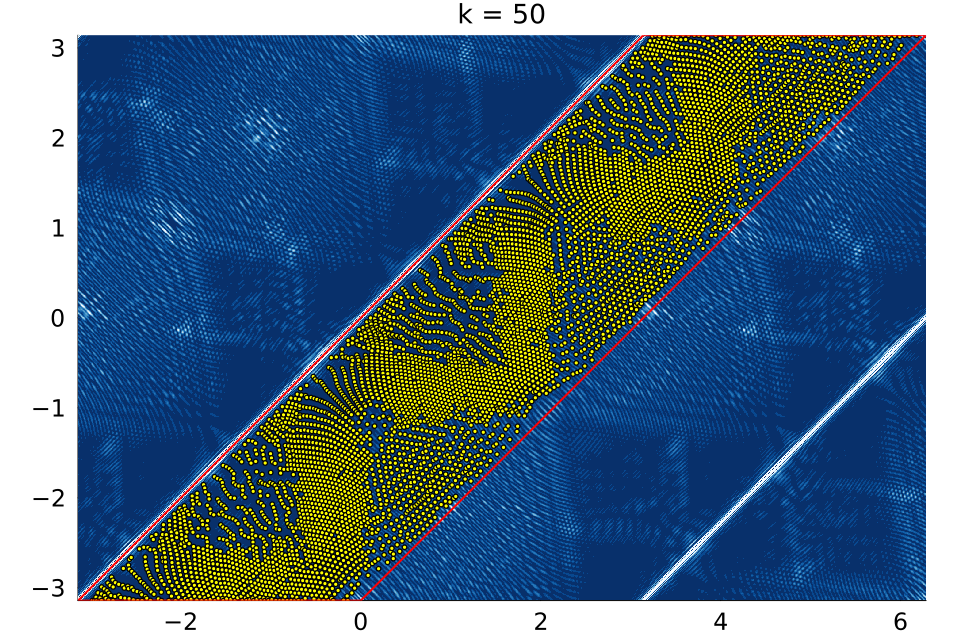} \\
     \includegraphics[width=0.48\textwidth]{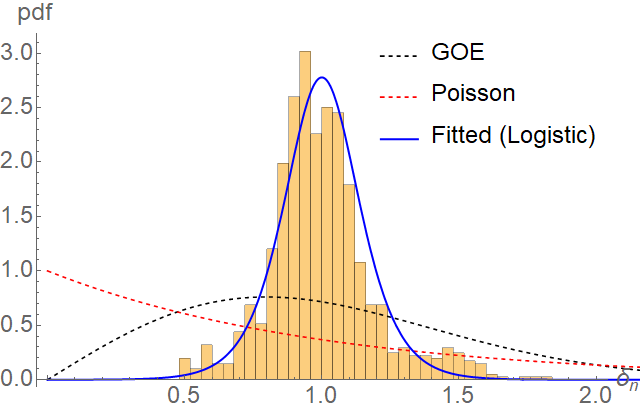}
     \includegraphics[width=0.48\textwidth]{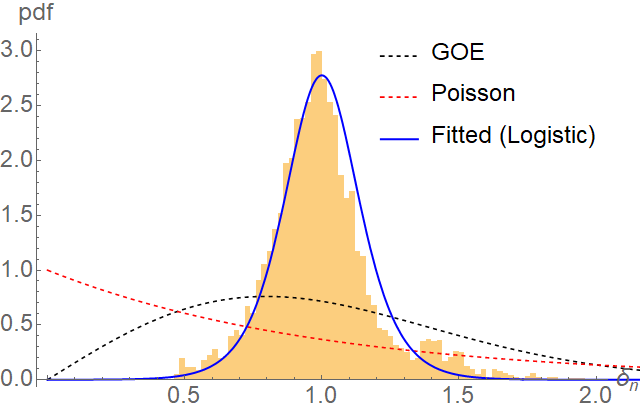} \\
    \includegraphics[width=0.48\textwidth]{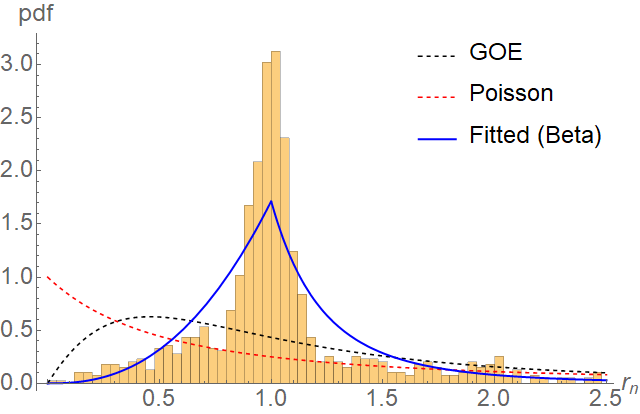}
     \includegraphics[width=0.48\textwidth]{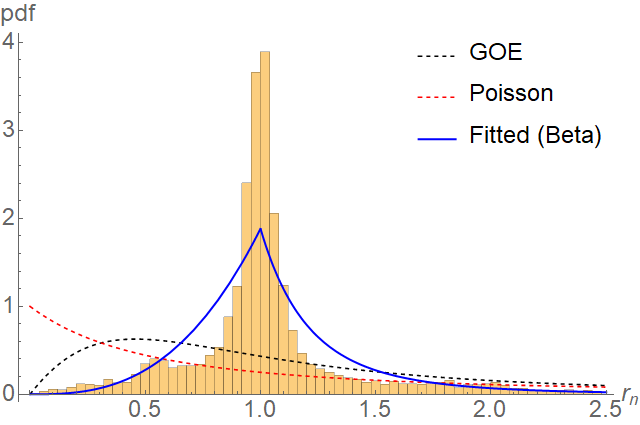} \\
     \caption{Peaks of the \textbf{three-disk} pinball amplitude for $k=25$ (left) and $k=50$ (right) and their distributions. Top row: The function and its peaks. Middle row: Distributions of nearest neighbor spacings. Bottom row: Distributions of path spacing ratios.}
     \label{fig:pinballspacings}
 \end{figure}

\begin{figure}[ht!]
     \centering
     \includegraphics[width=0.48\textwidth]{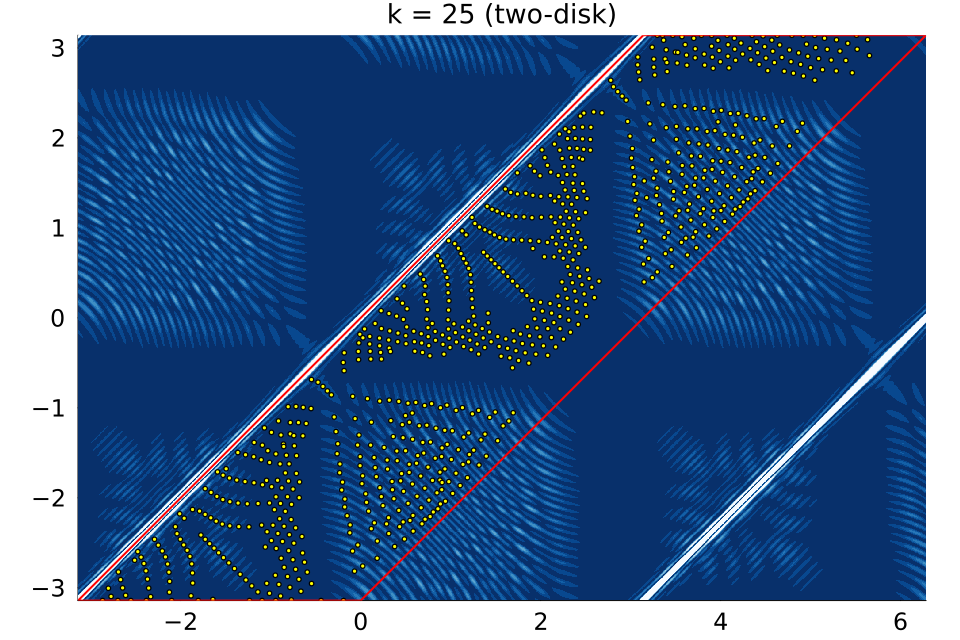}
     \includegraphics[width=0.48\textwidth]{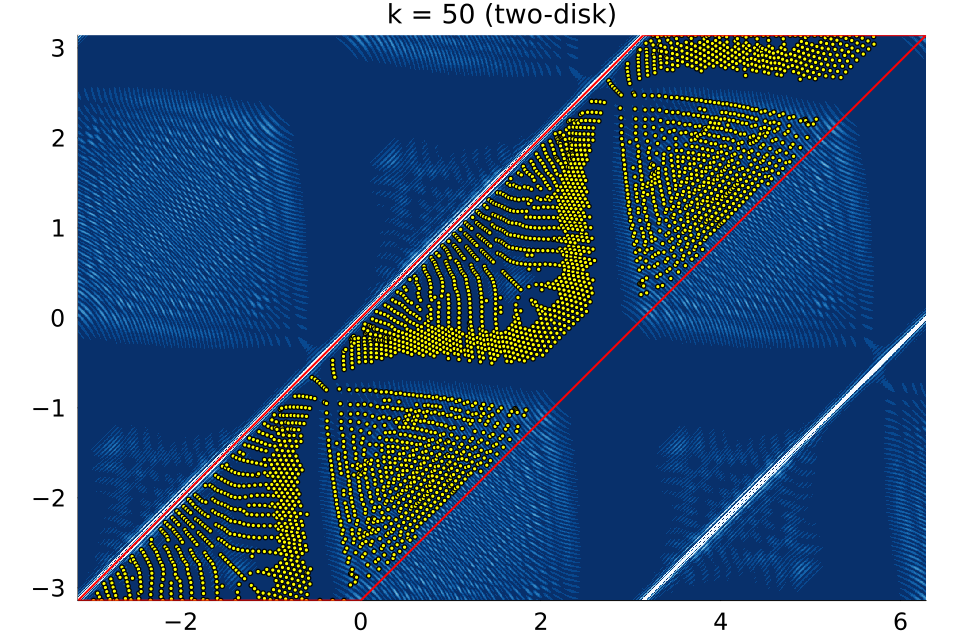} \\
     \includegraphics[width=0.48\textwidth]{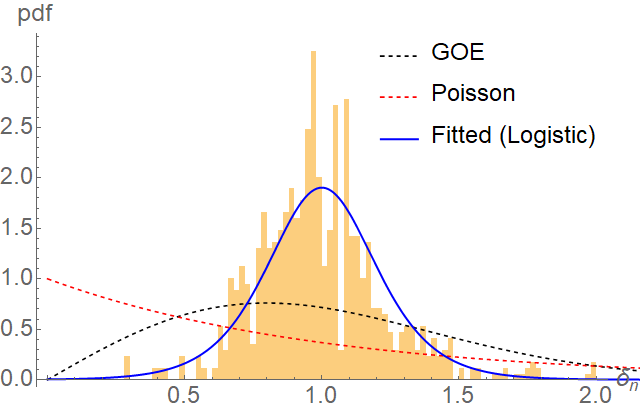}
     \includegraphics[width=0.48\textwidth]{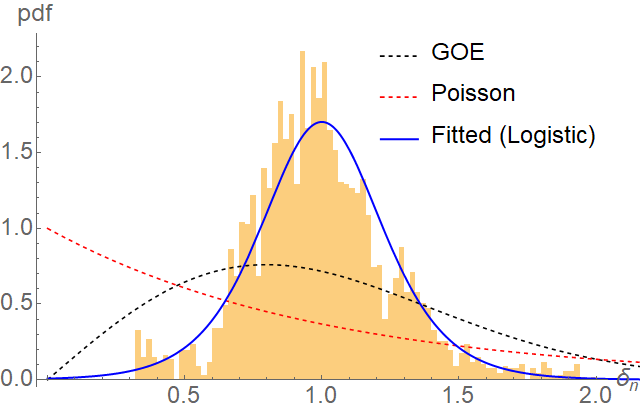} \\
    \includegraphics[width=0.48\textwidth]{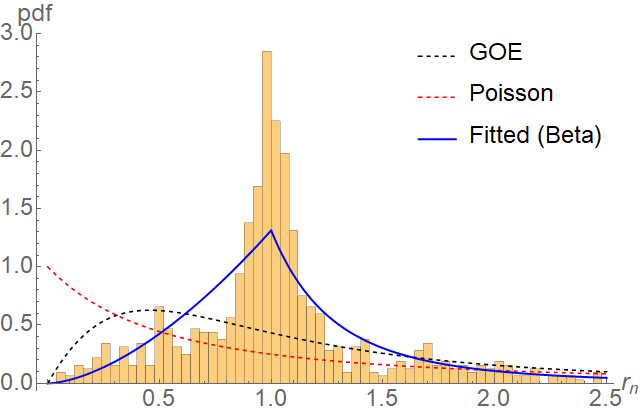}
     \includegraphics[width=0.48\textwidth]{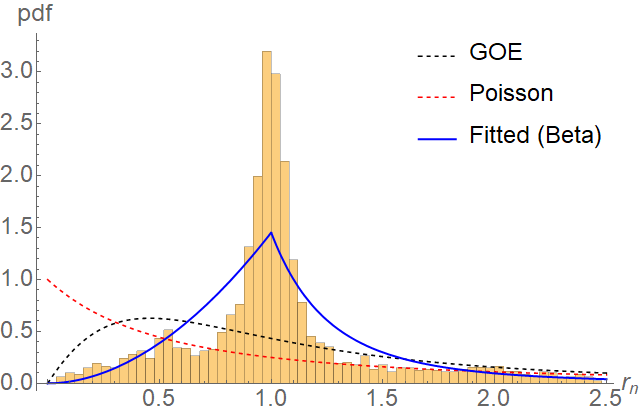} \\
     \caption{Peaks of the \textbf{two-disk} pinball amplitude for $k=25$ (left) and $k=50$ (right) and their distributions. Top row: The function and its peaks. Middle row: Distributions of nearest neighbor spacings. Bottom row: Distributions of path spacing ratios.}
     \label{fig:pinballspacings_twodisk}
 \end{figure}

Lastly, we can compare the two-disk system, whose $S$-matrix is not COE, but gives a Poisson distribution of eigenvalue spacings. This is not expected to match with the RMT model, but the result is that we again find a logistic peak, but wider. While the three-disk system had $\sigma \approx 0.1$ like the RMT model, here we find $\sigma \approx 0.15$. See figure \ref{fig:pinballspacings_twodisk}. This is somewhat counterintuitive as we would have expected the chaotic system to have the wider distribution, implying more disorder. 

Though the chaotic three-disk system is observed to give a distribution closer to that found in the model where the $S$-matrix was taken to be purely random, the fact that the non-chaotic two-disk system also yields the logistic distribution of the spacings, albeit with a different width, makes it hard to distinguish chaotic behavior using this measure. It is an open question to what is related this width parameter, and if it can be said to be an indicator of chaotic behavior.

\clearpage
\section{Summary and Outlook}\label{s}

  To introduce the concept of two- and, by a straightforward generalization, higher multi-dimensional chaotic behavior in scattering processes, we focused on simple models exposing chaotic features. 
  
 First we considered pinball scattering both in the classical and in the quantum context.  We analyzed the classical scattering off the  three-disk pinball system and determined the classical  scattering angle and the number of collisions as a function of the incident angle and the impact parameter, which can also be expressed in terms of an angle. Our analysis showed the basic  features of classical  chaotic scattering, namely erratic behavior and a self-similar structure. The chaotic behavior was shown using two-dimensional plots of the scattering angle and the number of collisions in terms of the two angles. The ``topography" of these figures was analyzed and various structures were identified.
 
 We then  considered the quantum scattering off the pinball system and computed the $S$-matrix for fully symmetric configurations and for asymmetric ones. For the former we found that the eigenvalues of the $S$-matrix follow a Poisson distribution. For the latter cases we found out that for a given asymmetric configuration there is a large enough wave-number $k$ for which the distribution is chaotic, namely a COE distribution. This confirmed the conjecture that the scattering can be described by a random unitary $S$-matrix.
 
We then computed the amplitude for various different configurations. We examined its dependence on the wave-number $k$, in particular near the resonances, and, in order to highlight the chaotic features of the three-disk pinball scattering, we compared our results with those for the two-disk pinball scattering, that is known to be non-chaotic. One can transition continuously between these two systems by shrinking the radius of one of the three disks to zero. We have examined also other possibilities to interpolate between chaotic configurations where the $S$-matrix eigenvalues are distributed as in COE, and non-chaotic ones where they are Poisson.

One of our aims was to introduce a measure to observe chaotic behavior directly from the scattering amplitudes, that would be applicable in cases where the full $S$-matrix is not known. The measure of chaotic behavior using the extrema of the scattering amplitude, introduced in \cite{Bianchi:2022mhs,Bianchi:2023uby},  was generalized to two and higher dimensions in this work.

To that end, we set out to analyze the spacings between peaks of amplitudes that depend on two variables. We first briefly reviewed the landmark case of the phase-shifts of the leaky torus \cite{Gutzwiller:1983}, where the time delay function is a series of resonances located at the non-trivial zeros of the Riemann zeta function, and whose spacings are distributed as in the GUE.

We proposed a simple toy model that generalizes the leaky torus time delay function to a function of two random variables related to the eigenvalues of two random matrices. This function can also be interpreted as the electric potential produced by a set of charges located at random positions on a plane in three spatial dimensions.

We introduced several measures to analyze the distribution of extrema of such a function, which exposes randomly distributed extrema. We proposed four different measures to describe the level spacings and applied each of them to the toy model mentioned above. In particular we considered the spacings between nearest neighbors and introduced a simple algorithm for ordering the points on a path, allowing us to define a measure similar to the ratios of consecutive spacings.

The corresponding distributions in two dimensions were close to that of the GOE for the spacings and spacing ratios. We interpreted the distribution by identifying the ``effective repulsion" in the toy model, which was also illustrated by means of analytic calculations. Lastly, we considered a case with no disorder in the ``eigenvalues'', in that they were taken to be integers, but the pairing between them was made ``chaotic'' by using a random permutation. In this case a GOE distribution was found for the distribution of the spacings, too.

We defined and computed also the two-dimensional Scattering Form Factor (ScFF), introduced in \cite{Bianchi:2024fsi} for scattering processes in analogy with the Spectral Form Factor (SFF). Unlike the simple SFF, the ScFF can be a function of multiple variables, associated with each of the kinematic variables in the problem. We have shown how the ScFF can shed light on the underlying distribution of the positions of extrema.

We applied the methods developed for the toy model on the quantum scattering from a pinball in in section \ref{taoteofsa}. This was compared to a model wherein  the
$S$-matrix of the pinball is taken to be purely a random matrix from the COE. In these cases one does not get a COE distribution for the spacings and ratios but rather a distribution that is peaked around the average value of one, and can be modeled as either a logistic peak for nearest neighbor spacings, or a Beta distribution for the spacing ratios.

We finished by conjecturing a possible map between multi-dimensional chaotic systems and random tensor theory. 

The idea of multi-dimensional chaotic behavior is a new concept and naturally there are many open questions we could not fully answer in the present work, and many directions for future research. Here we list several of them:
    \begin{itemize}
   \item 
   Obviously one can discuss  higher than two-dimensional  chaotic behavior. For instance the scattering and decay amplitudes involving highly excited string (HES) states are functions of several kinematical factors.
   \item 
   In the present investigation we have focused on chaotic scattering processes. It is quite probable that there are erratic functions of several variables describing chaotic behaviors that are not related to scattering processes. 
   \item 
   In the context of the pinball, one can consider generalizations to any number of disks and also replace the disks with balls in three space dimensions. Quite remarkably the dynamics is integrable. The quantum pinball system in three dimensions could have a richer structure than the two-dimensional case. We did not pursue this interesting direction of investigation.
   \item 
   In section \ref{aopop} we encountered  distributions of spacings and ratios which were peaked around unity. It could be that this results from a distribution of peaks that is a mixture of a periodic and a chaotic one. A reliable method for disentangling  periodic and chaotic eigenvalues is needed.
   %for instance using an ansatz for the properly normalized distribution function  of the form:  $f= w f_{per} + (1-w)f_{chao}$.
   \item 
   We found that for both the chaotic $S$-matrix and the pinball, the distributions of the spacings between the peaks and their ratios are distributed according to the logistic and Beta distributions respectively. This was not the case for the toy model, where we found the GOE distributions. On difference between the two systems was that the toy model was not described by a unitary $S$-matrix. An obvious question is to what extent these distributions are generic, and in which physical cases they apply. We will need to explore other systems with  two-dimensional chaotic behavior to check it.  
   \item 
   Classical scattering from two disks, unlike from three disks, is  non-chaotic. On the other hand  in the quantum scattering also for the two-disk case peaks and their ratios admitted a logistic and a Beta distributions, albeit with different parameters from the random $S$-matrix case. This behavior should be further explored.  
   \item 
In recent years increasing attention has been payed to the study of quantum chaos in  QFTs in 1+1 dimension \cite{Brandino:2010sv,Srdinsek:2020bpq,Delacretaz:2022ojg,Negro:2022hno,Sonnenschein:2025jlc}. A natural question is to look for multi-dimensional chaotic processes in QFTs in 1+1 and higher dimensions. 
\item 
An important tool that  has emerged in the recent study of quantum chaos in quantum mechanical and QFT systems is Krylov complexity (see \cite{Nandy:2024evd,Rabinovici:2025otw} and references therein). Can one develop a Lanczos method also for multi-dimensional chaos?
In particular in \cite{Bhattacharya:2024szw} Krylov complexity methods were implemented for the measure of spacings between peaks of scattering amplitudes proposed in  \cite{Bianchi:2022mhs}. This probably can also be generalized to higher-dimensional chaotic behaviors.
\item 
Recently, in \cite{Ageev:2025yiq} and \cite{Ageev:2025iiy}, the measure of \cite{Bianchi:2022mhs} was used for computing correlators in quenched QFTs in particular in AdS. 
Again a generalization to higher dimensions and to other types of correlators  should be possible.
 
   \item 
We have started our journey of chaotic processes in string theory with the proposal for a quantitative  measure \cite{Bianchi:2022qph} for the observed erratic behavior of scattering and decay amplitudes of  HES states \cite{Gross:2021gsj}. The latter are in fact generically functions of several kinematical parameters, and can be computed analytically at tree level. In a sequel to the present investigation we intend to present the results of the study of multi-dimensional chaotic behavior in such processes.  
    
   \item
Given that the Wigner-Dyson \cite{Dyson:1962es} map of the energy eigenvalues to those of  Gaussian random matrices \cite{Bohigas:1983er} provides a measure of quantum chaos, it is tempting to conjecture a relation of higher-dimensional chaos to Random Tensor Theory. Indeed it looks plausible that $d$ dimensional chaotic patterns be related to random tensors $T_{i_1 i_2 ... i_{d{+}1}}$ of rank $d+1$. 

In the case of two-dimensional chaotic behaviour, we noticed the possibility of reproducing one-dimensional chaotic behaviour simply `projecting out' one of the random variables. This would correspond to reducing three-tensors to matrices. One can get a matrix out of a three-tensor by reducing to different ``planes".  For any projection of the three-tensor, we can compute the eigenvalues of the corresponding matrix and we expect to recover the one-dimensional chaotic behavior for generic projections. For instance, the scattering amplitudes discussed in (\ref{tddotqs}) depend on two angles $x=\theta$ and $y=\phi$. Fixing one of the two variables {\it e.g.} $y=y_0$ the peaks of the function ${\cal F}(x)={\cal A}( x,y_0)$ form a set $\{x_i\}$. One can compute the spacings $\delta_i = x_{i+1}-x_{i}$ and spacing ratios $r_i\equiv {\delta_i}/{\delta_{i-1}}$ that can be compared with the corresponding distributions of the eigenvalues associated with the random matrix derived from the random three-tensor corresponding  to setting $y=y_0$.

We would like to further conjecture that the  higher-dimensional extrema like curves be related to eigenvectors and eigenvalues of the RTT without any reduction to matrices. For instance one can consider eigenvalues $h$ and eigenvectors $v_i$ of a (totally symmetric)  three-tensor in the following way
\begin{equation}
T_{ijk} v^i v^j = h v_k \qquad {\rm with} \qquad |v|=\sqrt{v_iv^i}=1
\end{equation}
The distributions of the eigenvectors were studied recently (see \cite{Sasakura:2023crd} and references therein). In particular  real eigenvector/value distributions of Gaussian
random three-tensors have been explicitly computed by expressing them as partition functions of
quantum field theories with quartic interactions. Our conjecture is that these eigen-vectors/values correspond to extrema objects  of the two-dimensional patterns of chaotic processes.

\end{itemize}

%\clearpage

\section*{Acknowledgments}

%{\bf copied from our last paper}

We thank R. Benzi, G. Salina, G. Parisi,  A. Gaikwad, N. Shrayer, V. Niarchos and E. Kiritsis for useful comments and discussions.

MB would like to thank the MIUR PRIN contract 2020KR4KN2 ``String Theory as a bridge between Gauge Theories and Quantum Gravity"  for partial support. MB and DW thank the the INFN project ST\&FI ``String Theory and Fundamental Interactions'' for partial support. The work of MF is supported by the European MSCA grant HORIZON-MSCA-2022-PF-01-01 ”BlackHoleChaos” No.101105116 and partially supported by the H.F.R.I call “Basic research Financing (Horizontal support of all Sciences)” under the National Recovery and Resilience Plan “Greece 2.0” funded by the European Union – NextGenerationEU (H.F.R.I. Project Number: 15384.). The work of JS was supported in part by a grant 01034816 titled “String theory reloaded- from fundamental questions to applications” of the “Planning and budgeting committee”. DW was supported by an INFN postdoctoral fellowship.

Many of the numerical computations in this work were carried out using Julia \cite{doi:10.1137/141000671}, and in particular the DynamicalBilliards package \cite{Datseris2017}.

\section*{Supplementary materials}
Some code, numerical data, and other supplementary materials to this paper, including some animated figures, are available at \cite{stringchaosgithub:2510}.

\appendix

\section{Supplementary figures} \label{app:supplfig}

\begin{figure}[t!]
    \centering
    \includegraphics[width=0.48\linewidth]{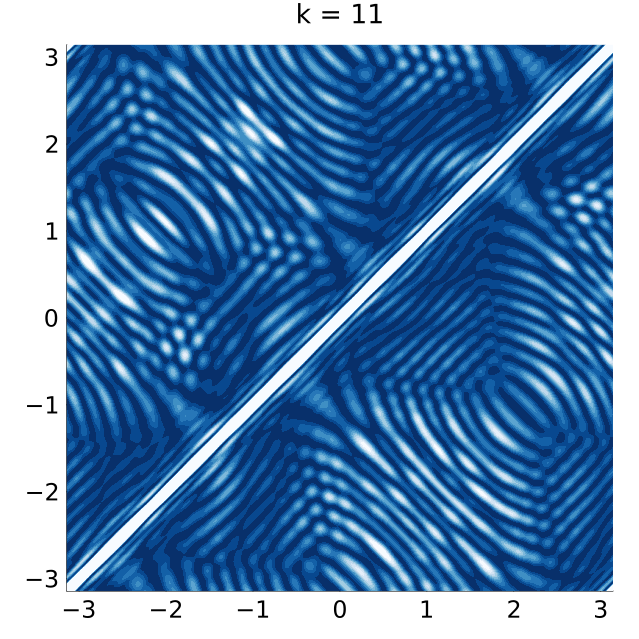} 
    \includegraphics[width=0.48\linewidth]{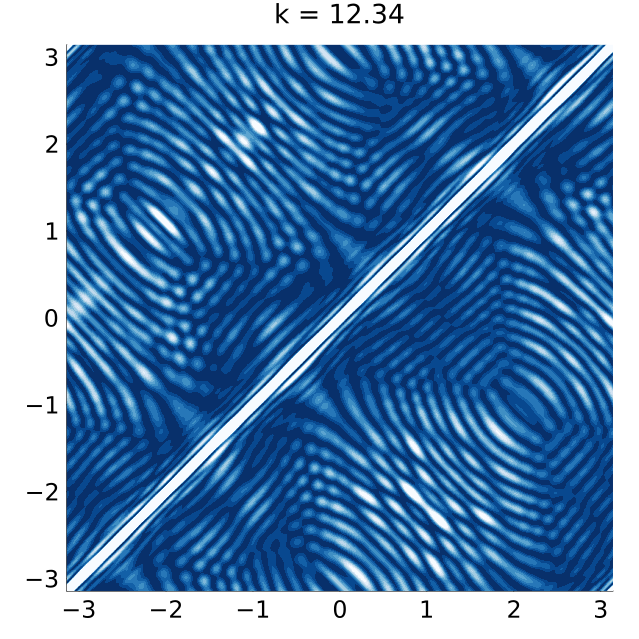} \\
    
    \includegraphics[width=0.48\linewidth]{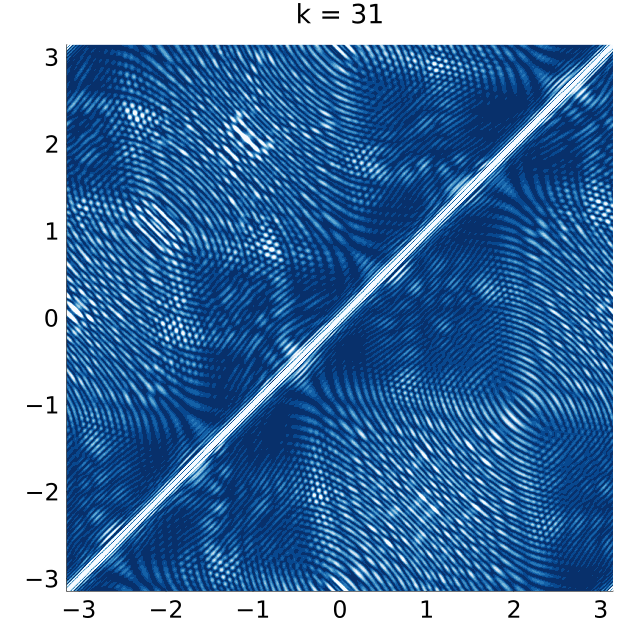}\includegraphics[width=0.48\linewidth]{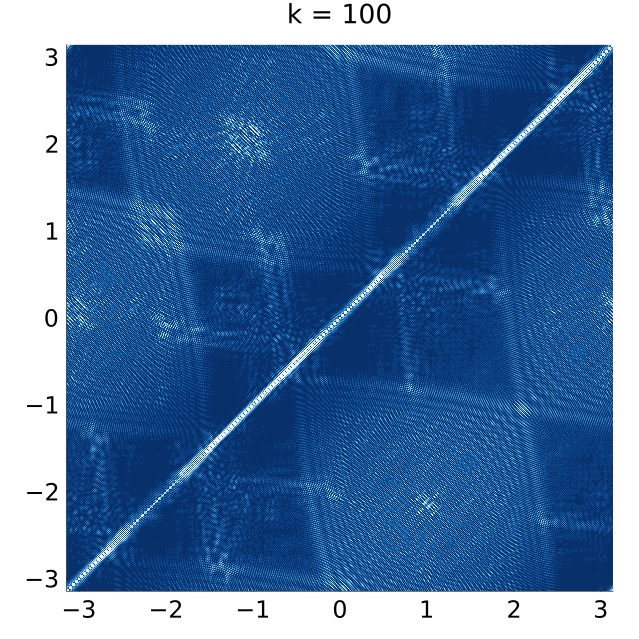}
    \caption{The differential cross-section for the system with $R_i = (1,1.2,0.8)$ and $L=3$ as a function of both angles, for different values of $k$. One value was chosen to be near the resonance located at $k \approx 12.336 -0.327i$.}
    \label{fig:Kdependence}
\end{figure}

In this appendix we include several supplementary figures to those found in section \ref{tddotqs}, plotting the differential cross-section of the pinball system as a function of the incoming and outgoing angles for different configurations.

In figure \ref{fig:Kdependence} we plot the function for the asymmetric configuration for different values of $k$, showing the effect of increasing $k$. At larger $k$ we see ``higher resolution'' images with a larger number of peaks.

\begin{figure}[tp!]
    \centering
    \includegraphics[width=0.48\textwidth]{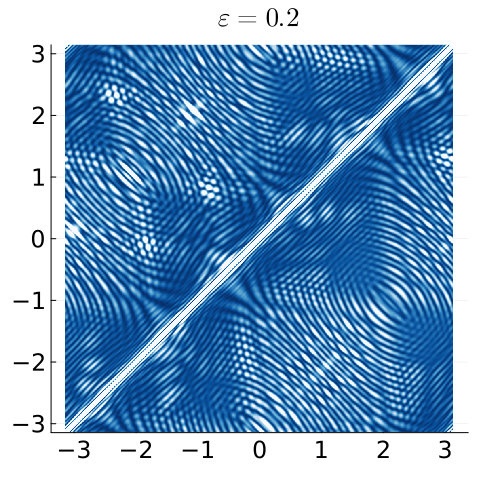}
    \includegraphics[width=0.48\textwidth]{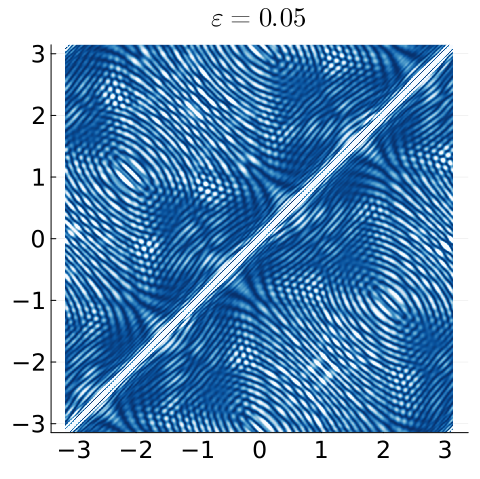} \\
    \includegraphics[width=0.48\textwidth]{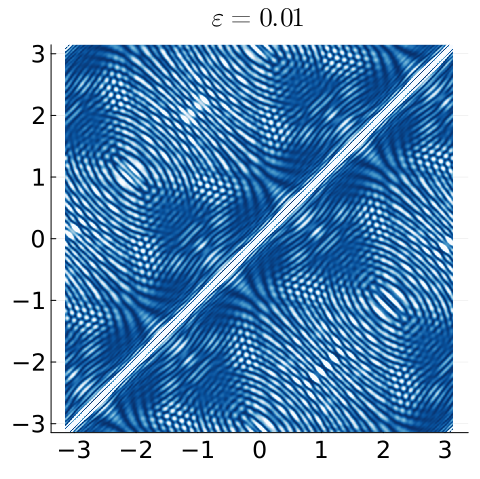}
    \includegraphics[width=0.48\textwidth]{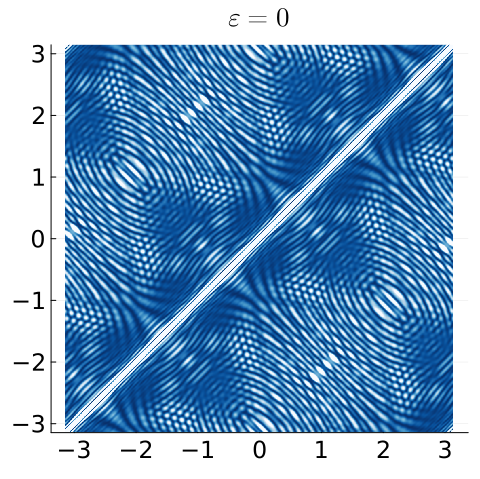}
    \caption{The differential cross-section for the setup with $R_1 = 1$, $R_2 = 1+\epsilon$, $R_3 = 1-\epsilon$, for various values of $\epsilon$, at $k=25$.     \label{radiir}}
\end{figure}

In figure \ref{radiir} we gradually remove the asymmetry in the radii by taking $R_1=1$, $R_2=1+\epsilon $ and $R_3=1-\epsilon$, and decreasing $\epsilon$ from 0.2 to 0, the fully symmetric system. Other than restoring the rotation symmetry $(\phi,\phi\pri)\to(\phi+2\pi/3,\phi\pri+2\pi/3)$, there are no obvious differences.

\begin{figure}[tp!]
    \centering
    \includegraphics[width=0.48\linewidth]{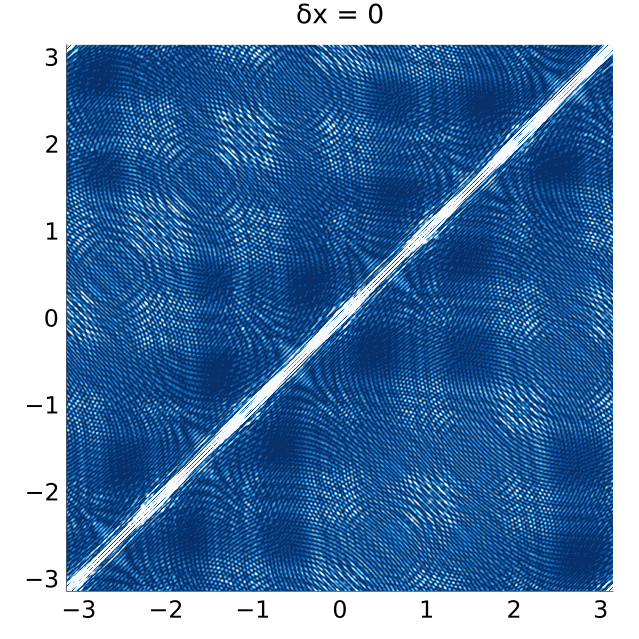} 
    \includegraphics[width=0.48\linewidth]{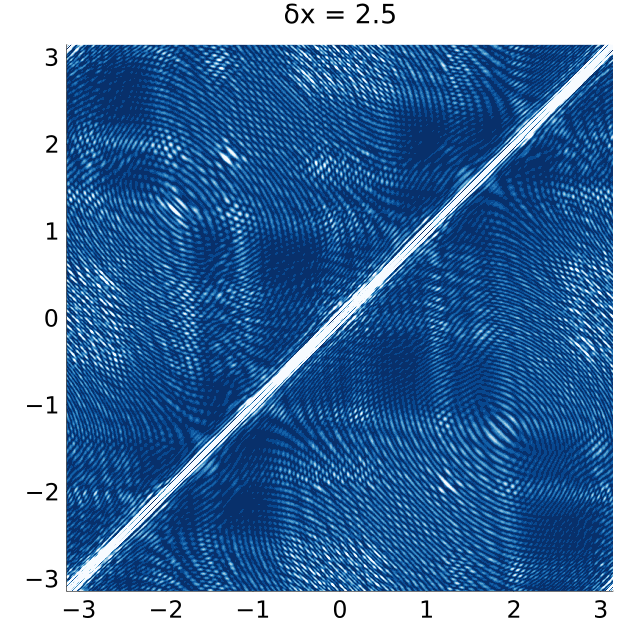} 
    \\
    \includegraphics[width=0.48\linewidth]{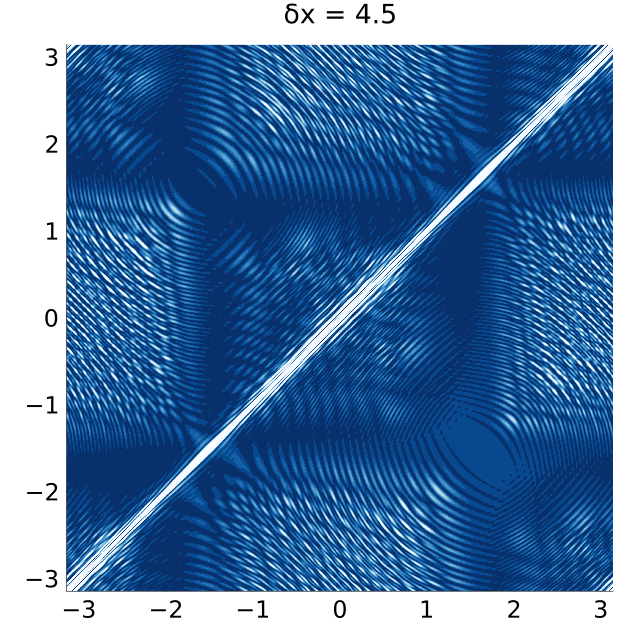} 
    \includegraphics[width=0.48\linewidth]{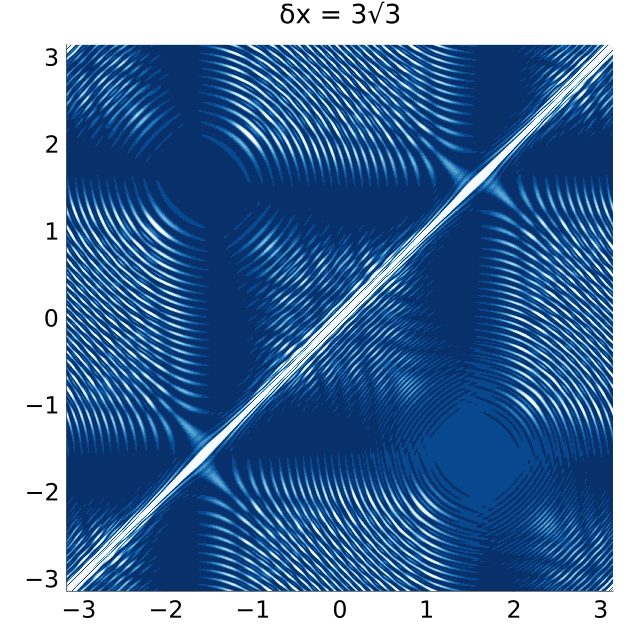} 
    \caption{The differential cross-section for an asymmetric setup with $R_i=(1,1.2,0.8)$. In the initial configuration the three disks are centered on the equilateral triangle with $L=6$. Then, $R_1$ is continuously moved to the left by $\delta x$ until all three disks are on a straight line, which is a non-chaotic system.}
    \label{asysetups}
\end{figure}

In figure \ref{asysetups} we plot another continuous transition for a chaotic to a non-chaotic system. We start from an initial configuration with the three disks of different radii are centered on the equilateral triangle. Then, one of the disks, in our case the first, is continuously moved to the left by $\delta x$ until the centers of all three disks lie on a straight line, which is a non-chaotic system. Note that in this case, unlike the others, we take the original triangle to have sides of length $L=6$, such that the disks do not overlap in the final configuration.

\section{Explicit computation of all spacings distribution for Poisson eigenvalues}\label{ecoasdfp}
In this appendix we review the computation of the distribution of all spacings in the Poisson-Poisson case, by explicitly performing the integral of eq. \eqref{eq:CDFintegral}. We will find the well known result for the distribution of distances of randomly chosen points on the square.

For the Poisson distribution the analysis of the distribution of all spacings drastically simplifies, as all eigenvalues are uniformly distributed: $x_i \sim U(0,N)$, and $y_i \sim U(0,N)$, and all completely independent of each other, so 
\begin{equation}
f_X(x_1,\ldots, x_N) = \prod_{i=1}^N u(x_i)\,,\qquad f_Y(y_1,\ldots y_N) = \prod_{i=1}^N u(y_i)  \end{equation}
where
\begin{equation} u(x) = \begin{cases} \frac{1}{N} & 0 \leq x \leq N \\ 0 & \text{else} \end{cases}\end{equation} 

The integral over $\vlam_3,\ldots\vlam_N$ equals one, so we do not need to write it anymore.

Then to compute $F(\delta)$ we need to compute the overlap of the circle ${\cal C}_\delta(\vlam_1)$ with the square $[0,N]\times[0,N]$ for each point $\vlam_1$, and then integrate the result over $\vlam_1$.

As a warm-up, we can perorm the 1D calculation. The CDF of $\delta_{ij}$ is then given by an integral that we can compute fairly easily:
\begin{align} F_{1D}(\delta) &= \frac{1}{N}\int_{0}^N dx \int_{x-\delta}^{x+\delta} dx^\prime u(x^\prime) \\
 &= \frac{1}{N^2}\int_0^N dx \left(\min(x+\delta,N)-\max(x-\delta,0)\right) \\
 &= \frac{1}{N^2}\left(\int_0^{N-\delta}(x+\delta)\,dx + \int_{N-\delta}^N N\, dx - \int_\delta^{N}(x-\delta) \,dx\right) \\
 &= \frac{\delta}{N}\left(2-\frac{\delta}{N}\right)
\end{align}
such that the PDF of $\delta_{ij} =|x_i-x_j|$ is 
\begin{equation} p_{1D}(\delta) = \frac{2}{N}(1-\frac{\delta}{N}) \end{equation}
which we can verify by measuring the distribution of $\delta_{ij}$ for $N$ Poisson distributed (1D) variables. This distribution obviously shows no repulsion.

Now the 2D version is:
\begin{align} F_{2D}(\delta) &= \frac{1}{N^2}\int_{0}^N dx \int_{x-\delta}^{x+\delta} dx^\prime u(x^\prime) \int_0^N dy \int_{y-\Delta}^{y+\Delta} dy^\prime u(y^\prime) \\
 &= \frac{1}{N^2} \int_0^N dx \int_{x-\delta}^{x+\delta} dx^\prime u(x^\prime) \Delta \left(2-\frac{\Delta}{N}\right) \label{eq:F2DPP2}
\end{align}
where we used the 1D integral to get to the second line. Recall that here
\begin{equation}\Delta^2 = \delta^2 - (x-x^\prime)^2 \end{equation}
We have also implicitly assumed in writing eq. \eqref{eq:F2DPP2} that $\delta\leq N$, which is not necessarily the case in 2D, where the maximum possible spacing is $\sqrt{2}N$. We will return to $\delta\geq N$ later.

The calculation is similar to the 1D case, in that we need to break up the integral depending on the values of $x$ and $\delta$. But now the dependence on $x$ and $x^\prime$ is non-trivial. Using the indefinite integral
\begin{align} i(x^\prime;x,\delta) &\equiv \int dx^\prime \frac{\Delta}{N}\left(1-\frac{\Delta}{N}\right) \\ \nonumber &= \frac{1}{N^2}\left((x-x^\prime)\delta^2 -\frac13(x-x^\prime)^3\right) - \frac{1}{N}\left((x-x^\prime)\Delta + \delta^2 \arctan\frac{x-x^\prime}{\Delta}\right)\end{align}
we can write an expression which can then be evaluated:
\begin{equation} F_{2D}(\delta) = \frac{1}{N^2}\int_0^N dx \bigg(i[x^\prime=\min(x+\delta,N)] - i(x^\prime=\max[x-\delta,0])\bigg) \end{equation}
that yields
\begin{equation} F_{2D}(\delta\leq N) = \pi\left(\frac{\delta}{N}\right)^2 - \frac83 \left(\frac{\delta}{N}\right)^3 + \frac12\left(\frac{\delta}{N}\right)^4\end{equation}
so that the PDF turns out to be
\begin{equation}
    p_{2D}(\delta\leq N) = \frac{1}{N}\left[2\pi\left(\frac{\delta}{N}\right) - 8 \left(\frac{\delta}{N}\right)^2 + 2\left(\frac{\delta}{N}\right)^3\right]
\end{equation}
As already mentioned, so far we assumed that $\delta\leq N$. The full range of $\delta$ is $0\leq\delta\leq\sqrt{2}N$. In the range $N\leq\delta\leq\sqrt{N}$ the calculation is a little more involved, but we can get the answer analytically using the same method. It is:
\begin{equation}
F_{2D}(\delta\geq N) = \frac13 + (\pi-2)\frac{\delta^2}{N^2}-\frac12
\frac{\delta^4}{N^4} + \frac43\frac{\sqrt{\delta^2-N^2}}{N}\left[1+2\frac{\delta^2}{N^2}\right]-4\frac{\delta^2}{N^2}\arctan{\frac{\sqrt{\delta^2-N^2}}{N}}
\end{equation}
with the PDF
\begin{equation} p_{2D}(\delta\geq N) = \frac{1}{N}\left[(2\pi-4)\left(\frac{\delta}{N}\right)-2\left(\frac{\delta}{N}\right)^3+8\frac{\delta}{N}\frac{\sqrt{\delta^2-N^2}}{N} - 8 \frac{\delta}{N}\arctan\frac{\sqrt{\delta^2-N^2}}{N}\right] \end{equation}

The average of this distribution
\begin{equation}
    \langle \delta \rangle = \int_0^{\sqrt{2}N} ds s\, p_{2D}(s) = \frac{2+\sqrt{2}+5\ln(1+\sqrt2)}{15} N \approx 0.521 N
\end{equation}
is known as the ``mean line segment length" of the square. This result and distribution are well known in the field of geometric probability. In fact generalizations exist in the literature for various other shapes, e.g. in \cite{Alagar:1976}. See also \cite{Weisstein} for more references.
\clearpage

\bibliographystyle{JHEP}
\bibliography{SACS}

\end{document}